\documentclass[12pt]{article}
\usepackage{amsmath,amssymb}
\usepackage{epsfig}

\renewcommand{\thefootnote}{\alph{footnote}}

\begin{document}
 
\thispagestyle{empty}
 
\hfill \parbox{4cm}{TPR-97-06}
\begin{center}
 
{\huge The Configuration Space of Low-dimensional Yang-Mills Theories}
 
\vspace{1.5cm}
 
{\large T.~Pause\footnote{e-mail: thomas.pause@physik.uni-regensburg.de
  }\footnote{Supported   by the DFG-Graduiertenkolleg 
  Erlangen-Regensburg ``Physics of Strong Interactions''}
  and  T.~Heinzl\\  \emph{Universit\"at Regensburg, Institut
  f\"ur Theoretische Physik,}\\  \emph{93040 Regensburg, Germany}}

\end{center}
 
\vspace{1.5cm}
 
\begin{abstract}
  We discuss the construction  of the physical configuration space for
  Yang-Mills quantum  mechanics and Yang-Mills  theory on  a cylinder.
  We explicitly eliminate  the redundant degrees  of freedom by either
  fixing a  gauge  or  introducing gauge  invariant   variables.  Both
  methods are shown to be equivalent if the  Gribov problem is treated
  properly  and the  necessary boundary identifications  on the Gribov
  horizon are performed.  In addition, we  analyze the significance of
  non-generic  configurations  and  clarify the   relation between the
  Gribov problem and coordinate singularities.
\end{abstract}
 
\vfill
\newpage
 
\renewcommand{\thefootnote}{\arabic{footnote}}
\setcounter{footnote}{0}
\setcounter{page}{2}

\section{Introduction}
\label{a1}
%
Gauge field theories are  at the heart  of  the standard model  of the
fundamental  interactions.   The weak  coupling phase  of the model is
rather well understood in terms of standard perturbation theory.  This
is sufficient for  the  electro-weak theory  where for the  physically
relevant scales weak and electromagnetic couplings are small.  For the
strong interactions, however,  the  situation is different.   At small
momentum transfer, or large distances, the  associated gauge theory of
color $SU(3)$, quantum chromodynamics (QCD), is in the strong coupling
phase and perturbation theory no  longer works.  One therefore has  to
develop  nonperturbative techniques,  the most  elaborate   one at the
moment         being          lattice           gauge           theory
\cite{wilson:74,creutz:83,muenster:95}.

An alternative approach, based on Bjorken's idea of the femto-universe
\cite{bjorken:82}  has been initiated  by L\"uscher \cite{luescher:83}
and    was    later  elaborated  by    van    Baal  and  collaborators
\cite{koller:88}.   In this approach, one  formulates  QCD in a finite
volume which  in a first step is  kept sufficiently small so that, due
to   asymptotic freedom, perturbation  theory   is still valid.   Upon
enlarging the volume, nonperturbative effects come into play, however,
as is  believed, in a controllable   manner.  Technically, one  uses a
Hamiltonian formulation of QCD  or, neglecting quarks, pure Yang-Mills
theory  in the     Coulomb gauge   \cite{christ:80}.   The   way   the
nonperturbative    effects      show up    is    conceptually   simple
\cite{vanbaal:92a}.   For small  volumes, the  wave functionals behave
essentially as those  in QED, i.e.  they  are concentrated around  the
classical   vacuum.    For  larger  volumes,   the  effective coupling
increases,  the wave functionals start to  spread out in configuration
space and  become sensitive to its  boundaries and nontrivial geometry
\cite{heuvel:94}.  It  is therefore  crucial for the  understanding of
these effects to learn as much as possible  about the structure of the
configuration space.  Let us illuminate this reasoning with an example
from quantum mechanics.   For a particle  in an infinitely deep square
well of size $d$ there  is a gap between  the ground and first excited
state of order $1/d^2$.  Obviously, the existence of the finite energy
gap   is directly related  to the  finite volume  of the configuration
space.   Similar arguments have been given  by  Feynman to explain the
origin  of   the mass  gap for   Yang-Mills  theory in  2+1 dimensions
\cite{feynman:81}      and      are currently   being  re-investigated
\cite{karabali:96}.

Let us discuss the  case of non-Abelian gauge  theories \cite{yang:54}
in more  detail.   The   configuration  space $\mathcal A$     of pure
Yang-Mills    theory  is given    in  terms   of    the gauge   fields
(``configurations'') $A(x)$, which under the action of the gauge group
$\mathcal G$ transform as
\begin{equation}
  ^U{\!A} = \, U^{-1} A \, U + i \: U^{-1} \mathrm{d}U
  \quad\text{with}\quad U \in {\mathcal G} \; .
  \label{Gl-101}
\end{equation}
The  set  of   all gauge   equivalent  points $^U{\!A}$    of a  given
configuration  $A$  constitutes  the orbit of  $A$.   Gauge invariance
requires    physical  quantities to  take  the    same value for every
configuration $^U{\!A}$  on the orbit   of  $A$.  In this sense,   the
description  of gauge theories    in terms of   the potentials  $A$ is
somewhat uneconomic as there   is a  huge redundancy associated   with
these variables. One  way to  see  this is the  infinite volume factor
they  contribute to the    path  integral measure.   It is   therefore
desirable to   find   the  set   of all   gauge    \emph{inequivalent}
configurations, i.e. the space of gauge orbits
\begin{equation}
  \mathcal M := \mathcal A/\mathcal G \; ,
  \label{Gl-102}
\end{equation}
which  we  will refer to  as   the \emph{physical} configuration space
$\mathcal M$. The interesting question, of course, is, how to actually
find $\mathcal  M$. A first hint  can be obtained from (\ref{Gl-102}),
which can naively be ``solved'' for $\mathcal A$ yielding
\begin{equation}
  \mathcal A \sim \mathcal M \times \mathcal G \; .
  \label{Gl-103}
\end{equation}
Though at this point it is unclear in which sense this identity really
holds, it nevertheless suggests   that the large  configuration  space
$\mathcal   A$ of gauge potentials  should   be decomposed into  gauge
invariant  quantities   from $\mathcal  M$    and  gauge variant  ones
parameterizing group elements $ U \in  \mathcal G$.  The decomposition
indicated  in  (\ref{Gl-103}) can   explicitly be  achieved  using the
transformation law  (\ref{Gl-101}):   parameterize the group  elements
$U[\varphi]$    with an appropriate    collection  of angle  variables
$\varphi$ such  that $U[\varphi \!  = \!   0] = 1  \!\!\!\: \text{l}$,
then   pick a  representative $\tilde A$    on any  orbit  and rewrite
(\ref{Gl-101}) as
\begin{equation}
  A [\tilde A, \varphi] \, := \, ^U{\! \tilde A} = 
  U^{-1}[\varphi] \, \tilde A \: U[\varphi] + 
  i \: U^{-1}[\varphi] \: \mathrm{d}U [\varphi] \; .
  \label{Gl-104}
\end{equation}
Thus, any gauge potential  $A$  carries an (implicit) label  $\varphi$
which determines the position of $A$ on  its orbit, in particular $A =
\tilde A$ for $\varphi =   0$.  The identity (\ref{Gl-104}) defines  a
map $(\tilde A, \varphi) \mapsto  A$ which provides (at least locally)
the  decomposition   of an arbitrary  configuration   $A$ into a gauge
invariant  representative $\tilde  A $  and the   gauge variant angles
$\varphi$.  In  general, this map  will be  a transformation  from the
cartesian  coordinates  $A$ to \emph{curvilinear} coordinates $(\tilde
A, \varphi)$ \cite{christ:80,cheng:87}.

Usually,  the representative  $\tilde A$  of the  orbit is  chosen via
\emph{gauge fixing}, i.e. by defining functionals  $\chi$ on $\mathcal
A$ such that
\begin{equation}
  \chi[\tilde A \,] = 0 \; ,
  \label{Gl-105}
\end{equation}
This defines a  hypersurface $\Gamma_\chi \in {\mathcal A}$ consisting
of all the representatives $\tilde A$ (or fields in the gauge $\chi \!
=  \! 0$).   There are  two  requirements that  have to be   met by an
admissible gauge  fixing:   \emph{existence}    and \emph{uniqueness}.
Existence   means that on \emph{any}  orbit  there is a representative
satisfying the  gauge condition.   Thus, for  any  $A \in  \mathcal A$
there has to be a solution $U(\varphi)$ of the  equation $^U{\!A}\!  =
\!   \tilde A $ with  $\chi[\tilde A\,] \! =  \! 0$.  The criterion of
uniqueness is satisfied if  on  each orbit  there is  only  \emph{one}
representative obeying the gauge  condition.   If, on the other  hand,
there  are (at least)  two    gauge equivalent fields, $\tilde   A_1$,
$\tilde  A_2$,  satisfying the gauge   condition,   the gauge  is  not
completely fixed.  Instead, there is a residual gauge freedom given by
the gauge transformation $V$ connecting the copies, $\tilde  A_2 \!  =
\!  ^V{\!  \tilde  A}_2$.  In terms  of the angles $\varphi$ existence
and uniqueness mean  that there is one and  only one solution $\varphi
\!  =  \!  0$ such  that $\chi[A(\varphi)] = 0$.   As  shown by Gribov
\cite{gribov:78}, for   infinitesimal $\varphi$  this  amounts to  the
condition that the Faddeev-Popov determinant,
\begin{equation}
  \bigtriangleup_{\scriptscriptstyle \text{FP}} := 
    \left| \; \frac{\; \delta \chi \; }{\; \delta \varphi \; }
    \; \right|_{\varphi = 0} \; ,
\label{Gl-106} 
\end{equation}
should be  non-vanishing. In this paper,  we  will concentrate  on the
transformation (\ref{Gl-104}).  Therefore, it is more natural to study
the Jacobian  $\det         J$   of  (\ref{Gl-104})     instead     of
$\bigtriangleup_{\scriptscriptstyle \text{FP}}$. The relation of  both
quantities is obtained via the chain rule,
\begin{equation}
  \det J \: \Big|_{\varphi=0}:= 
  \left| \frac{\delta A}
              {\delta (\tilde A, \varphi)}\right|_{\varphi=0} =
  \left| \frac{\delta A}
              {\delta (\tilde A, \chi)}\right|_{\chi=0} \!\!\!\cdot
  \left| \frac{\delta (\tilde A, \chi)}
              {\delta (\tilde A, \varphi)}\right|_{\varphi=0} =
  \left| \frac{\delta A}{\delta (\tilde A, \chi)}\right|_{\chi=0}
  \!\!\! \cdot \bigtriangleup_{\scriptscriptstyle \text{FP}} \; .
  \label{Gl-107}
\end{equation}
In what follows we will always work in a Hamiltonian formulation using
the Weyl  gauge,  $A_0  =  0$,  which  allows  for a   straightforward
quantization  \cite{jackiw:80}.   The discussion above  remains valid;
one merely has to replace $A$ by its three-vector part $\vec{A}$.

For   QED,  the construction   of   the physical  configuration  space
$\mathcal M$ is  rather straightforward, as  gauge transformations are
basically  translations that  preserve   the cartesian  nature  of the
coordinates. Explicitly, (\ref{Gl-104}) becomes
\begin{equation}
  \vec{A}\, [\vec{A}_\perp, \varphi\,] = 
  \vec{A}_\perp + \nabla \varphi \; .
  \label{Gl-108}
\end{equation}
Thus, a natural representative  $\tilde A$ is  given by the transverse
photon field $\vec{A}_\perp$ (Coulomb gauge), and the angle $\varphi =
\nabla \cdot \vec{A}/ \Delta$ is in one-to-one correspondence with the
gauge   variant  longitudinal gauge   field  (for fields  vanishing at
spatial  infinity).  The  physical  configuration  space consisting of
transverse gauge potentials  is Euclidean,  i.e.  flat and  unbounded.
This gives    another explanation of why there  is no mass gap for the
photon so that it stays massless \cite{feynman:81}.

The situation becomes much    more complicated for  non-Abelian  gauge
theories.  At variance  with QED the decomposition (\ref{Gl-104})  now
involves curvilinear coordinates.  It   turns  out that in this   case
(\ref{Gl-103}) does not  hold in a global sense  as was first shown by
Gribov  \cite{gribov:78}  and  Singer \cite{singer:78}.   To  be  more
specific consider the following example, which we will refer to as the
Christ-Lee  model    \cite{christ:80,cheng:87,prokhorov:91,heinzl:97}.
This  model   describes the motion  of   a particle in  a   plane with
coordinates  $x$  and $y$  which is   the large  configuration  space,
$\mathcal A  = \mathbb{R}^2 $.   Let the  gauge transformations be the
rotations  around the origin.   If we introduce polar coordinates, the
radius $r$ and the angle $\varphi$, it is obvious  that the radius $r$
is gauge invariant whereas $\varphi$, parameterizing the rotations, is
gauge variant.  The decomposition of  $\mathcal A$ (denoting  $A \!  =
\!  (x,y)$) is thus given by the transformation
\begin{equation}
  A(r, \varphi) = (r \, \cos \varphi, r \, \sin \varphi ) \; .
  \label{Gl-109}
\end{equation}
Accordingly, the physical configuration space is the non-negative real
line
\begin{equation}
  \mathcal{M}  = \mathbb{R}_0^+ = \mathbb{R}^2 / SO(2) \; .
  \label{Gl-110}
\end{equation}
Let  us  assume now that we  are  not as smart as   to guess the gauge
invariant  variable  and proceed in   a  pedestrian's manner via gauge
fixing.  We gauge  away  $y$, $\chi(A)  := y   = 0$,  and  immediately
realize  that this  gauge  selects \emph{two}  representatives on each
orbit  at $\pm x$.  There  is a \emph{discrete} residual gauge freedom
between the copies,  $x   \to - x$,   which constitutes  the  ``Gribov
problem'' for the example at  hand.  If we calculate the Faddeev-Popov
determinant,
\begin{equation}
  \bigtriangleup_{\scriptscriptstyle \text{FP}}  = 
  \left|\; \frac{\; \partial \chi \;}
                {\; \partial \varphi \;} \; \right|_{\varphi = 0}
  = \: x \; ,
  \label{Gl-111}
\end{equation}
we find  that it vanishes at $x=0$, the  ``Gribov horizon'',  which is
just the  point separating the two  gauge equivalent regions $x>0$ and
$x<0$.  Only if we  fix the gauge completely by  demanding that $x$ be
non-negative we again have the non-negative real line as the  physical
configuration space and can identify $x$ with the radius $r$. Denoting
the representative satisfying $\chi \! = \! 0$ as $\tilde A = (r,0)$,
we obtain the transformation analogous to (\ref{Gl-104}),
\begin{equation}
  \begin{matrix}   
  A(r, \varphi) = \tilde A(r) \: U(\varphi) = 
  \big( \, r \;\;  0 \, \big) \\ \quad 
  \end{matrix} \!
  \begin{pmatrix} 
     \quad \cos \varphi \;& \sin \varphi \\
       -   \sin \varphi \;& \cos \varphi
  \end{pmatrix} \; ,
  \label{Gl-112}
\end{equation}
which is equivalent to   the  decomposition (\ref{Gl-109}).   In  this
simple case,   the  Jacobian of (\ref{Gl-112})  is    identical to the
Faddeev-Popov  determinant   (\ref{Gl-111}).    We   point  out   that
$\mathcal{M}$ has a boundary point, the origin, which is a fixed point
under the  action of the gauge  group. In  field theory such partially
gauge  invariant configurations  \cite{yabuki:96} are called reducible
\cite{singer:78}.  For our simple   models, however, we will  use  the
term ``non-generic''  instead,  to  describe configurations  that  are
invariant  under subgroups of     $\mathcal  G$. Note  that    in  the
Christ-Lee model a single  coordinate system suffices to  parameterize
the whole physical configuration space $\mathcal{M}$. This is not true
in general as will be discussed in a moment.

The  example  above also raises another   question. For  $SU(2)$ gauge
field  theory  several types of  gauge  invariant  variables have been
proposed    \cite{simonov:85,haagensen:95a}).   In  the  case  of  the
Christ-Lee model we were able  to ``guess'' a gauge invariant variable
and after that found a gauge fixing and a representative corresponding
to  this particular choice of  a gauge invariant  variable.  One might
therefore ask whether it is generally true that to any construction of
gauge invariant   coordinates  there corresponds  a particular   gauge
fixing.  We will address this question in the following sections.

It  may also   happen that  the residual gauge   freedom is continuous
instead of  just   discrete.  In this  case   there are  whole  orbits
contained in the  gauge fixing hypersurface, $\Gamma_\chi$, which  are
located  at the Gribov  horizon.  A prominent   example is provided by
axial-type gauges,  $n \cdot A =0$,  where the residual  gauge freedom
consists of all  gauge transformations independent of  $n \cdot x$. To
proceed, one  generally   has   to  impose   \emph{additional}   gauge
conditions to eliminate  the continuum of  Gribov copies. In  this way
one identifies gauge equivalent points on  the Gribov horizon.  As the
latter  seems  to constitute (part of)  the  boundary of  the physical
configuration space the     described procedure  is  referred  to   as
``boundary                                           identifications''
\cite{vanbaal:95a,zwanziger:93,zwanziger:96}.     It is  due  to these
identifications   that    the  nontrivial  topology  of   the physical
configuration  space comes into play,  indicated  by the fact that one
needs more than one coordinate system to cover  $\mathcal M$.  We will
discuss several  examples where boundary identifications are necessary
and explicitely show how they are related to the topology of $\mathcal
M$.

In  general, we expect the features  discussed above  to also arise in
Yang-Mills field theory.  Of course there are additional complications
due to the infinite number of degrees of freedom  and the necessity of
renormalization. Nevertheless, since  Gribov's original work there has
been much progress in determining the physical configuration space, in
particular by  using the  Coulomb gauge.  In  this  particular  case a
certain distance functional turned  out to be  a very powerful tool to
characterize                           $\mathcal                    M$
\cite{vanbaal:92a,zwanziger:93,franke:82,weisberger:83}.   Due  to the
complicated   nature of the  functional,  however,  the  set of  gauge
inequivalent configurations is only approximately known.  A variant of
the  method  also   seems to   work  for the    maximal abelian  gauge
\cite{thooft:81} used to analyze the condensation of abelian monopoles
and confinement due to a dual Meissner effect. Within lattice studies,
in  particular,  the  influence    of   Gribov  copies  on  the   dual
superconductor scenario has been studied \cite{bali:96,hart:97}.

At the moment, however, it is unclear how the same configuration space
(which of course, by construction,  has a gauge invariant meaning) can
be  obtained  in  different  gauges.  The   method with the   distance
functional, for  example,   does   not work  in    axial-type  gauges.
Furthermore, for   the    maximal   abelian   gauges,   the   physical
configuration space has not been  determined. We therefore consider it
worthwhile  to go back  to  quantum mechanics and  a  finite number of
degrees  of  freedom. In the spirit   of a recently  presented soluble
gauge  model \cite{friedberg:96}  we  will  address  the question   of
finding  the physical configuration space via   (i) different types of
gauge  fixings, (ii)   constructing gauge  invariant variables without
gauge fixing and (iii) relating these two methods.

The paper is  organized as follows.  In Section  \ref{a2} we discuss a
simple version of $SU(2)$ Yang-Mills quantum mechanics where the gauge
group is reduced to $SO(2)$.    We will explicitly show the   relation
between the gauge   fixing method and   the method of gauge  invariant
variables.     We  will    also   perform  the    necessary   boundary
identifications  and visualize  the resulting  physical  configuration
space   $\mathcal M$  by     means of   a  suitable  embedding    into
$\mathbb{R}^3$.  Section \ref{a3}  is  mainly devoted to the  study of
non-generic configurations for  the structure group $SO(3)$.   It will
be shown, how these configurations give  rise to a genuine boundary of
the physical configuration space $\mathcal M$.  As in Section \ref{a2}
we will  compare the spectra of the  Hamilton operators defined on the
gauge fixing surface and on  $\mathcal M$, showing the equivalence  of
both.   In Section \ref{a4}  we will discuss $SU(2)$ Yang-Mills theory
on a cylinder,  which also reduces to a  quantum mechanical model.  We
will apply the methods used in the preceding sections to construct the
physical configuration  space of this  model and study the non-generic
configurations.
%
%
\section{$SO(2)$ Yang-Mills theory of constant fields}
\label{a2}
%
%
The  first  model we  want  to  discuss is   defined by the Lagrangian
\begin{equation}
  {\mathcal L}^{2 \! \times \! 2} = \frac{1}{2\,g^2} \sum_{i,a=1}^2 
    \big( \dot a^a_i - a_0 \, \epsilon^{ab}a^b_i \big)
    \big( \dot a^a_i - a_0 \, \epsilon^{ac}a^c_i \big) 
    - {\mathcal V}^{2 \! \times \! 2}(a_i^a) \;,
  \label{Gl-301}
\end{equation}
with the antisymmetric   tensor $\epsilon^{ab}$.  The  special form of
the kinetic  term  in  (\ref{Gl-301}) stems  from  the  covariant time
derivative in $SU(2)$ Yang-Mills theory for spatially constant fields.
Since the lower indices  $i, j, \ldots$ and the  upper indices  $a, b,
\ldots$ of the  basic variables $a_i^a$ only  take the  values $1$ and
$2$ each, we will call  our model the  ``$2 \!  \times \!  2$-model''.
For the time being we interpret ${\mathcal L}^{2 \!  \times \!  2}$ as
the Lagrangian   describing   the motion  of   two  ``particles'' with
position vectors $\vec a_1 \!  = \!  a_1^a \,  \hat e_a$ and $\vec a_2
\!  = \!   a_2^a \, \hat  e_a$ in  a ``color'' plane  with orthonormal
basis  vectors   $\hat e_1, \hat   e_2$  under   the influence of  the
potential ${\mathcal V}^{2  \!  \times \!  2}$\cite{prokhorov:91}.  We
choose the potential ${\mathcal V}^{2 \!  \times \!  2}$ such, that it
is invariant under
\begin{equation}
  a^a_i \mapsto a^b_i \: U^{ba} \; ,
  \label{Gl-302a}
\end{equation}
where we parameterize the rotation matrix $U \in SO(2)$ by
a time-dependent angle $\phi(t)$
\begin{equation}
  U \big[ \phi(t) \big] :=
  \begin{pmatrix} \quad \cos \phi(t) & \;\; \sin \phi(t) \\
                    -   \sin \phi(t) & \;\; \cos \phi(t)
  \end{pmatrix} \; .
  \label{Gl-303}
\end{equation}
For example, we may take a Yang-Mills type potential
\begin{equation}
  {\mathcal V}^{2 \! \times \! 2}_{\scriptscriptstyle \text{YM}} =
  \frac{1}{2 g^2} \, \left(a_1^1 a_2^2 - a_1^2 a_2^1\right)^2
  \label{Gl-304a}
\end{equation}
or the harmonic oscillator form
\begin{equation}
  {\mathcal V}^{2 \! \times \! 2}_{\text{osc}} =
  |\vec a_1|^2 + |\vec a_2|^2 \; .
  \label{Gl-304b}
\end{equation}
Having chosen  such  a potential, we find,   that ${\mathcal  L}^{2 \!
  \times \!  2}$ is invariant  under the  combination of  the  $SO(2)$
transformations (\ref{Gl-302a}) and
\begin{equation}
  a_0 \mapsto a_0 - \partial_t \phi
  \label{Gl-302b}
\end{equation}
Hence ${\mathcal L}^{2  \!  \times \!  2}$  in fact describes  a gauge
model   with the   abelian gauge  group    $SO(2)$.   Interpreting the
transformations (\ref{Gl-302a}) as rotations  of the coordinate system
($\hat e_1, \hat e_2$), we realize that gauge invariance in our simple
model means, that   the physical motion  of  the two ``particles''  at
positions   $\vec a_1$,  $\vec  a_2$ has   to be   independent  of the
(time-dependent) orientation  of the  coordinate  axes.  We will  find
that the correct implementation   of  this condition  will  eventually
spoil  our interpretation   of $\vec   a_1$ and   $\vec a_2$  as   the
coordinates of independent particles.

As   pointed  out  in   the   introduction,   invariance under   gauge
transformations  (\ref{Gl-302a}) and (\ref{Gl-302b}) implies, that the
space $\mathcal A$ of all    configurations ($a_i^a, a_0$)    contains
redundant (unphysical) degrees   of   freedom.  We will realize    the
reduction of   $\mathcal   A$ to  the   physical   configuration space
$\mathcal M$ using  a   Hamiltonian formalism. Denoting  the   momenta
canonically  conjugate  to the   coordinates $a_i^a$ by  $e^{ia}$ (the
canonical momentum for $a_0$ vanishes) we get
\begin{equation}
  {\mathcal H} = \frac{g^2}{2} \sum_{i,a=1}^2 e^{ia} e^{ia} 
  - a_0 \, {\mathcal G} +
  {\mathcal V}^{2 \! \times \! 2}_{\scriptscriptstyle \text{YM}} \;,
  \quad {\mathcal G} = \epsilon^{ab} \, a^a_i \,e^{ib} \; ,
  \label{Gl-305}
\end{equation}
where we have  put in the potential  (\ref{Gl-304a}). The condition of
gauge invariance is now expressed by  the Gau{\ss} constraint equation
${\mathcal G}=0$, following from  the Lagrangian equations  of motion.
In the particle picture we interpret $\mathcal G$ as the total angular
momentum,  which   has to  vanish by   gauge invariance.  The variable
$a_0$,  besides  being  the   Lagrange multiplier   of  the constraint
$\mathcal G$, may be interpreted as the angular velocity of a rotating
coordinate system \cite{lee:81}.  Because the physical quantities have
to be independent  of the rotation  of the  coordinate system,  we are
allowed  to set  $a_0=0$ (``body-fixed  frame'' \cite{lee:81}).   This
amounts to applying the gauge (fixing) transformations (\ref{Gl-302a})
and (\ref{Gl-302b}) with angle
\begin{equation}
  \phi(t) = \varphi + \int_0^t a_0(\tau) \: \mathrm{d}\tau \; .
  \label{Gl-306}
\end{equation}
For  Yang-Mills field theory this  would correspond to  the Weyl gauge
$A_0=0$, which does not fix  gauge  transformations constant in  time.
We will denote the group  of time-independent gauge transformations as
${\mathcal G}_0$.  In  our  model these residual   transformations are
parameterized   by    the  undetermined   time-independent integration
constant $\varphi$ in (\ref{Gl-306}), which provides   the orientation
of the coordinate system at $t=0$,
\begin{equation}
  \begin{pmatrix} \tilde a_1^1 & \;\tilde a_1^2 \\ 
                  \tilde a_2^1 & \;\tilde a_2^2 \end{pmatrix} = 
  \begin{pmatrix} a_1^1 & \; a_1^2 \\ 
                  a_2^1 & \; a_2^2 \end{pmatrix}
  \begin{pmatrix} \quad \cos \varphi & \; \sin \varphi \\
                  -  \sin \varphi & \; \cos \varphi \end{pmatrix}\; .
  \label{Gl-307a}
\end{equation}
The ``Weyl-gauge'' Hamiltonian,
\begin{equation}
  {\mathcal H}^{2 \! \times \! 2} = \frac{g^2}{2} 
  \sum_{i,a=1}^{2} e^{ia} e^{ia} +
  {\mathcal V}^{2 \! \times \! 2}_{\scriptscriptstyle \text{YM}} \; ,
  \label{Gl-307}
\end{equation}
only  depends  on  the variables  $e^{ia}$   and $a_i^a$.   Assuming a
Euclidean metric on $\mathcal A$, the $a_i^a$ form a pre-configuration
space   ${\mathcal A}_0$  homeomorphic   to  $\mathbb{R}^4$   with the
Euclidean metric
\begin{equation}
  \label{Gl-308a}
  g =  g_{ij}^{ab} \: \mathrm{d}a_i^a \mathrm{d}a_j^b \quad 
  \text{with} \quad g_{ij}^{ab} = \delta_{ij}\: \delta^{ab} \; .
\end{equation}
Therefore  we can   canonically quantize our   model by  replacing the
Poisson brackets with quantum mechanical commutators,
\begin{equation}
  \left[ e^{ia}, a^b_j \right] =
  -\, \mathrm{i} \: \delta^i_{\,j} \, \delta^{ab} \; .
  \label{Gl-308}
\end{equation}
Accordingly, we promote functions    on the phase space to   operators
acting on  a Hilbert space, in particular  ${\mathcal  G} \mapsto \hat
G$.   Within   the Hamiltonian   formalism,  the  Gau{\ss}  constraint
equation,  ${\mathcal G}     = 0$,  can   only    be  realized  weakly
\cite{dirac:50} on  the  Hilbert   space  of physical  states   $|\Psi
\rangle_{\text{phys}}$,
\begin{equation}
  \hat G \; |\Psi \rangle_{\text{phys}} = 0 \; .
  \label{Gl-309}
\end{equation}
In  the Schr\"odinger representation the  Hamilton  operator acting on
wave  functions $\Psi(a)= \langle  a | \Psi \rangle$  is  given by the
Laplacian on  the Euclidean pre-con\-fi\-gu\-ra\-tion space ${\mathcal
  A}_0$ and the Yang-Mills potential
\begin{equation}
  \hat H^{2 \! \times \! 2} = -\frac{\;g^2}{2} \; 
    \frac{\partial^2}{\partial a^a_i \, \partial a^a_i} + 
    {\mathcal V}^{2 \! \times \! 2}_{\scriptscriptstyle \text{YM}}(a)
  \;.
  \label{Gl-310}
\end{equation}
As discussed in  the introduction there are  several ways to eliminate
the residual gauge symmetry and thus obtain the physical configuration
space  $\mathcal   M$    on   which  the   physical    wave  functions
$\Psi_{\text{phys}}(a)=\langle a     | \Psi \rangle_{\text{phys}}$ are
defined. To  begin with,  we will  analyze a  gauge condition which we
will refer to as ``axial gauge''.
%
%
\subsection{Axial gauge}
%
%
Since   we have to  eliminate  one gauge  degree  of freedom, the most
straightforward condition is to set one of  the $a_i^a$ equal to zero.
Thus, we demand the ``axial gauge'' condition
\begin{equation}
  \chi_{\text{ax}}(\tilde a^a_i) := \tilde a_1^2 = 0 \; ,
  \label{Gl-311}
\end{equation}
which    determines   a    three-dimensional   gauge  fixing   surface
$\Gamma_{\text{ax}} \subset {\mathcal   A}_0$. For any   configuration
$a=(a_1^1, a_1^2,   a_2^1,   a_2^2)$   there  is  a   gauge   (fixing)
transformation which  maps $a$ onto  a point $(\tilde a_1^1, 0, \tilde
a_2^1, \tilde a_2^2)$ on  $\Gamma_{\text{ax}}$.   The inverse of  this
map is given explicitly by
\begin{equation}
  \begin{pmatrix} a_1^1 & \; a_1^2 \\ 
                  a_2^1 & \; a_2^2 \end{pmatrix} = 
  \begin{pmatrix} \tilde a_1^1 &  \;      0      \\
                  \tilde a_2^1 & \; \tilde a_2^2  \end{pmatrix} 
  \begin{pmatrix} \cos \varphi & \;  -   \sin \varphi \\
                  \sin \varphi & \;\quad\cos \varphi \end{pmatrix}
  . \label{Gl-312}
\end{equation}
We  interpret  equation (\ref{Gl-312}) as  the  transformation $a^a_i(
\tilde a_i^a, \varphi)$ from  the Euclidean coordinates $a_1^1, a_1^2,
a_2^1, a_2^2$ to curvilinear coordinates $ \tilde a_1^1, \tilde a_2^1,
\tilde a_2^2, \varphi$.   Multiplying equation (\ref{Gl-312}) on  both
sides  from  the left with  $U(\bar  \varphi)$, we find that $\varphi$
gets shifted to  $\varphi  -   \bar \varphi$, whereas  the   variables
$\tilde  a_i^a$     remain unchanged.  Therefore    the transformation
(\ref{Gl-312}) explicitly   realizes the separation  of gauge  variant
from gauge  invariant degrees of freedom. If  this map was one-to-one,
we would have found  an homeomorphism ${\mathcal A}_0 \cong  {\mathcal
  M}  \times {\mathcal G}_0$, where  we  have identified the  physical
configuration   space $\mathcal  M$   with the  gauge  fixing  surface
$\Gamma_{\text{ax}}$ given in terms of  the gauge invariant  variables
$\tilde  a_i^a$.  However,  it  has been shown that   in general it is
impossible to  write   ${\mathcal A}_0$  as  a trivial    fibre bundle
${\mathcal M} \times {\mathcal  G}_0$ \cite{singer:78}.  Hence, let us
study the map (\ref{Gl-312}) in more detail  by examining its Jacobian
matrix $J$, in particular the zeros of the Jacobian $\det J$ evaluated
at $\varphi=0\,$,
\begin{equation}
  \det J = \left| \frac{\partial(a_1^1, a_1^2, a_2^1, a_2^2)}
       {\partial (\tilde a_1^1, \tilde a_2^1, \tilde a_2^2, \varphi)}
          \right| = - \, \tilde a_1^1 \;.
  \label{Gl-313}
\end{equation}
We find that $\det J$ vanishes for $\tilde  a_1^1 = 0$ indicating that
the map (\ref{Gl-312}) may not be one-to-one. In the following we will
demonstrate   how this is related  to  the existence of residual gauge
copies. As in  the case of the Christ-Lee  model $\det J$ is  equal to
the    Faddeev-Popov determinant   $\bigtriangleup_{\scriptscriptstyle
  \text{FP}}$ modulo a possible sign change.

Intuitively the  gauge condition (\ref{Gl-311}) means  that  we rotate
the coordinate system in color space such  that  the vector $\vec a_1$
is collinear  to the     $\hat e_1$-axis.    There are clearly     two
possibilities for this to happen: one  where $\vec a_1$ is parallel to
$\hat e_1$ and the other, where $\vec a_1$ is anti-parallel.  In terms
of  the transformation    (\ref{Gl-312}) we    find   that a     given
configuration $a \in {\mathcal A}_0$ may be represented by two sets of
coordinates $(\tilde   a_1^1, \tilde a_2^1,  \tilde  a_2^2, \varphi)$,
since
\begin{equation}
  \begin{pmatrix} \tilde a_1^1 & \; 0 \\ 
                  \tilde a_2^1 & \; \tilde a_2^2 \end{pmatrix}\!
  \begin{pmatrix} \cos \varphi & \;   -  \sin \varphi \\
                  \sin \varphi & \;\quad \cos \varphi \end{pmatrix} =
  \begin{pmatrix}-\tilde a_1^1 & \; 0 \\
                 -\tilde a_2^1 & \; -\tilde a_2^2  \end{pmatrix}\!
  \begin{pmatrix} \cos(\varphi \!-\! \pi) & \! 
                 -\sin(\varphi \!-\! \pi)\\
  \sin(\varphi \!-\! \pi) & \! \quad \cos(\varphi \!-\! \pi)
  \end{pmatrix} .
  \label{Gl-314}
\end{equation}
So if we discard the gauge variant variable $\varphi$, there are gauge
equivalent configurations $(\tilde a_1^1, \tilde a_2^1, \tilde a_2^2)$
and  $(-\tilde a_1^1,   -\tilde a_2^1, -\tilde  a_2^2)$  related  by a
discrete residual   gauge  symmetry   with  the corresponding   matrix
$U(\varphi\!=\!  \pi)$. We  may  resolve this  problem by  restricting
$\tilde a_1^1$ to positive values
\begin{equation}
  \tilde a_1^1 > 0 \;.
  \label{Gl-315}
\end{equation}
But what happens for $\tilde a_1^1=0$, which implies, that there is no
vector  $\vec a_1$ to  rotate?  The gauge condition (\ref{Gl-311}) and
$\det    J  =  -   \tilde   a_1^1  = 0$     define  a hypersurface  in
$\Gamma_{\text{ax}}$, usually called  the ``Gribov horizon''.  For the
axial gauge, this is   the plane $  H  = \{ 0,0,\tilde   a_2^1, \tilde
a_2^2\} \subset \Gamma_{\text{ax}} \subset  {\mathcal A}_0$.  From the
discussion above, we conclude, that  the Gribov horizon $H$  separates
regions on $\Gamma_{\text{ax}}$ (``Gribov copies''), which are related
by discrete residual gauge transformations.   After the restriction to
one Gribov  copy   (the ``reduced gauge  fixing  surface''), demanding
$\tilde a_1^1 >0\,$, the Gribov horizon seems to constitute a boundary
of the  configuration space.  In order  to  see if this   is in fact a
boundary of the physical configuration space $\mathcal M$, let us have
a closer  look at configurations on the  Gribov horizon.  We find that
every point   on  the gauge  orbit of  a   horizon configuration  $a =
(0,0,a_2^1,  a_2^2)$ does   not  only   satisfy the   gauge  condition
(\ref{Gl-311}) but also $\tilde a_1^1=0\,$:
\begin{equation}
  \begin{pmatrix} 0 &\;0\\ \tilde a_2^1 & \;\tilde a_2^2 \end{pmatrix}
   = \begin{pmatrix} 0 & \; 0 \\ a_2^1 & \; a_2^2 \end{pmatrix}
   \begin{pmatrix} \quad\cos \varphi &   \; \sin \varphi \\
                    - \sin \varphi & \;\cos \varphi \end{pmatrix}\; .
  \label{Gl-315a}
\end{equation}
Hence, the  Gribov horizon consists of  complete gauge  orbits. In the
generic  case  these orbits are   non-degenerate  and give  rise to  a
\emph{continuous} residual gauge symmetry  on the gauge fixing surface
$\Gamma_{\text{ax}}$.   Therefore  the   complete  reduction   of  the
configuration space  requires an \emph{additional} gauge condition for
the  points   on  the Gribov   horizon.  We   note   that  for horizon
configurations the ``$2 \!  \times \!2$-model'' is reduced to the ``$1
\! \times \!  2$-model''  of Christ and Lee \cite{christ:80} discussed
in  the introduction.  So by analogy  to (\ref{Gl-112}) we may proceed
with fixing the   continuous residual gauge  symmetry   by imposing an
additional gauge  condition   on  the configurations   on   the Gribov
horizon:
\begin{equation}
  \chi'_{\text{ax}}(\tilde a) = \tilde a_2^2 = 0 
  \quad\text{for}\quad \tilde a_1^1 = \tilde a_1^2 = 0 \; .
  \label{Gl-316}
\end{equation}
As in the Christ-Lee model we  take into account the residual discrete
gauge symmetry, $\tilde  a_2^1 \mapsto -\tilde  a_2^1$, by restricting
the remaining degree of freedom to positive values,
\begin{equation}
  \tilde a_2^1 \geq 0 \; .
  \label{Gl-317}
\end{equation}
Note, that in the picture of independent particles, the problem arises
because  the total  angular    momentum is not  well-defined,   if one
particle is  at the origin. Nevertheless,  it is possible to implement
Gau{\ss}' law by requiring the  angular momentum of the other particle
to vanish.  This is exactly,  what we have  done in (\ref{Gl-316}) and
(\ref{Gl-317}).
 
Now that we have  eliminated   all gauge symmetries,   let us  try  to
identify  the physical  configuration space  $\mathcal M$.  The  axial
gauge condition   $\chi_{\text{ax}}$  (\ref{Gl-311})     reduces   the
pre-configuration space ${\mathcal A}_0$  to a three dimensional gauge
fixing surface  $\Gamma_{\text{ax}}$  parameterized by the coordinates
$\tilde  a_i^a$.   One  might  be tempted   to  regard  this space  as
Euclidean.   In order  to  check  this,  let us calculate  the  metric
$g_\Gamma$  on  the gauge  fixing surface $\Gamma_{\text{ax}}$. Taking
the Euclidean metric  (\ref{Gl-308a}) on  ${\mathcal A}_0$, we  obtain
$g_\Gamma$ by projecting tangent vectors in $T{\mathcal A}_0$ onto the
horizontal subspace defined via Gau{\ss}' law  as shown by Babelon and
Viallet  \cite{babelon:79,babelon:81}. The  projection onto  the gauge
fixing  surface $\Gamma_{\text{ax}}$ with  coordinates $(\tilde a_1^1,
\tilde a_2^1, \tilde a_2^2)$ finally yields
\begin{equation}
   g_\Gamma = \frac{1}{\tilde a \cdot \tilde a} 
    \begin{pmatrix} \tilde a \cdot \tilde a & 0 & 0 \\
     0 &  \tilde a_1^1 \tilde a_1^1 + \tilde a_2^1 \tilde a_2^1 
       &  \tilde a_2^1 \tilde a_2^2 \\
     0 &  \tilde a_2^1 \tilde a_2^2 
       &  \tilde a_1^1 \tilde a_1^1 + \tilde a_2^2 \tilde a_2^2
    \end{pmatrix}
  \label{Gl-317b}
\end{equation}
with $\tilde a \!\cdot\!   \tilde  a := \tilde  a_1^1 \tilde  a_1^1  +
\tilde  a_2^1 \tilde   a_2^1  +  \tilde  a_2^2  \tilde  a_2^2$.    The
corresponding   scalar curvature  is  given by   $R  =  6 /  (\tilde a
\!\cdot\!  \tilde a)$, which is  different  from the zero curvature in
the Euclidean case and even singular at the origin.  We therefore have
to conclude that the gauge  fixing surface $\Gamma_{\text{ax}}$ is not
Euclidean.
\begin{figure}
  \begin{center}
    \leavevmode
    \epsfig{file=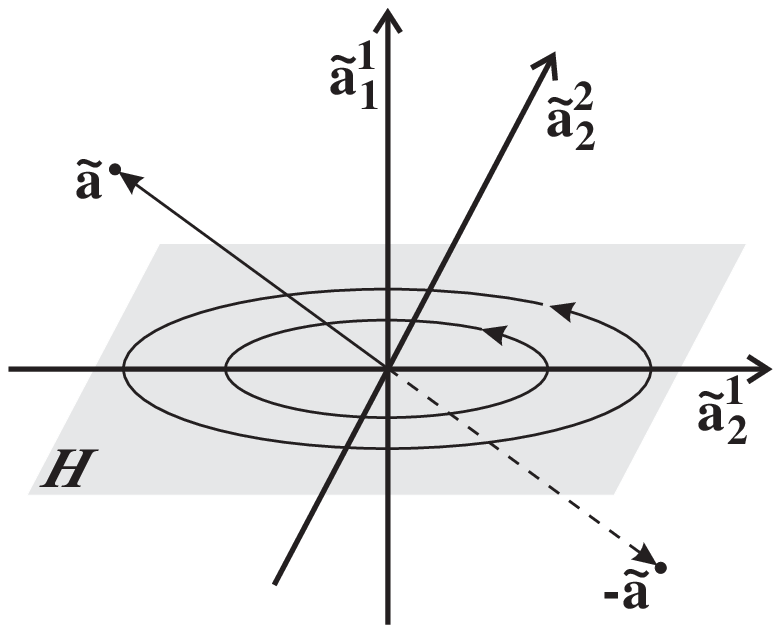,height=40mm}
  \end{center}
  \caption{The gauge fixing surface $\Gamma_{\text{ax}}$ embedded 
    into $\mathbb{R}^3$ with the  residual discrete symmetry $\tilde a
    \mapsto -   \tilde a$ and the   continuous symmetry on  the Gribov
    horizon ($H$) at $\tilde a_1^1 = 0$}
  \label{fig-21}
\end{figure}

To    get   some  intuition   for   what    happens    let us    embed
$\Gamma_{\text{ax}}$ parameterized  by the  coordinates $\tilde a_i^a$
into   $\mathbb{R}^3$  like depicted in     Fig.  \ref{fig-21}.  Note,
however, that unlike  for  a  Euclidean space  the  geodesics in  this
picture would   no longer be straight   lines,  due to  the nontrivial
metric  (\ref{Gl-317b}).  We  have    also sketched the  gauge  orbits
corresponding to the continuous residual  gauge symmetry on the Gribov
horizon  $H$  at  $\tilde a_1^1=0$  and  the  discrete residual  gauge
symmetry  which interchanges $\tilde    a_i^a$ with  $-\tilde  a_i^a$.
Condition (\ref{Gl-315}) eliminates the latter symmetry by restricting
the  gauge fixing surface to the  upper half space, whereas the former
symmetry reduces  the  Gribov horizon to  the half  line $\tilde a_2^1
\geq  0$   in  accordance  with  the  conditions   (\ref{Gl-316})  and
(\ref{Gl-317}).  How can  the upper half  space and  the half  line be
glued together to form the  physical configuration space $\mathcal M$,
which according  to     Singer \cite{singer:78} should be   a   smooth
manifold, if non-generic configurations are discarded?

The answer is that  we have to  reconsider the additional gauge fixing
on the Gribov horizon.   Conditions (\ref{Gl-316}) and  (\ref{Gl-317})
imply, that we have to identify every  point on a residual gauge orbit
with one point  on the positive $\tilde  a_2^1$-axis.  But we might as
well  identify such an orbit with  the point ($0,  0, -|\tilde a_2| \!
:= \!   -\sqrt{\tilde a_2^1\tilde a_2^1+\tilde a_2^2\tilde a_2^2}$) on
the   negative  $\tilde  a_1^1$-axis   in    a continuous   way.  This
identification is   most  easily   performed  by  choosing   spherical
coordinates on  $\Gamma_{\text{ax}}$ and doubling the azimuthal angle.
Just imagine the plane  $\tilde a_1^1  = 0$ to  be  the surface of  an
opened umbrella. What we will do in the following is nothing but close
the umbrella. We parameterize the gauge  fixing surface with spherical
coordinates   $r\geq  0$,  $\vartheta\in[\,0,\pi]$ and   $\psi\in[\,0,
2\pi[\:$,
\begin{equation}
  \tilde a_1^1 = r \, \cos\frac{\vartheta}{2}, \quad 
  \tilde a_2^1 = r \, \sin\frac{\vartheta}{2} \, \cos\psi
  \quad \text{and} \quad 
  \tilde a_2^2 = r \, \sin\frac{\vartheta}{2} \, \sin\psi \; .
  \label{Gl-318}
\end{equation}
Writing $\vartheta / 2$ instead  of $\vartheta$ guarantees that within
the   given  range of  $\vartheta$   we only parameterize  the reduced
configuration space  defined by     (\ref{Gl-315}).  With these    new
coordinates it is   possible to define  an embedding   of the physical
configuration space $\mathcal M$  into $\mathbb{R}^3$, such that there
are no residual gauge  symmetries.  Let $x_1$,  $x_2$ and $x_3$ denote
cartesian coordinates in  $\mathbb{R}^3$.  Then  we  map any point  in
$\Gamma_{\text{ax}}$  with coordinates    ($r, \vartheta,   \psi$)  to
$\mathbb{R}^3$ via
\begin{alignat}{2}
  \label{Gl-319}
  x_1 &= r \,\sin\vartheta \,\cos\psi 
      &&= 2 \: \tilde a_1^1 \tilde a_2^1 \: / \: r \;, \nonumber \\
  x_2 &= r \,\sin\vartheta \,\sin\psi
      &&= 2 \: \tilde a_1^1 \tilde a_2^2 \: / \: r \;, \\
  x_3 &= r \,\cos\vartheta 
      &&=( \tilde a_1^1 \tilde a_1^1 - \tilde a_2^1 \tilde a_2^1 - 
           \tilde a_2^2 \tilde a_2^2) \: / \: r 
  \;, \nonumber
\end{alignat}
where  we  have also  specified the  transformation  in  terms  of the
original variables  $\tilde a_i^a$  ($r^2 \!:=\!  \tilde  a_i^a \tilde
a_i^a$).   Since any  point with  $\tilde a_1^1=0$  gets indeed mapped
onto  the negative $x_3$-axis to   the point ($0,0,-|\tilde a_2|$), we
have    accomplished  the identifications on    the  Gribov horizon as
required  by the continuous  residual gauge symmetry.   In fact, these
identifications   are nothing   but  the ``boundary  identifications''
discussed in the literature \cite{zwanziger:93,vanbaal:95a}, which are
known to indicate a  nontrivial topology of the physical configuration
space $\mathcal M$. Since there are no residual gauge symmetries left,
we can now  identify the  space  obtained via  (\ref{Gl-319}) with the
physical configuration space $\mathcal  M$.  Notice, that apart from a
singular point at the origin, $r \! = \! 0$, the space $\mathcal M$ is
a smooth manifold (in particular without boundary).
\begin{figure}
  \begin{center}
    \leavevmode
    \epsfig{file=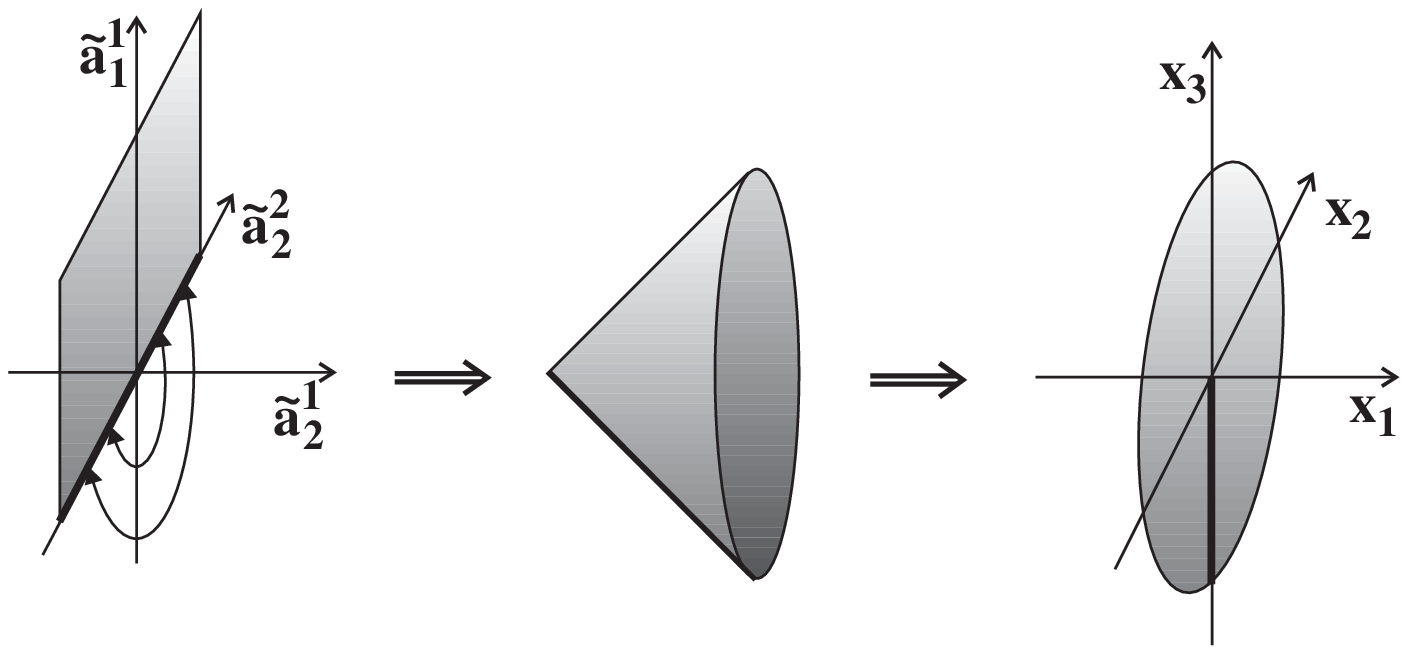,height=50mm}
  \end{center}
  \caption{Boundary identifications on the Gribov horizon and choice
           of an embedding for a subset of $\Gamma_{\text{ax}}$}
  \label{fig-22}
\end{figure}

To    make this procedure  more  transparent   we have represented  it
graphically in   Fig.   \ref{fig-22}  focusing   on a  half  plane  in
$\Gamma_{\text{ax}}$.  To the left    we have  drawn the  half   plane
($\tilde a_1^1 \! \geq \!  0, \tilde a_2^1 \!  = \! 0, \tilde a_2^2$).
After  identification of the gauge  equivalent  configurations ($0, 0,
\tilde a_2^2$) and ($0, 0, -\tilde a_2^2$)  the half plane becomes the
surface of a cone, which by a suitable embedding in $\mathbb{R}^3$ may
be represented as a plane.  This ``plane'', however, has non-vanishing
curvature with a singularity at the origin corresponding to the tip of
the cone.   This is also  true for  the  physical configuration  space
$\mathcal M$ as a whole, as indicated by the scalar curvature which in
spherical  coordinates is given  by $R=6 /  {r^2}$.   We note that the
origin $r \! = \! 0$ of the  physical configuration space $\mathcal M$
becomes a singular point analogous to the tip of a cone, because it is
a  fixed point    under   the operation  of  boundary  identification.
Therefore,  $\mathcal   M$  has  the     structure   of  an   orbifold
\cite{nakahara:92}.   However, the  deeper  reason for  the origin  to
become a singular  point of $\mathcal M$ lies  in the fact, that it is
the only   (non-generic)  configuration with   a  nontrivial stability
group: $a\!=\!0$ is invariant under the entire gauge group $SO(2)$.

If we calculate the Jacobian  for the transformation (\ref{Gl-312}) in
terms of  the     coordinates  $x_1,  x_2,  x_3,   \varphi$,     using
(\ref{Gl-319}), we obtain
\begin{equation}
\left| \frac{\partial (a_1^1, a_1^2, a_2^1, a_2^2)}
            {\partial (x_1, x_2, x_3, \varphi)} \right|
  = -\frac{1}{4}\: r\; ,
  \label{Gl-321}
\end{equation}
with  $r^2=x_1^2+x_2^2+x_3^2$. Thus, in   agreement with our  previous
considerations, there is  only one zero of  the Jacobian left, the one
corresponding to the  non-generic configuration, $a\!=\!0$.

We will study possible physical consequences of the orbifold structure
of $\mathcal M$ in more  detail at the end  of this section. Before we
can  do so we  have to determine the Hamiltonian  on $\mathcal M$ from
$\hat H^{2 \!  \times \!  2}$ (\ref{Gl-310})  defined on the Euclidean
pre-configuration space ${\mathcal A}_0$.  This is most easily done in
spherical coordinates combining the transformations (\ref{Gl-312}) and
(\ref{Gl-318}).  To  find  the Laplacian  on  $\mathcal   M$ in  these
coordinates we need the Jacobian matrix and its determinant
\begin{equation}
  \det J =  -\frac{1}{4}\,r^3  \sin\vartheta \;.
  \label{Gl-322}
\end{equation}
Note the factors  $r$  reflecting the  non-generic singularity at  the
origin (\ref{Gl-321})   and $r^2\sin\vartheta\,$ owing  to the  use of
spherical coordinates. In particular we  can now interpret the zero at
$\vartheta  \! = \! \pi$ as  a pure coordinate singularity without any
physical significance.   Independent  of the  parameterization of  the
gauge  fixing     surface  $\Gamma_{\text{ax}}$   or     the  physical
configuration space $\mathcal M$, the Gau{\ss} constraint is given by
\begin{equation}
  \label{Gl-322a}
  \mathrm{i} \: \frac{\partial}{\partial \varphi} \:
  |\Psi \rangle_{\text{phys}} = 0 \; .
\end{equation}
We solve (\ref{Gl-322a}) by requiring the wave functions not to depend
on the gauge variant  variable $\varphi$, so  that we can  discard all
terms  in $\hat H$ containing  derivatives with  respect to $\varphi$.
Thus we obtain the physical Hamiltonian in the axial gauge,
\begin{equation}
  \hat H^{2\! \times \! 2}_{\text{ax}} = 
  - \frac{g^2}{2 r^3} \frac{\partial}{\partial r}
          r^3     \frac{\partial}{\partial r} \! - \!
  \frac{2g^2}{r^2} \!
  \left( \frac{1}{\sin\!\vartheta}\frac{\partial}{\partial \vartheta}
              \sin\!\vartheta     \frac{\partial}{\partial \vartheta} 
  \! + \! 
  \frac{1}{\sin^2\!\vartheta} \frac{\partial^2}{\partial \psi^2}
  \! \right) \! + \! 
  \frac{r^4}{8\,g^2} \,\sin^2\!\vartheta\:\sin^2\!\psi \:. 
  \label{Gl-323}
\end{equation}
This Hamiltonian  only depends  on the  gauge invariant  variables $r,
\vartheta,  \psi$ and acts on wave   functions defined on the physical
configuration space $\mathcal M$.

In the next   subsection, we  will  compare the  results  obtained  by
choosing the gauge condition (\ref{Gl-311}) with  those, which we will
get  from  a  different  procedure   related to  the  method of  gauge
invariant variables.
%
%
\subsection{Polar Representation}
%
%
We  notice  that the   Hamiltonian  (\ref{Gl-307})  has an  additional
symmetry   generated  by   ${\mathcal   J} \!:=   \epsilon^{ij}\,a^a_i
\,e^{ja}$, which we write as
\begin{equation}
  \label{Gl-324}
  \begin{pmatrix} a_1^1 & \; a_1^2 \\ a_2^1 & \; a_2^2 \end{pmatrix}
  \mapsto \begin{pmatrix} \cos\gamma & \;-\sin\gamma \\
           \sin\gamma & \;\quad\cos\gamma  \end{pmatrix}
  \begin{pmatrix} a_1^1 & \; a_1^2 \\ a_2^1 & \; a_2^2 \end{pmatrix}
\end{equation}
with  $\gamma \in [\,0, 2\pi[\:$.    Apart  from being useful in   the
diagonalization of the Hamiltonian (as $[\:\hat H,\hat  J \,]\, = 0$),
this  symmetry can   be  further exploited  to   represent the  matrix
$(a^a_i)$ as
\begin{equation}
  \begin{pmatrix} a_1^1 & \; a_1^2 \\ a_2^1 & \; a_2^2 \end{pmatrix} =
  \begin{pmatrix} \cos\gamma & \;-\sin\gamma \\   
                  \sin\gamma & \;\quad\cos\gamma  \end{pmatrix}
  \begin{pmatrix} \lambda_1 & \; 0 \\ 0 & \; \lambda_2 \end{pmatrix}
  \begin{pmatrix} \cos \varphi & \;  -  \sin \varphi \\
                  \sin \varphi & \;\quad\cos \varphi \end{pmatrix} \;.
  \label{Gl-325}
\end{equation}
The representation (\ref{Gl-325}) is known  as the polar decomposition
of an arbitrary quadratic matrix into  one diagonal and two orthogonal
matrices \cite{aitken:61}.   This  decomposition  has  been frequently
applied  to   classical     and   quantum     Yang-Mills     mechanics
\cite{martin:94,matinyan:81,savvidy:85,palumbo:87,halperin:95}.
Actually,  upon  inserting the  transformation (\ref{Gl-325}) into the
classical  Lagrangian (\ref{Gl-301}), going  to the Weyl gauge $a_0=0$
and setting  $\varphi=\gamma=0$, one would  obtain the ``xy-model'', a
well-known  playground      for     studying    non-linear    dynamics
\cite{chang:84,martins:89}.  For   the case of  field  theory, Simonov
proposed   the         closely   related  ``polar     representation''
\cite{simonov:85}, whereas   Goldstone and  Jackiw  applied  the polar
decomposition   within     the    electric     field    representation
\cite{goldstone:78}.

By   analogy  with  (\ref{Gl-312})  we    rewrite the   representation
(\ref{Gl-325}) as
\begin{equation}
  \begin{pmatrix} a_1^1 & \; a_1^2 \\ a_2^1 & \; a_2^2 \end{pmatrix} =
  \begin{pmatrix} 
    \lambda_1 \: \cos\gamma & \;   -  \lambda_2 \: \sin\gamma \\   
    \lambda_1 \: \sin\gamma & \; \quad\lambda_2 \: \cos\gamma
  \end{pmatrix}
  \begin{pmatrix} \cos \varphi & \;  -  \sin \varphi \\
                  \sin \varphi & \;\quad\cos \varphi \end{pmatrix} \;
  \label{Gl-326}
\end{equation}
and interpret (\ref{Gl-326})  as the transformation to gauge invariant
variables $\lambda_1$,  $\lambda_2$, $\gamma$  and  the  gauge variant
coordinate  $\varphi$. The crucial  point of writing (\ref{Gl-325}) in
this form is, that  (\ref{Gl-326}) can also  be interpreted as a gauge
fixing   transformation.   The  corresponding \emph{non-linear}  gauge
condition \cite{palumbo:91}
\begin{equation}
  \chi_{\text{pr}}(\tilde a_i^a) = 
  \tilde a_1^1 \tilde a_1^2 + \tilde a_2^1 \tilde a_2^2 = 0
  \label{Gl-327}
\end{equation}
can easily be read off from the first matrix on the right hand side of
(\ref{Gl-326}), where we denoted  the corresponding matrix elements by
$\tilde a_i^a$ as in (\ref{Gl-312}).  Hence the variables $\lambda_1$,
$\lambda_2$  and     $\gamma$   form     a   parameterization  $\tilde
a_i^a(\lambda_1,   \lambda_2, \gamma)$   of the  gauge  fixing surface
$\Gamma_{\text{pr}}$ defined  by   $\chi_{\text{pr}}(\tilde a_i^a)=0$.
Expression  (\ref{Gl-326}) provides   an  explicit   example  for  the
equivalence  between the method of gauge  fixing and  the use of gauge
invariant variables from the outset.

\begin{figure}
  \begin{center}
    \leavevmode
    \epsfig{file=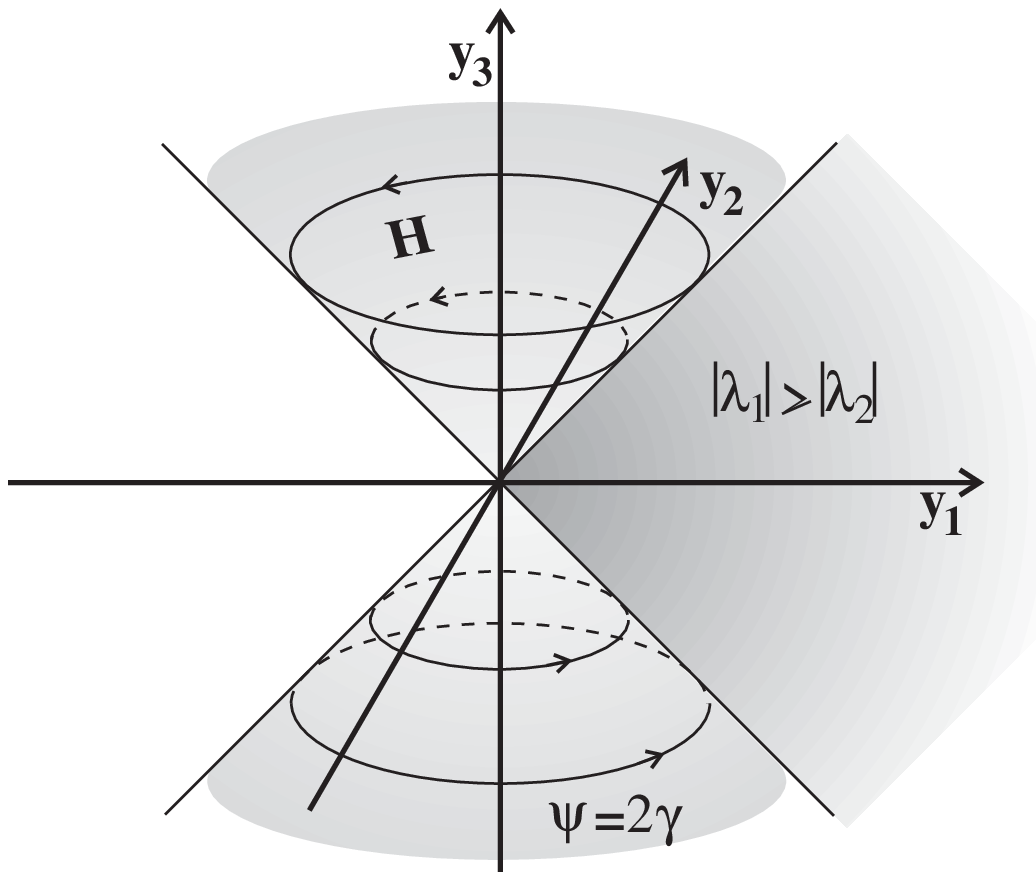,height=50mm}
  \end{center}
  \caption{Embedding of the reduced gauge fixing surface $\tilde
    \Gamma_{\text{pr}}$        into          $\mathbb{R}^3$:   $\tilde
    \Gamma_{\text{pr}}$ is  the  complement of   the double cone  with
    symmetry axis along the $y_3$-direction,  the Gribov horizon ($H$)
    being its boundary, containing complete gauge orbits.}
  \label{fig-23}
\end{figure}

From our experience with the axial  gauge we anticipate the appearance
of residual gauge symmetries.  The calculation of the Jacobian for the
transformation  (\ref{Gl-326}) yields
\begin{equation}
  \left| \frac{\partial(a_1^1, a_1^2, a_2^1, a_2^2)}
  {\partial (\lambda_1, \, \lambda_2, \, \gamma, \, \varphi)} \right|
  =  \lambda_1^2-\lambda_2^2 \; ,
  \label{Gl-328}
\end{equation}
which is  zero   for   $\lambda_1=\pm  \lambda_2$.   Gauge  equivalent
configurations may  be detected  by   investigating whether there  are
gauge copies $a^a_i \! = \! \tilde a_i^b\, U^{ba}(\varphi)$ of $\tilde
a_i^a(\lambda_1,\lambda_2,\gamma)$    in    the  same     gauge,  i.e.
$\chi_{\text{pr}}(a^a_i) = 0\,$,
\begin{equation}
  \label{Gl-329}
  \chi_{\text{pr}}
    \big( a^a_i(\lambda_1,\lambda_2, \gamma, \varphi) \big) =
    \frac{1}{2}\: (\lambda_2^2-\lambda_1^2) \: \sin(2\,\varphi) = 0
  \; .
\end{equation}
We find that, apart from the zeros  of the Jacobian (\ref{Gl-328}), we
have additional gauge  copies related by  $U(\varphi \!=\!  n\,  \pi /
2)$,  corresponding to discrete residual gauge  symmetries.  As in the
case  of the axial  gauge  we eliminate  these discrete  symmetries by
restricting the values of  the gauge invariant variables to $\lambda_1
\geq |\lambda_2|$  and $\gamma \in  [\,0,  \pi[\:$  where  we have  to
identify  the  points   $(\lambda_1,  \lambda_2,  0)\sim   (\lambda_1,
\lambda_2,   \pi)$.  The  reduced    gauge   fixing   surface  $\tilde
\Gamma_{\text{pr}}$ can be embedded in $\mathbb{R}^3$ as shown in Fig.
\ref{fig-23}, where the  shaded   region defined by  $\lambda_1   \geq
|\lambda_2|$ is rotated around the $y_3$-axis.  The explicit embedding
is  given by $y_1=\lambda_1   \cos\psi$, $y_2=\lambda_1  \sin\psi$ and
$y_3=\lambda_2$ with $\psi=2\gamma$.

Let  us   turn   to  the  configurations    $\lambda_1=\pm  \lambda_2$
corresponding to the zeros of  the Jacobian (\ref{Gl-328}). The set of
these configurations constitutes the  Gribov horizon $H$, which in the
embedding of Fig. \ref{fig-23} forms the surface of a double cone with
the $y_3$-axis as its symmetry  axis.  Recalling the discussion of the
axial gauge we anticipate the existence of a continuous residual gauge
symmetry on this  surface.   And   in  fact we  find,  constructing  a
relation similar to (\ref{Gl-315a}),  that on  the Gribov horizon  $H$
there  are  gauge   orbits  in   the   form  of  circles  (cf.    Fig.
\ref{fig-23}).  This  continuous  residual gauge symmetry is  directly
related to the fact, that  the gauge invariant  variable $\psi$ is not
well  defined for $\lambda_1=\pm  \lambda_2$.  As  in the  case of the
axial gauge we have to fix this residual gauge symmetry by imposing an
\emph{additional} gauge condition.  Geometrically, we have to identify
all the points  on a gauge orbit with  one point.  Choosing this point
to  be   on the $y_3$-axis  of  $\mathbb{R}^3$,   we  may realize this
boundary   identification by a   similar   ``doubling of an  azimuthal
angle''  as in  the   case of   the  axial gauge.   Define the   angle
$\vartheta \in [-\pi   / 2,\pi / 2  \,  ]$  via  $\lambda_1 =  r\:\cos
\vartheta /  2$ and $\lambda_2 =  r\: \sin \vartheta   / 2$.  Then map
every point of $\Gamma_{\text{pr}}$ to $\vec x\in \mathbb{R}^3$ via
\begin{equation}
  x_1=r\:\cos\vartheta\,\cos\psi \:, \quad
  x_2=r\:\cos\vartheta\,\sin\psi \:, \quad
  x_3=r\:\sin\vartheta \: ,
  \label{Gl-330}
\end{equation}
where $x_1, x_2, x_3$ are cartesian coordinates in $\mathbb{R}^3$.  As
there are no residual symmetries left, we have thus found an embedding
of the physical configuration space  $\mathcal M$ into $\mathbb{R}^3$.
In other words,   (\ref{Gl-330}) defines a coordinate  system covering
all of  $\mathcal  M$.  Once again   the physical configuration  space
$\mathcal M$  has the structure of an  orbifold  with a singularity at
the  origin.   The  Jacobian  of  the  resulting  gauge transformation
$a_i^a(\vec x, \varphi)$ is proportional to $r$. As in the case of the
axial gauge we  conclude that the only zero  of the Jacobian which  is
not due   to  incomplete  gauge   fixing  or coordinate  singularities
corresponds   to   the   non-generic   configuration   $r\!=\!0$.  The
significance  of   this  configuration also   follows from  the scalar
curvature $R  \!=\!  6 / {r^2}$, which  we can calculate for the polar
representation analogously to the axial gauge.

Expressed  in terms of the  spherical coordinates $r, \vartheta, \psi$
the Yang-Mills potential is given by
\begin{equation}
  {\mathcal V}^{2 \! \times \! 2}_{\scriptscriptstyle \text{YM}}
  = \frac{1}{8\, g^2}\: r^4 \, \sin^2 \! \vartheta \; ,
  \label{Gl-332}
\end{equation}
Notice,   that in polar  representation  ${\mathcal V}^{2 \! \times \!
  2}_{\scriptscriptstyle \text{YM}}$ does not depend on $\psi$, due to
the additional symmetry (\ref{Gl-324}) for constant fields in the Weyl
gauge.  The corresponding  generator  $\hat J =  \epsilon^{ij}\,  \hat
a_i^a \, \hat e^{ja}$ is given in terms of spherical coordinates by
\begin{equation}
  \hat J = - \: \mathrm{i} \: \frac{\partial}{\partial\psi}\;,
  \label{Gl-333}
\end{equation}
which commutes with the  Hamiltonian.    We also note the   difference
between ${\mathcal J}$ and the Gau{\ss} constraint  $\mathcal G$.  For
the gauge symmetry generated by $\mathcal  G$ we require invariance of
the  wave function  under   the corresponding transformations of   its
arguments.    This  implies    that  wave     functions  are   trivial
representations of the  gauge group.  In  the case  of ${\mathcal J}$,
however, the    wave  function may     transform in    an    arbitrary
representation of the corresponding symmetry group.  Hence there is no
restriction  on  the configuration  space  coming  from the additional
symmetry.

In order  to establish the   equivalence of  the physical  Hamiltonian
obtained from the    polar   representation  with the    axial   gauge
Hamiltonian (\ref{Gl-323}) we   just  need  to  redefine  the  angular
variables $\vartheta$ and   $\psi$,  which parameterize  the  physical
configuration space $\mathcal M$ by choosing another axis as the polar
axis.   It  is  a  then  a trivial exercise   to  show  that one  gets
precisely the same results as in the axial gauge.
%
%
\subsection{Natural coordinates}
\label{a23}
%
%
In the     preceding subsection  we    have already   demonstrated the
equivalence of the choice of a  gauge condition and the transformation
to gauge invariant  variables.  Still, we want to consider yet another
set of  gauge invariant coordinates, which  we will  call ``natural'',
because  they  are the  most obvious  ones  for our problem. Since the
physical observables must  not    depend on  the orientation   of  the
coordinate axes with respect to the vectors $\vec  a_1$ and $\vec a_2$
a natural choice of  gauge invariant variables  are the lengths  $r_1,
r_2$ of the two vectors and the angle $\psi$ between them.  Once again
we can write the transformation to this  set of coordinates as a gauge
transformation
\begin{equation}
  \begin{pmatrix} a_1^1 & \;a_1^2\\ a_2^1 &\; a_2^2 \end{pmatrix} =
  \begin{pmatrix} 
    r_1\:\cos\frac{\psi}{2} & \;    - r_1\:\sin\frac{\psi}{2} \\   
    r_2\:\cos\frac{\psi}{2} & \;\quad r_2\:\sin\frac{\psi}{2} 
  \end{pmatrix}
  \begin{pmatrix} \cos \varphi & \;  -  \sin \varphi \\
                  \sin \varphi & \;\quad\cos \varphi \end{pmatrix} \;.
  \label{Gl-334}
\end{equation}
The Jacobian of  the map $(r_1, r_2,  \psi, \varphi) \mapsto (a_i^a)$
is given by
\begin{equation}
  \left| \frac{\partial(a_1^1, a_1^2, a_2^1, a_2^2)}
    {\partial (r_1, r_2, \psi, \varphi)} \right|
  = -\: r_1 \, r_2 \; .
  \label{Gl-336}
\end{equation}
As    in the   familiar example   of    the transformation  to   polar
coordinates in the plane, where the polar  angle is not defined at the
origin, the angle $\psi$  is not well defined  when $r_i$ is zero  for
one $i$. This is analogous to the case $\lambda_1  = \pm \lambda_2$ in
the polar   representation  where the angle  $\psi$   was  not defined
either.    Likewise,  we  observe, that   (\ref{Gl-334})   can also be
interpreted   as the   gauge   transformation, relating  an  arbitrary
configuration   $(a_i^a)$ with $(\tilde  a_i^a)$,   where the  $\tilde
a_i^a$  satisfy a certain  gauge  condition.  In  our case,  the gauge
condition can easily  be read off from the  first matrix on  the right
hand side of (\ref{Gl-334}), and we get
\begin{equation}
  \chi_{\text{nc}}(\tilde a_i^a) = 
  \tilde a_1^1 \tilde a_2^2 + \tilde a_1^2 \tilde a_2^1 = 0 \; ,
  \label{Gl-335}
\end{equation}
which  is again non-linear.   Let us  for  the moment assume, that the
variables $r_i$ and   $\psi$ just define   a coordinate system on  the
gauge   fixing     surface   $\Gamma_{\text{nc}}$     corresponding to
(\ref{Gl-335}), ignoring  e.g.  the  significance   of the $r_i$'s  as
positive  lengths. So let  us take $r_i  \in \mathbb{R}$ and $\psi \in
[0, 4\pi[\:$.  Then, using the language of gauge  fixing, the zeros of
the Jacobian (\ref{Gl-336})  indicate the existence of Gribov horizons
on the gauge fixing surface $\Gamma_{\text{nc}}$, separating different
Gribov   regions related by   discrete residual gauge symmetries.  How
these residual symmetries  can be found  has been  demonstrated before
(e.g.  in (\ref{Gl-329})).  Thus, we only present  the results of this
analysis, which   will  be needed later  on.  The  discrete symmetries
relating different Gribov regions follow from
\begin{alignat}{2}
  a_i^a(r_1, r_2, \psi, \varphi \! + \! \pi/2) & =
  a_i^a(-r_1, r_2, \psi \! + \! \pi, \varphi) \;\; & \Rightarrow \;\;
    (r_1, r_2, \psi) &\sim (-r_1, r_2, \psi \!+\!\pi) \;,\nonumber \\
  a_i^a(r_1, r_2, \psi, \varphi \! + \! \pi/2) & = 
  a_i^a(r_1, -r_2, \psi \! + \! \pi, \varphi) \;\; & \Rightarrow \;\;
    (r_1, r_2, \psi) &\sim (r_1, -r_2, \psi \!+\!\pi) \;,\nonumber  \\
  a_i^a(r_1, r_2, \psi, \varphi \! + \! \pi) & =
  a_i^a(r_1, r_2, \psi\!+\!2\pi, \varphi) \;\; & \Rightarrow \;\;
    (r_1, r_2, \psi) &\sim (r_1, r_2, \psi \!+\! 2\pi )\;,\nonumber \\
  a_i^a(r_1, r_2, \psi, \varphi \! + \! \pi) & =
  a_i^a(-r_1, -r_2, \psi, \varphi) \;\; & \Rightarrow \;\;
    (r_1, r_2, \psi) &\sim (-r_1, -r_2, \psi) \;, 
  \label{Gl-337}
\end{alignat}
whereas the continuous symmetries are (for arbitrary $\psi_0$)
\begin{alignat}{2}
  & a_i^a(r_1, 0, \psi, \varphi\!+\!\psi_0 /2) =
    a_i^a(r_1, 0, \psi\!+\!\psi_0, \varphi) \;\; \Rightarrow \;\; &
    (r_1, 0, \psi) & \sim (r_1, 0, \psi\!+\!\psi_0) \;,\nonumber  \\
  & a_i^a(0, r_2, \psi, \varphi\!+\!\psi_0 /2) =
    a_i^a(0, r_2, \psi\!+\!\psi_0, \varphi) \;\; \Rightarrow \;\; &
    (0, r_2, \psi) & \sim (0, r_2, \psi\!+\!\psi_0) \; ,
  \label{Gl-337a}
\end{alignat}
for  the case, where $r_1$  or $r_2$  vanishes.   We can eliminate the
residual symmetries (\ref{Gl-337}) by restricting $\Gamma_{\text{nc}}$
to one Gribov region (the reduced gauge fixing  surface) via $r_i \geq
0$ and $\psi \in [0, 2\pi[\:$, $\psi$ being a  polar angle living on a
circle,  since $\psi \sim \psi +  2\pi$.  From these considerations we
see the interpretation of the gauge invariant variables as lengths and
angle  between  $\vec  a_1$ and  $\vec a_2$  re-emerge  again. We also
notice,  that the existence of  the  Gribov horizon and the continuous
symmetries (\ref{Gl-337a}) for $r_i=0$  is equivalent to the breakdown
of  the  corresponding coordinate  system,  indicating a topologically
nontrivial structure   of the  physical configuration space  $\mathcal
M$.

Again,   there   are different   possibilities   how   to  embed   the
appropriately    restricted    gauge       fixing    surface   $\tilde
\Gamma_{\text{nc}}$   parameterized   by  $(r_1,    r_2,  \psi)$  into
$\mathbb{R}^3$.  The necessary identifications on  the boundary of the
reduced gauge  fixing surface may  be realized by the parameterization
$r_1=r\sin    \vartheta / 2$   and   $r_2=r\cos  \vartheta  / 2$  with
$\vartheta \in [\:0, \pi]\,$.  As demonstrated before, this ``doubling
of   the azimuthal   angle''   $\vartheta$ explicitly   realizes   the
identification of  a residual gauge  orbit with  one  point.  For  the
total gauge transformation we obtain from (\ref{Gl-334})
\begin{equation}
  \begin{pmatrix} a_1^1 & \; a_1^2 \\ a_2^1 & \; a_2^2 \end{pmatrix} =
  \begin{pmatrix} 
      r\:\sin\frac{\vartheta}{2}\:\cos\frac{\psi}{2} & 
   \;-r\:\sin\frac{\vartheta}{2}\:\sin\frac{\psi}{2} \\   
      r\:\cos\frac{\vartheta}{2}\:\cos\frac{\psi}{2} &
   \;\quad r\:\cos\frac{\vartheta}{2}\:\sin\frac{\psi}{2} 
  \end{pmatrix}
  \begin{pmatrix} \cos \varphi & \;  -  \sin \varphi \\
                  \sin \varphi & \;\quad\cos \varphi \end{pmatrix}
  \label{Gl-338}
\end{equation}
with the Jacobian
\begin{equation}
  \left| \frac{\partial(a_1^1, a_1^2, a_2^1, a_2^2)}
              {\partial (r, \vartheta, \psi, \varphi)} \right|
  = \frac{1}{4}\: r^3 \: \sin\vartheta \;.
  \label{Gl-339}
\end{equation}
This is  the same expression as  for  the Jacobian in the  axial gauge
(\ref{Gl-322}).  Once again we  can  find another set of  coordinates,
such that  the Jacobian is  $r/4$.  This  implies, that the additional
factor $r^2 \sin \vartheta$  in (\ref{Gl-339}) is only  due to the use
of  spherical coordinates.  Not surprisingly,   the metric, the scalar
curvature and  the Hamiltonian expressed  in spherical coordinates are
equivalent  to the corresponding expressions  in  the gauges discussed
before. Of course, the physical  configuration space $\mathcal M$ once
again is   an orbifold. We conclude that (modulo re-parameterizations)
there  is  indeed a unique Hamiltonian   on the physical configuration
space $\mathcal  M$   independent of the   method  chosen to determine
$\mathcal M$, provided  the Gribov problem  is correctly resolved.  In
terms of spherical coordinates  \emph{the} Hamiltonian for the  ``$2\!
\times \!  2$-model'' is given by
\begin{equation}
  \hat H^{2\! \times \! 2}_{\mathcal M} = 
  - \frac{g^2}{2 r^3} \, \frac{\partial}{\partial r} \,
          r^3  \,   \frac{\partial}{\partial r}  
  + \frac{2g^2}{r^2} \, \hat L^2
  + \frac{r^4}{8\,g^2} \,\sin^2\!\vartheta\:\sin^2\!\psi \:. 
  \label{Gl-340}
\end{equation}
The  angular part  of the  Hamiltonian (\ref{Gl-340})  is  the angular
momentum   operator    $\hat L^2$  well-known    from standard quantum
mechanics.  There are, however,  two important differences compared to
an ordinary quantum mechanical problem in $\mathbb{R}^3$ formulated in
terms of spherical      coordinates.   The  part of    the    Jacobian
(\ref{Gl-339}) related to the gauge fixing leads to additional factors
$r$ and   $1/r$  in the   radial part  of the  Laplacian.   This  is a
remainder  of the fact, that  the original problem  was  posed in four
dimensions. In addition,  there is a  relative factor $4$  between the
radial  and   the  angular   part,   which is  due    to the  boundary
identifications on    the  Gribov horizon revealing    the topological
structure of the physical configuration space $\mathcal M$.
%
%
\subsection{Quantum mechanics}
%
%
In order to study the physical implications of the gauge reduction and
the special structure  of  the physical configuration space  $\mathcal
M$, we have to solve the Schr\"odinger equation for  our model.  Since
the   Yang-Mills   potential     ${\mathcal    V}^{2\!    \times    \!
  2}_{\scriptscriptstyle  \text{YM}}$  is  quartic of  the form  $(z\,
r)^2$ (\ref{Gl-340}), to  get  quantitative results we  would  have to
apply  numerical or semi-classical  methods,  like those used for  the
``xy-model''   \cite{chang:84,martins:89,dahlqvist:92}.   Although the
configuration  space of the  ``xy-model''  has infinite volume and the
lines   of  minimal  potential extend   to    infinity,  it has   been
demonstrated that  this  toy  model has   a discrete  energy  spectrum
\cite{simon:83}.  Using similar arguments, it is possible to show that
this is  also  the case  for the ``$2\!    \times \!   2$-model'' with
${\mathcal  V}^{2\!   \times \!  2}_{\scriptscriptstyle  \text{YM}}$ .
However, in order to get  analytical results in a straightforward way,
we proceed by choosing  the  harmonic oscillator potential  ${\mathcal
  V}^{2\!  \times \!  2}_{\text{osc}}$ (\ref{Gl-304b}) instead.

We want to discuss the physical  implications of the identification of
gauge copies   on the Gribov  horizon,  which   eventually leads to  a
singular point of the physical configuration space $\mathcal M$. Hence
we will compare the solutions of the Schr\"odinger equation defined on
some gauge fixing surface  $\Gamma_{\chi}$ with the results we  obtain
from  the    Hamiltonian (\ref{Gl-340})    defined  on    the physical
configuration  space  $\mathcal   M$.  Notice,  that  the   first case
corresponds  to  the usual treatment of  gauge   theories when a gauge
condition is  chosen to eliminate gauge degrees  of  freedom.  For the
comparison we will use the  system of natural coordinates discussed in
the previous   subsection.  In  these coordinates  the   Schr\"odinger
equation on the gauge fixing surface $\Gamma_{\text{nc}}$ reads
\begin{equation}
  \left( \sum_{i=1}^2 \left( \frac{1}{r_i}
  \frac{\partial}{\partial r_i} r_i \frac{\partial}{\partial r_i} +
  \frac{1}{r_i^2} \frac{\partial^2}{\partial\psi^2}- \frac{r_i^2}{g^2}
  \right) + \frac{2}{g^2} \: E \right) \Psi(r_1, r_2, \psi) = 0\;.
  \label{Gl-343}
\end{equation}
Since  the wave  function  $\Psi(r_1,   r_2,  \psi)$  is defined    on
$\Gamma_{\text{nc}}$,   the coordinates   take  the values   $r_i  \in
\mathbb{R}$  and  $\psi  \in [0,  4\pi  [\:$.   We will implement  the
residual gauge symmetries listed in (\ref{Gl-337}) and (\ref{Gl-337a})
as symmetry conditions on the wave function
\begin{align}
  \label{Gl-344}
  \Psi(r_1, r_2, \psi) &= \Psi(-r_1, r_2,\psi+\pi) \;,\\
  \Psi(r_1, r_2, \psi) &= \Psi(r_1, -r_2,\psi+\pi) \;,\nonumber\\
  \Psi(r_1, r_2, \psi) &= \Psi(r_1, r_2,\psi+2\pi) \;,\nonumber\\
  \Psi(r_1, r_2, \psi) &= \Psi(-r_1, -r_2,\psi) \; ;\nonumber \\
  \label{Gl-344a}
  \Psi(r_1, 0, \psi) &= \Psi(r_1, 0,\psi')\;,\\
  \Psi(0, r_2, \psi) &= \Psi(0, r_2,\psi')\; ;\nonumber
\end{align}
for   arbitrary  $\psi$ and $\psi'$.    The  form of the Schr\"odinger
equation  (\ref{Gl-343})  allows  for a   separation ansatz $\Psi(r_1,
r_2,\psi)=R_1(r_1)\:R_2(r_2)\:Y(\psi)$.   The angular  wave   function
$Y_l(\psi)$ is  given by the  exponential $\exp(\mathrm{i}\,l\,\psi)$,
where the third  of  the  conditions (\ref{Gl-344}) restricts   $l$ to
integers.  The radial  wave functions are given  in terms  of Laguerre
polynomials $L_{n_i}^{|l|}$ where  $n_i$ can only  take  values in the
non-negative integers ($n_i = 0, 1, 2, \ldots$)  for the wave function
to remain finite for  $r_i\rightarrow \infty$.  The  complete solution
of  the Schr\"odinger equation normalized with  respect to the measure
following from the Jacobian (\ref{Gl-336}) is given by
\begin{equation}
  \Psi(r_1,\! r_2,\! \psi) \!=\!
  {\textstyle \frac{2}{g}
  \sqrt{\frac{n_1!\:n_2!}{(n_1+|l|)!\,(n_2+|l|)!}} }\, 
  \left({\textstyle \frac{r_1\, r_2}{g} } \right)^{\!|l|}\: 
  \mathrm{e}^{-\frac{1}{2g}(r_1^2+r_2^2)}
  \, L_{n_1}^{|l|}\!\!\left( {\textstyle\frac{r_1^2}{g} } \right)
  \, L_{n_2}^{|l|}\!\!\left( {\textstyle\frac{r_2^2}{g} } \right)
  \, \mathrm{e}^{\mathrm{i}\,l\,\psi} .
  \label{Gl-348}
\end{equation}
This solution  explicitly  realizes the conditions (\ref{Gl-344})  and
(\ref{Gl-344a}).  For example for odd  $l$ we  have $Y_l(\psi+\pi) = -
Y_l(\psi)$ and $R_{n_1}^{|l|}(-r_1) = - R_{n_1}^{|l|}(r_1)$, such that
the total  wave    function remains  unchanged under   $\psi   \mapsto
\psi+\pi$,  $r_1  \mapsto -r_1$.  As far    as the continuous residual
gauge symmetries are concerned, we note  that the radial wave function
$R_{n_i}^{|l|}(r_i)$ vanishes  at $r_i=0$ for $l  \neq 0$.  For $l=0$,
the  radial  function  $R_{n_i}^{|l|}(r_i)$ remains   finite, but  the
angular part $Y_0(\psi)$  is  a constant, such   that the total   wave
function is constant along a gauge orbit on the Gribov horizon defined
by $r_i=0$. The energy spectrum is given by
\begin{equation}
  E_\nu = 2 \: g \: (\nu+1) \quad\text{with}\quad \nu = n_1+n_2+|l|\;.
  \label{Gl-350}
\end{equation}
So the ground state energy  is $E_0=2\, g$  and we have an equidistant
level  spacing with $\Delta  E=2\,g$. The  degeneracy  of a state with
energy $E_\nu$ is given by
\begin{equation}
  g_\nu=(\nu+1)^2
  \label{Gl-351}
\end{equation}
and is  a  consequence  of choosing   the highly  symmetric  potential
${\mathcal V}^{2\!  \times \!  2}_{\text{osc}}$ (\ref{Gl-304b}).

Let us compare these  findings   with the physical  Hamiltonian  $\hat
H^{2\!  \times \!    2}_{\mathcal M}$  (\ref{Gl-340}).  We solve   the
corresponding Schr\"odinger equation
\begin{equation}
  \left( \frac{\partial^2}{\partial r^2} + 
  \frac{3}{r}\frac{\partial}{\partial r} - \frac{4}{r^2}\:\hat L^2 -
  \frac{r^2}{g^2} + 
  \frac{2}{g^2}\:E \right) \Psi(r, \vartheta, \psi) = 0
  \label{Gl-352}
\end{equation}
with another separation ansatz, where  the angular dependence is given
by   the standard eigenfunctions  of  the total angular momentum $\hat
L^2$.  We obtain
\begin{equation}
  \Psi(r, \!\vartheta, \!\psi) \!=\!
  {\textstyle \frac{1}{g}
  \sqrt{\frac{2}{\pi} \frac{(l-m)!}{(l+m)!} 
  \frac{n!\;(2l+1)}{(2l+n+1)!}}}\: 
  \left( {\textstyle \frac{r^2}{g} } \right)^{\!l}\;
  \mathrm{e}^{-\frac{1}{2g} r^2} \; 
  L_n^{2l+1}\!\left( {\textstyle\frac{r^2}{g} } \right)\;
  P_l^m(\cos\vartheta) \; \mathrm{e}^{\mathrm{i}m\psi}\; ,
  \label{Gl-353}
\end{equation}
which has been  normalized with respect  to the measure induced by the
Jacobian (\ref{Gl-339}).  The  wave  function   $\Psi$  lives on   the
physical  configuration space $\mathcal  M$. Therefore, there are only
the usual boundary conditions for the spherical coordinates $\psi$ and
$\vartheta$, restricting  the quantum numbers  to $l=0,1,2,\ldots$ and
$-l \leq m \leq l$.   Since for the  unreduced gauge theory we require
the  wave function to  be regular at every  point of the configuration
space  $\mathcal A$, we  have to demand  regularity of $\Psi$ at every
point of the physical   configuration space $\mathcal M$ (representing
one  orbit in $\mathcal A$) as well.  This  also includes the singular
point at  $r \! = \!  0$, where $L_n^{2l+1}$ has  to be finite. Hence,
the radial   quantum  number  $n$    has to   be a   positive  integer
($n=0,1,2,\ldots$). The energy spectrum is then given by
\begin{equation}
  E_\nu= 2\:g \:(\nu + 1) \quad\text{with}\quad \nu = n + l\;
  \label{Gl-354}
\end{equation}
in total agreement with the result (\ref{Gl-350}). The degeneracy is
\begin{equation}
  g_\nu = \sum_{\lambda=0}^{\nu}(2\lambda+1) = (\nu+1)^2,
  \label{Gl-355}
\end{equation}
which again coincides with the result  obtained before.  We would also
like to   mention, that on   the  $\hat x_3$-axis  ($\vartheta=0,\pi$)
corresponding to the Gribov horizon ($r_i\!=\! 0$) on the gauge fixing
surface $\Gamma_{\text{nc}}$, the wave  function vanishes for $m\neq0$
just as the wave function (\ref{Gl-348}) does in the case $l\neq0$.

We have thus shown   that the spectra  of the Hamiltonians in the  two
different frameworks are identical,  where on  one  hand $\hat H$  was
defined on the  gauge fixing surface  $\Gamma_{\text{nc}}$, and on the
other hand $\hat H$ was the  Hamiltonian on the physical configuration
space $\mathcal M$.   This equivalence, however, depends crucially  on
the  correct   implementation of  the residual   gauge symmetries when
working  on the  gauge   fixing  surface $\Gamma_{\text{nc}}$.   These
residual symmetries have to be imposed as adequate symmetry conditions
on the  wave function (cf.  (\ref{Gl-344}) and  (\ref{Gl-344a})).  For
the Hamiltonian defined on  the physical configuration space $\mathcal
M$ we have to require regularity of the wave function  not only at the
regular   (``generic'') configurations,  but   also at    the singular
(``non-generic'') point  $r =  0$.  This  distinguishes  our treatment
from    general  discussions   of  quantum   mechanics   on  orbifolds
\cite{emmrich:90}, where in certain cases singular  values of the wave
function may be allowed at singular points of the configuration space.

Let  us end this  section on the ``2x2-model''  by summarizing what we
have  obtained so  far.  We  have  seen that  every choice  of a gauge
condition $\chi$ corresponds to a gauge fixing transformation
\begin{equation}
  \begin{pmatrix} a_1^1 & \; a_1^2 \\ a_2^1 & \; a_2^2 \end{pmatrix} =
  \begin{pmatrix} \tilde a_1^1 & \;\tilde a_1^2\\
                  \tilde a_2^1 & \;\tilde a_2^2 \end{pmatrix}
  \begin{pmatrix} \cos \varphi & \;  -  \sin \varphi \\
                  \sin \varphi & \;\quad\cos \varphi \end{pmatrix}\; ,
  \label{Gl-356} 
\end{equation}
where the $\tilde a_i^a$ satisfy  $\chi(\tilde a_i^a) = 0$.  The gauge
condition $\chi$, defining a certain hypersurface $\Gamma_\chi$ in the
space of all   (Weyl)  gauge configurations, the   ``pre-configuration
space''   ${\mathcal  A}_0$,     may   be  realized   by   a  suitable
parameterization $\tilde a_i^a(r_i)$ of $\Gamma_\chi$, where the $r_i$
are   gauge invariant coordinates.   On  the other hand,  every set of
gauge invariant variables given in terms  of a map $a_i^a(r_j)$ can be
related to  a gauge transformation (\ref{Gl-356}) and  thus to a gauge
condition $\chi$. Stated more precisely, every transformation to gauge
invariant variables  $r_i$  may also   be  considered as  defining   a
hypersurface $\Gamma  \in {\mathcal A}_0$   via the  map $r_i  \mapsto
a_i^a(r_i)$.      This surface  can   be    described by an   equation
$\chi(a_i^a)=0$,  which    we interpret  as   a gauge  condition, thus
establishing the connection  between gauge fixing and  gauge invariant
coordinates.   However, the solution   of $\chi(a_i^a)=0$ may  yield a
larger hypersurface $\Gamma_\chi  \supset \Gamma$ containing different
(Gribov) regions   related  by  residual  gauge  transformations  (cf.
Subsection \ref{a23}).  This also   happens, if we   do not know   the
domains of the chosen gauge invariant variables  from the beginning as
was the case for the polar representation.

Analyzing gauge equivalent configurations  on the gauge fixing surface
$\Gamma_\chi$, we have to distinguish discrete and continuous residual
gauge  symmetries.   We may  leave these   symmetries as  they are and
define wave  functions on $\Gamma_\chi$.   In  this case,  however, we
have   to  translate  the   residual  gauge  symmetries into  symmetry
conditions  imposed on the     wave  function defined on   the   space
$\Gamma_\chi$.   If,  on the  other hand,  we  want to   construct the
physical  configuration  space   $\mathcal  M$, we  have    to fix the
\emph{discrete} residual  symmetries relating different Gribov regions
by  restricting the   values  of the  gauge invariant  variables which
parameterize $\Gamma_\chi$ to one  Gribov  region (the  reduced  gauge
fixing  surface).  The \emph{continuous}  residual gauge symmetries on
the Gribov   horizon    can be implemented    by  choosing appropriate
\emph{additional} gauge    conditions.   This   corresponds  to    the
identification  of points on the Gribov  horizon such that in the end,
the  former  boundary of the  reduced  gauge fixing surface completely
vanishes. Due to this identification, the physical configuration space
$\mathcal   M$ cannot be  identified with  $\Gamma_\chi$.  Instead, we
have  to consider $\Gamma_\chi$ as  defining a chart for $\mathcal M$,
which  is  only valid   locally,   the Gribov horizon  indicating  the
breakdown of the respective coordinate system.

For  our simple  model we were  able  to demonstrate  the procedure of
boundary identifications  explicitly, since the physical configuration
space is only three dimensional and can be embedded in $\mathbb{R}^3$,
admitting a coordinate system covering all of $\mathcal M$. As we will
see in Section \ref{a4}, this is not possible in general. Although the
gauge group of our model was abelian, the physical configuration space
$\mathcal M$ turned out to be nontrivial.   In fact $\mathcal M$ has a
cone-like  structure with a singular point  at the zero configuration.
As predicted by  Shabanov  et al.  \cite{prokhorov:91}, the   complete
reduction of all gauge symmetries implies a  mixing of the coordinates
$a_i^a$, thus invalidating our  picture of two  particles moving in  a
plane.   This  prohibits an  ansatz for  the total wave  function as a
product of one-particle  functions. We also  point out, that,  even in
the case of a total reduction of all gauge symmetries, there remains a
zero   of the Jacobian  which   corresponds to the zero configuration.
This configuration is peculiar due to the fact that it is non-generic,
which means that it   is a fixed  point under  gauge  transformations.
Since the  gauge   group  $SO(2)$ does   not  contain  any  nontrivial
continuous  subgroups, the    zero   configuration  is actually    the
\emph{only} non-generic configuration, which   we have in the   ``$2\!
\times \!   2$-model''.    Thus, in  order to  study  more interesting
examples  of  such configurations,   we   need  to consider  a  larger
structure group.  This will be done in the next section.
%
%
\section{$SO(3)$ Yang-Mills theory of constant fields}
\label{a3}
%
%
The   main object  of    this    section  is to     study  non-generic
configurations. Therefore we extend the ``$2 \!  \times \!  2$-model''
of Section \ref{a2} by letting   the two ``particles'' $\vec a_1$  and
$\vec a_2$ move  in a three-dimensional  ``color space'' instead  of a
plane.   Hence  we will   write $\vec   a_i = a_i^a   \hat  e_a$ using
orthonormal basis vectors $\hat  e_1,  \hat e_2,  \hat e_3$  in  color
space  $\mathbb{R}^3$  and  call this model   the  ``$2 \!   \times \!
3$-model''.  The Lagrangian of our model is
\begin{equation}
  {\mathcal L}^{2 \! \times \! 3} = \frac{1}{2\,g^2}
  \sum_{i=1,2} \sum_{a=1}^3
  \big( \dot a^a_i + \epsilon^{abc} \: a_0^b \: a^c_i \big)
  \big( \dot a^a_i + \epsilon^{ade} \: a_0^d \: a^e_i \big) 
  - {\mathcal V}^{2 \! \times \! 3}(a_i^a) \; ,
  \label{Gl-401}
\end{equation}
where the   potential ${\mathcal V}^{2   \!  \times \!  3}$   shall be
invariant under the transformations
\begin{equation}
  a^a_i \mapsto a^b_i \: U^{ba} \quad \text{and} \quad 
  a_0^a \mapsto a_0^b \: U^{ba} + u_0^a\; .
  \label{Gl-402}
\end{equation}
The matrix $U \in SO(3)$  is an ordinary  $3 \!  \times \! 3$ rotation
matrix, parameterized by three  time-dependent angles $\phi^a(t)$. The
inhomogeneous  part  $u_0^a$  in the  transformation  law  for $a_0^a$
originates  from the term $\mathrm{i}  \,  U^{-1}\partial_t U$ in  the
general gauge transformation (\ref{Gl-101}).   For the time being
we will choose the Yang-Mills type potential
\begin{equation}
  {\mathcal V}^{2 \! \times \! 3}_{\scriptscriptstyle \text{YM}} 
  = \frac {1}{2g^2}\: |\, \vec a_1 \times \vec a_2|^2 \;.
  \label{Gl-403}
\end{equation}
A simplified version of ${\mathcal L}^{2 \! \times \!  3}$ with such a
potential has been studied by Levit  et al. \cite{levit:91}. Note that
${\mathcal V}^{2  \! \times \!   3}_{\scriptscriptstyle \text{YM}}$ is
proportional to the area squared of the parallelogram spanned by $\vec
a_1$ and $\vec a_2$.  Thus   the potential vanishes whenever the   two
vectors  are parallel or anti-parallel. Later  on we will also discuss
the  harmonic  oscillator potential  ${\mathcal   V}^{2 \!  \times  \!
  3}_{\text{osc}}$ as defined in (\ref{Gl-304b}).

It  is easy to check that  with such  a  gauge invariant potential the
Lagrangian ${\mathcal L}^{2 \!    \times  \!  3}$  (\ref{Gl-401})   is
invariant under the transformations  (\ref{Gl-402}), so that the  ``$2
\!  \times \!  3$-model'' is an  $SO(3)$ gauge model.  In fact, adding
a third ``particle''  $\vec a_3$ would  yield the Lagrangian  for pure
$SU(2)$ Yang-Mills  theory of constant  fields \cite{matinyan:81} (the
``$3 \!  \times \!  3$-model'').  However, taking three instead of two
``particles''  does  not  substantially   change  the  problem of  the
reduction to  the physical configuration  space $\mathcal M$.  So, for
simplicity  we will stick to  two particles,  which has the additional
advantage,  that  we  can take  over   some  of the   results from the
discussion  of the  ``$2 \!  \times  \!   2$-model''  in  the previous
section.

Passing   to the Hamilton   formalism  with the color-electric  fields
$e^{ia}$ as the canonical momenta for the variables $a_i^a$ we obtain
\begin{equation}
  {\mathcal H}^{2 \! \times \! 3} = \frac{g^2}{2} \: e^{ia} \, e^{ia}
  - a_0^a \: {\mathcal G}^a +
  {\mathcal V}^{2 \!\times\! 3}_{\scriptscriptstyle \text{YM}}(a_i^a)
  \label{Gl-404}
\end{equation}
with three Gau{\ss} constraints
\begin{equation}
  {\mathcal G}^a = \epsilon^{abc} \: a^b_i \: e^{ic} \;,
  \label{Gl-405}
\end{equation}
corresponding to  the three gauge  degrees of  freedom, $\phi^a$.  The
$a_0^a$  play the r\^ole  of  Lagrange multipliers of the  constraints
${\mathcal  G}^a$.  In analogy to the  ``$2 \!  \times \!  2$-model'',
they  can  be interpreted  as the components   of  an angular velocity
describing  the time dependent rotation   of the coordinate system  in
color space.  As   in   Section \ref{a2}  we  can  make  use   of  the
inhomogeneous transformation (\ref{Gl-402})  to  set  $a_0^a=0$  (Weyl
gauge).  Thus we  are left with   six coordinates $a_i^a$ forming  the
pre-configuration space ${\mathcal  A}_0$ equipped with the  Euclidean
metric
\begin{equation}
  g_{ij}^{ab} = \delta_{ij}\: \delta^{ab} \; .
  \label{Gl-406}
\end{equation}
Quantization   is    straightforward,  and using   the   Schr\"odinger
representation  with   wave functions depending    on the  coordinates
$a^a_i$ we obtain the Hamilton operator
\begin{equation}
  \hat H^{2 \! \times \! 3} = - \frac{g^2}{2} \! \bigtriangleup + \:
    {\mathcal V}^{2 \! \times \! 3}_{\scriptscriptstyle \text{YM}}(a)
  \label{Gl-408}
\end{equation}
with the Euclidean Laplace operator
\begin{equation}
  \bigtriangleup = \sum_{i=1}^2 \sum_{a=1}^3 \frac{\partial^2} 
  {\partial a^a_i \: \partial a^a_i }
  \label{Gl-409}
\end{equation}
on the pre-configuration space ${\mathcal  A}_0$.  In addition we have
to impose the constraints ${\mathcal  G}^a$ in operator form weakly on
the physical states
\begin{equation}
  \hat G^a\, | \Psi\rangle_{\text{phys}} = 0\; .
  \label{Gl-410}
\end{equation}
The  Weyl gauge $a_0^a  \! = \! 0$ does  not fix gauge transformations
(\ref{Gl-402}) with  a  time-independent  matrix  $U(\vec \varphi) \in
SO(3)$.  Let us  parameterize the rotation  matrices $U(\vec \varphi)$
as the  product of simple rotations around   the coordinate axes $\hat
e_a$
\begin{equation}
  U(\vec \varphi) = 
  U_1(\varphi^1) \: U_2(\varphi^2) \: U_3(\varphi^3) 
  \label{Gl-410a}
\end{equation}
with $U_a(\varphi)   =  \exp(\mathrm{i} \,  \varphi  \,  t^a)$ and the
generators $(t^a)^{bc} \!  = \!  -\mathrm{i} \, \epsilon^{abc}$ of the
adjoint representation  of $SU(2)$.  Like in  the preceding section we
interpret the transformations
\begin{equation}
  \begin{pmatrix} \tilde a_1^1 & \;\tilde a_1^2 & \;\tilde a_1^3\\ 
       \tilde a_2^1 & \; \tilde a_2^2 & \; \tilde a_2^3 \end{pmatrix}
  = \begin{pmatrix} a_1^1 & \; a_1^2 & \; a_1^3 \\
                    a_2^1 & \; a_2^2 & \; a_2^3 \end{pmatrix}
  \Big( U^{ab}(\vec\varphi) \Big)
  \label{Gl-411}
\end{equation}
as  the rotation of  the coordinate system  in  color space at a fixed
time $t$. Again, gauge invariance  requires the physical quantities to
be independent of the orientation of the coordinate axes.
%
%
\subsection{Planar gauge}
%
%
In order to establish the relation to  the ``$2\!  \times \!2$-model''
of Section \ref{a2} we rotate the coordinate  system in color space in
such a way,   that  the two  vectors  $\vec  a_1$ and $\vec   a_2$ are
confined to the plane spanned by $\hat  e_1$ and $\hat e_2$.  In terms
of gauge fixing, we impose the ``planar gauge'' conditions
\begin{equation}
  \chi_{\text{pl}}^1(\tilde a) = \tilde a_1^3 = 0
  \quad \text{and} \quad 
  \chi_{\text{pl}}^2(\tilde a) = \tilde a_2^3 = 0 \;.
  \label{Gl-412}
\end{equation}
The gauge fixing  transformation can be written as
\begin{equation}
  \begin{pmatrix} a_1^1 & \; a_1^2 & \; a_1^3 \\
                  a_2^1 & \; a_2^2 & \; a_2^3 \end{pmatrix} = 
  \begin{pmatrix} \tilde a_1^1 & \; \tilde a_1^2 & \; 0 \\
                  \tilde a_2^1 & \; \tilde a_2^2 & \; 0 \end{pmatrix}
  \: U^\dagger_2(\varphi^2) \; U^\dagger_1(\varphi^1) 
  \label{Gl-413}
\end{equation}
having Jacobian
\begin{equation}
  \left| \frac{ \partial (a_1^1, a_1^2, a_1^3, a_2^1, a_2^2, a_2^3) }
              { \partial (\tilde a_1^1,\tilde a_1^2,\tilde a_2^1,
                \tilde a_2^2,   \varphi^1,   \varphi^2 ) } 
  \right|_{\vec \varphi = 0} =  
  \: \tilde a_1^1\tilde a_2^2-\tilde a_1^2\tilde a_2^1 \;.
  \label{Gl-414}
\end{equation}
We expect   the zeros of  the  Jacobian to  indicate the  existence of
residual gauge symmetries, in addition to the remaining symmetry under
rotations $U_3(\varphi^3)$ around the $\hat e_3$-axis.  So let us look
for residual gauge symmetries by calculating
\begin{alignat}{2}
  \chi_{\text{pl}}^1(^U\tilde a)=&& 
  \:\tilde a_1^1 \, \cos \varphi^1 \: \sin \varphi^2 
  - \tilde a_1^2 \, \sin \varphi^1 = 0& \\
  \chi_{\text{pl}}^2(^U\tilde a)=&& 
  \:\tilde a_2^1 \, \cos \varphi^1 \: \sin \varphi^2 
  - \tilde a_2^2 \, \sin \varphi^1 = 0& \; .
  \label{Gl-415}
\end{alignat}
We    may  solve   these equations   choosing   $\varphi^1=n_1\pi$ and
$\varphi^2=n_2\pi$  ($n_i  \in \mathbb{N}$), which leads   to a set of
discrete  residual gauge transformations, corresponding to reflections
of the transformed vectors in the coordinates axes $\hat e_1$ or $\hat
e_2$.   These discrete symmetries   can   be eliminated by   requiring
$\tilde a_1^1 \geq 0$ and $\tilde a_2^2\geq 0$. If, on the other hand,
$\tilde a_1^1\tilde   a_2^2-\tilde a_1^2\tilde a_2^1=0$,  then  we are
free to  arbitrarily choose  one  rotation angle,  say $\varphi^1$, as
long as the other angle $\varphi^2$ satisfies
\begin{equation}
  \sin \varphi^2 = 
  \frac{\tilde a_1^2}{\tilde a_1^1} \tan \varphi^1 =
  \frac{\tilde a_2^2}{\tilde a_2^1} \tan \varphi^1 \; .   
  \label{Gl-416}
\end{equation}
In planar gauge the term $|\tilde a_1^1\tilde a_2^2-\tilde a_1^2\tilde
a_2^1|$ is  equal to the  area spanned by the  vectors  $\vec a_1$ and
$\vec a_2$.    We note, that   this gauge invariant quantity  can only
vanish if either one of the two vectors $\vec a_1$, $\vec a_2$ is zero
or both vectors are collinear.  The case of  a vanishing color  vector
has already  been discussed in a  similar context in Section \ref{a2}.
The latter, however,   corresponds to  a  new  feature of  the  ``$2\!
\times \! 3$-model''.  In fact, this is the  first indication of a new
type of  non-generic  configurations: collinear configurations,  where
$\vec a_1$   is  parallel or anti-parallel to   $\vec a_2$\footnote{In
  matrix  notation  $a_i=a_i^a   \sigma^a/2$   ($\sigma^a$ the   Pauli
  matrices):  $\:[a_1,a_2] = 0$.}.  The  freedom  to choose one  angle
$\varphi^1$ at will   corresponds to  a  nontrivial stability   group.
Intuitively we may understand this situation by observing that we need
only one  rotation,   say $U^\dagger_2(\varphi^2)$, to   transform the
parallel  vectors $\vec a_1$  and $\vec a_2$ into  the $(\hat e_1 \hat
e_2)$-plane,  whereas  rotations  around the  common  axis leave  this
configuration invariant, generating an $SO(2)$ stability group.

We  still need to fix  the  remaining  gauge  freedom with respect  to
rotations around  the $\hat e_3$-axis.  But for  this,  we can use the
results of  the last  section, because  in  planar gauge, the  ``$2 \!
\times  \!   3$-model''   is  reduced to    the  ``$2 \!    \times  \!
2$-model'', at   least on a formal   level.  Consider for  example the
classical Lagrangian (\ref{Gl-401}).  Putting  $a_1^3$ and $a_2^3$  to
zero  and identifying $a_0  \equiv   a_0^3$ we recover the  expression
(\ref{Gl-301})  for the  ``$2  \!   \times  \!   2$-model''; and   the
remaining  symmetry in planar gauge  is the $SO(2)$ symmetry discussed
in Section  \ref{a2}.   This equivalence  even  holds  on the  quantum
level.    In particular  the  Hamiltonian  $\hat   H^{2 \!  \times  \!
  3}_{\text   {pl}}$,   obtained   from plugging  the   transformation
(\ref{Gl-413}) into (\ref{Gl-408}) is
\begin{equation}
  \hat H^{2 \! \times \! 3}_{\text {pl}} = 
    -\frac{g^2}{2} \sum_{i,a=1}^2 
     \frac{\partial^2} {\partial \tilde a_i^a \partial \tilde a_i^a}
   + \frac{1}{2\,g^2} 
   \big( \tilde a_1^1\tilde a_2^2-\tilde a_1^2\tilde a_2^1 \big)^2
  \label{Gl-417a}
\end{equation}
and thus identical to the one  for the ``$2  \!  \times \!  2$-model''
(\ref{Gl-310}).   To  calculate $\hat  H^{2  \!   \times \!  3}_{\text
  {pl}}$ we have solved two of the  Gau{\ss} constraints, by demanding
physical  states not  to   depend on $\varphi^1$  and $\varphi^2$.  In
addition, from the Laplacian   in (\ref{Gl-417a}) we deduce  that  the
reduced configuration  space   in  planar gauge   is  still Euclidean.
Hence, we  can (with  some care) take  over most  of the results  from
Section \ref{a2}.  In  particular we can apply  the same gauge fixings
and related sets of gauge invariant coordinates to reduce the residual
gauge symmetry generated by   ${\mathcal G}^3$.  We might  also expect
the physical configuration space $\mathcal M$  to have a similar form.
However, as we have  already noticed, there  is the additional feature
of collinear configurations with   nontrivial stability group $SO(2)$.
%
%
\subsection{Natural coordinates}
\begin{figure}
  \begin{center}
    \leavevmode
    \epsfig{file=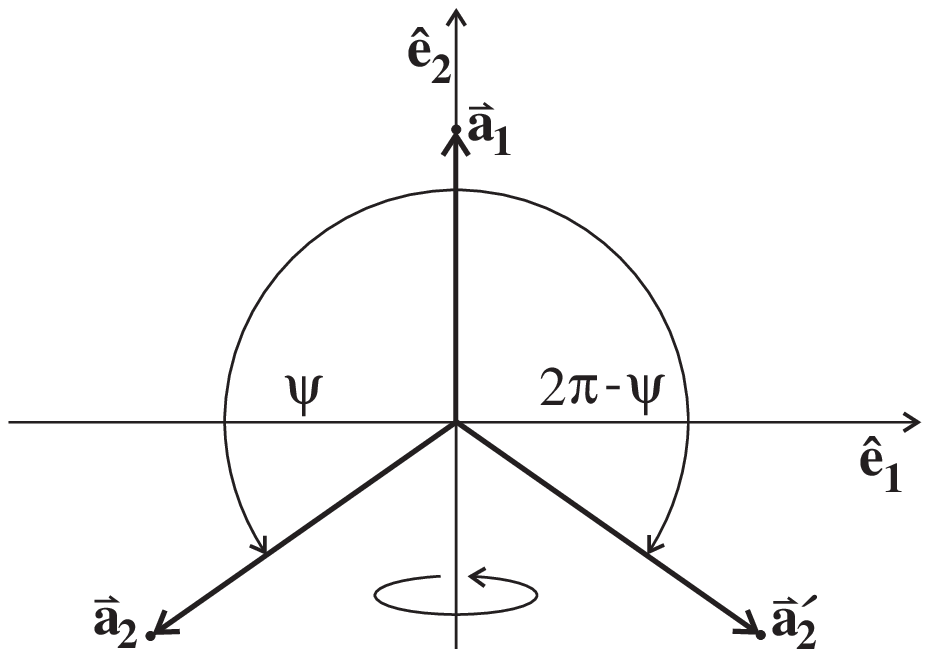,height=40mm}
  \end{center}
  \caption{The additional symmetry of the $2\!\times\!3$-model in 
    planar gauge  with respect to reflections  in the $\hat e_2$ axis:
    $\psi \sim 2\pi - \psi$}
  \label{fig-31}
\end{figure}

%
%
Having demonstrated the (formal) equivalence of the ``$2 \!  \times \!
3$-model'' in planar gauge with the ``$2 \!  \times \!  2$-model'', we
continue  the  reduction of the  former,  using the  system of natural
coordinates $r_1$,  $r_2$  and $\psi$.  Other  gauge choices just give
identical results (cf.  Section \ref{a2}).  As  before $r_1$ and $r_2$
correspond  to the lengths of the  vectors $\vec  a_1$ and $\vec a_2$,
whereas $\psi$ is the angle between them. However, $\psi$ is no longer
a polar angle defined on $S^1$.  Actually,  we have to identify $\psi$
and  $2\pi-\psi$, since  the  corresponding configurations are related
through a gauge transformation.   This is shown in  Fig. \ref{fig-31},
where the  transformation relating the  two configurations $(\vec a_1,
\vec a_2)$ and $(\vec a_1, \vec a'_2)$ is a rotation by an angle $\pi$
around the $\hat e_2$-axis.   Thus $\psi$ may  only take values in the
closed interval  $[\, 0,\pi]$ with  its boundary points $0$  and $\pi$
corresponding  to the non-generic  cases,  where $\vec a_1$  and $\vec
a_2$ are collinear.

The combination  of  the map   (\ref{Gl-413}),  rotating an  arbitrary
configuration to the planar  gauge  and the transformation to  natural
coordinates    (corresponding   to     the    gauge   condition      $
\chi_{\text{nc}}(\tilde   a)  = \tilde     a_1^1\tilde a_2^2 +  \tilde
a_1^2\tilde a_2^1$) yields
\begin{equation}
  \begin{pmatrix} a_1^1 & \; a_1^2 & \; a_1^3 \\
                  a_2^1 & \; a_2^2 & \; a_2^3 \end{pmatrix} =
  \begin{pmatrix} 
    r_1\:\cos\frac{\psi}{2} & \; -r_1\:\sin\frac{\psi}{2} & \; 0 \\ 
    r_2\:\cos\frac{\psi}{2} & \; \quad r_2\:\sin\frac{\psi}{2} & \; 0
  \end{pmatrix} U^\dagger(\vec \varphi) \;,
  \label{Gl-422}
\end{equation}
where  $U^\dagger(\vec     \varphi)     =  U^\dagger_3(\varphi^3)   \,
U^\dagger_2(\varphi^2) \,  U^\dagger_1(\varphi^1)$.  The   Jacobian of
(\ref{Gl-422}) is given by
\begin{equation}
  \left| \frac{ \partial( a_1^1, a_1^2, a_1^3, a_2^1, a_2^2, a_2^3 )}
         { \partial( r_1, r_2, \psi, \varphi^1,\varphi^2,\varphi^3 )} 
  \right|_{\vec \varphi = 0} = -\: (r_1 r_2)^2 \: \sin\psi\;.
  \label{Gl-423}
\end{equation}
In contrast to  the result (\ref{Gl-336}) of Section  \ref{a2} we have
an additional  factor $\sin\psi$ vanishing exactly  on the boundary of
the domain of $\psi$.

Upon quantizing (\ref{Gl-405}) the Gau{\ss} operators expressed in the
new   coordinates     $(r_1,   r_2,     \psi,     \varphi^a)$   become
\cite{christ:80,cheng:87}
\begin{equation}
  \hat G^a = (T^{-1}[\vec \varphi \,])^{ba} 
  \frac{\partial}{\mathrm{i}\,\partial\varphi^b}\;,
  \label{Gl-424}
\end{equation}
where the matrix $T[\vec \varphi]$ is defined via the relation
\begin{equation}
  \mbox{tr} \left( U[\vec \varphi\,]\: 
                   \mathrm{d}U^{-1}[\vec \varphi \,] \: t^a \right) 
  = \mathrm{i} \: T^{ab}[\vec \varphi \,]\, \mathrm{d}\varphi^a \;. 
  \label{Gl-425}
\end{equation}
The matrix $T$ is invertible as long as  $\det T = \cos(\varphi^2)\neq
0$, indicating a coordinate singularity of the chosen parameterization
of the gauge group elements $U(\vec\varphi)$.  We solve the conditions
(\ref{Gl-410}), requiring the physical states to be independent of the
gauge    variant     variables  $\varphi^a$.    Thus the   Hamiltonian
(\ref{Gl-408}), transformed to natural coordinates via (\ref{Gl-422}),
is
\begin{equation}
  \hat H^{2 \! \times \! 3}_{\text {nc}} = 
   -\frac{g^2}{2} \sum_{i=1}^2 \left(
    \frac{1}{r_i^2} \frac{\partial}{\partial r_i}
       r_i^2        \frac{\partial}{\partial r_i}
   +\frac{1}{r_i^2} \frac{1}{\sin\psi}\frac{\partial}{\partial \psi}
                     \sin\psi\frac{\partial}{\partial \psi} \right)
  + \frac{1}{2\, g^2} \, (r_1r_2\sin\psi)^2 \; .
  \label{Gl-427}
\end{equation}
The  metric on  the gauge fixing  surface  $\Gamma_{\text{nc}}$ may be
calculated in  terms of the  coordinates $r_1$,  $r_2$ and  $\psi$  as
demonstrated in Section \ref{a2},
\begin{equation}
  g_{\Gamma} =
  \begin{pmatrix}
    1 & \quad 0 & \;\, 0 \\ 0 & \quad 1 & \;\, 0 \\ 
    0 & \quad 0 & \:\frac{r_1^2\:r_2^2}{r_1^2+r_2^2}
  \end{pmatrix} \; ,
  \label{Gl-428}
\end{equation}
resulting in a non-vanishing   scalar curvature $R =  6/(r_1^2+r_2^2)$
which is identical  to  the one derived  for   the ``$2 \! \times   \!
2$-model''.

Suppose we had started from the gauge conditions $\chi_{\text{pl}}^1$,
$\chi_{\text{pl}}^2$ and $\chi_{\text{nc}}$.   Then we would  not know
anything about the  possible values of the  variables $r_1$, $r_2$ and
$\psi$  parameterizing   the    corresponding gauge   fixing   surface
$\Gamma_\chi$.  However, a similar analysis as  carried out in Section
\ref{a2} (cf. (\ref{Gl-337}) and (\ref{Gl-337a})) yields the following
residual symmetries
\begin{align}
  \label{Gl-428a}
  (r_1,r_2,\psi) &\sim ( r_1,-r_2, \psi+\pi)  \;,    \\
  (r_1,r_2,\psi) &\sim (-r_1, r_2, \psi+\pi)  \;,\nonumber \\
  (r_1,r_2,\psi) &\sim ( r_1, r_2, \psi+2\pi) \;,\nonumber \\
  (r_1,r_2,\psi) &\sim (-r_1,-r_2, \psi)      \;;\nonumber \\
  \label{Gl-428b}
  (r_1, 0 ,\psi) &\sim ( r_1,  0 , \psi') \;,\\
  ( 0 ,r_2,\psi) &\sim (  0 , r_2, \psi') \;;\nonumber 
\end{align}
with arbitrary $\psi$, $\psi'$ and
\begin{equation}
  \label{Gl-428c}
  (r_1,r_2,\psi) \sim ( r_1, r_2,2\pi -\psi) \;.
\end{equation}
We     recognize  the   discrete     (\ref{Gl-428a})  and   continuous
(\ref{Gl-428b})  residual gauge symmetries  that we have already found
for  the ``$2\!  \times  \!    2$ - model''  (cf.  (\ref{Gl-337})  and
(\ref{Gl-337a})),  but we also  have recovered the additional symmetry
(\ref{Gl-428c}) discussed before.  So   let us eliminate the  discrete
residual symmetries by restricting  the values of the gauge  invariant
variables to $r_1,r_2 \in \mathbb{R}^+_0$ and $\psi \in [\, 0,\pi]$.

The  continuous  residual gauge symmetries  implied by (\ref{Gl-428b})
may be reduced  by  analogy to the boundary   identification procedure
carried out in detail in  the previous section. Thus we reparameterize
$r_1=r\,\sin \vartheta /2$,  $r_2=r\,\cos \vartheta /2$ and  interpret
$r \in  \mathbb{R}^+_0$, $\vartheta  \in  [\,0, \pi]\:$ and  $\psi$ as
spherical coordinates in $\mathbb{R}^3$:
\begin{equation}
  x_1 = r\:\sin\vartheta\:\cos\psi,\quad
  x_2 = r\:\sin\vartheta\:\sin\psi \quad\text{and}\quad
  x_3 = r\:\cos\vartheta.
  \label{Gl-430}
\end{equation}
Due   to the additional   symmetry (\ref{Gl-428c}) and the restriction
$\psi \in [\,0, \pi]\:$  the physical configuration space $\mathcal M$
obtained   by  eliminating   all   residual   gauge   symmetries   and
parameterized  by the variables $r,  \vartheta,\psi$  now only forms a
half space, whereas in the ``$2\!   \times \! 2$-model'' it filled the
entire   $\mathbb{R}^3$.   Hence, the   physical configuration   space
$\mathcal  M$   has  a genuine  boundary   determined  by $\psi=0$ and
$\psi=\pi$ (or $x_2=0$). The Jacobian corresponding to the coordinates
$x_i$ is
\begin{equation}
  \left| \frac{\partial(\,a_1^1, a_1^2, a_1^3, a_2^1, a_2^2, a_2^3\:)}
    {\partial (x_1,x_2,x_3,\varphi^1,\varphi^2,\varphi^3)} \right|
  = \frac{1}{8}\: x_2 \: r^2  \;,
  \label{Gl-433}
\end{equation}
indicating  the singularities  at the origin $r^2=x_1^2+x_2^2+x_3^2=0$
and on the boundary $x_2=0$.  For practical purposes  and for the sake
of intuition we   introduce another set  of  angles $\Theta  \in [\,0,
\pi/4]$   and $\Phi \in  [\,0, 2\pi[\:$  on the physical configuration
space   $\mathcal M$ via $  x_1   = r\:\sin2\Theta\:\sin\Phi$, $x_2  =
r\:\cos2\Theta$ and $x_3 = r\:\sin2\Theta\:\cos\Phi$.  Choosing also a
different embedding  in    $\mathbb{R}^3$, where  we now    label  the
cartesian axes by $y_i$,
\begin{equation}
  y_1 = r\:\sin\Theta\:\sin\Phi \, , \quad
  y_2 = r\:\cos\Theta \, , \quad
  y_3 = r\:\sin\Theta\:\cos\Phi \, ,
  \label{Gl-435}
\end{equation}
the  physical configuration  space $\mathcal  M$ takes  the  form of a
cone.
\begin{figure}
  \begin{center}
    \leavevmode
    \epsfig{file=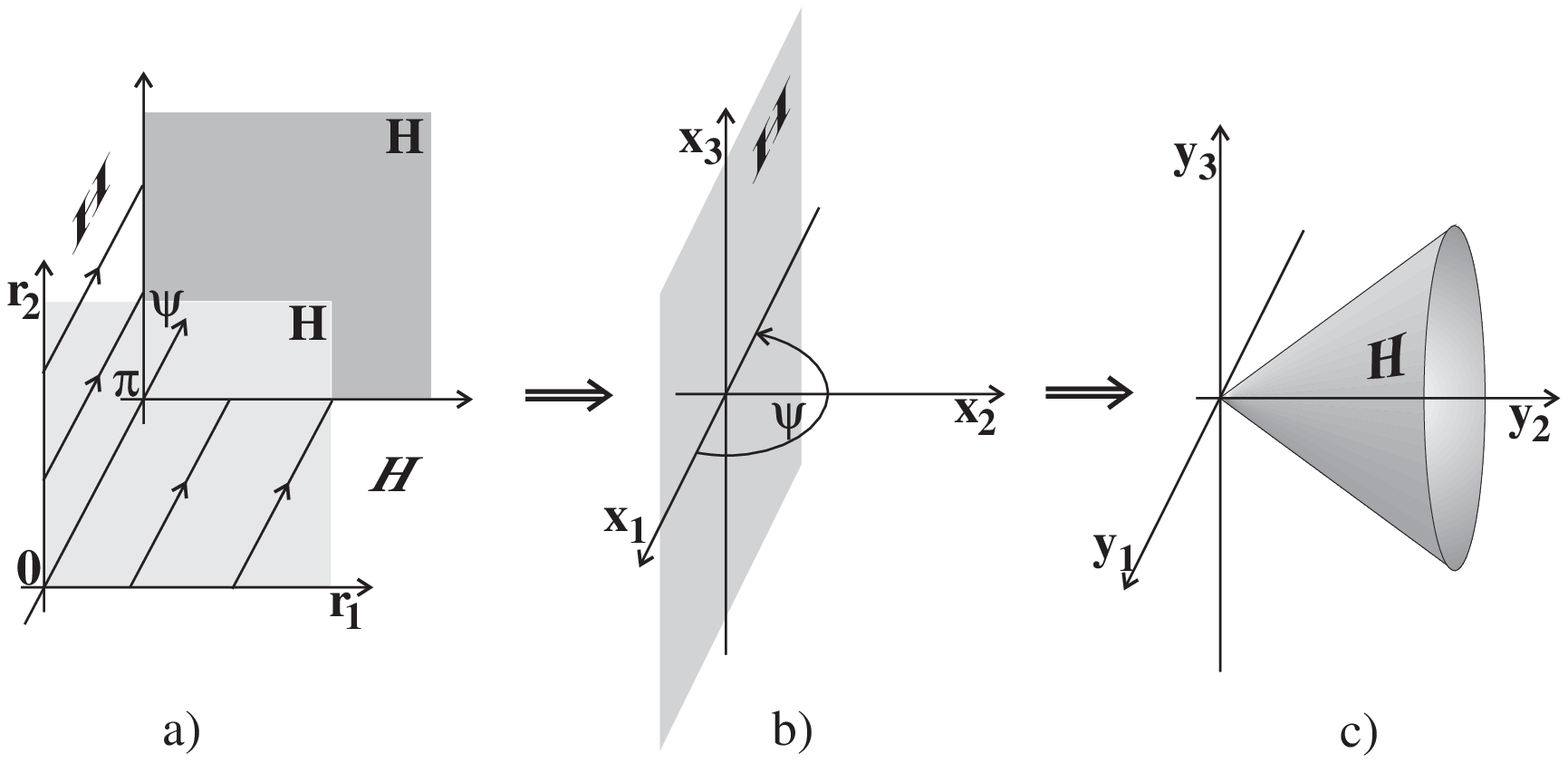,height=50mm}
  \end{center}
  \caption{Construction of the physical configuration space $\mathcal
    M$: a) Embedding of the  reduced gauge fixing surface, defined via
    $r_i \geq 0$ and $\psi \in [0, \pi]$ with residual gauge orbits on
    part of the Gribov  horizon $H$; b)  Embedding of $\mathcal M$ via
    (\ref{Gl-430}) and c) via (\ref{Gl-435}) with the remaining Gribov
    horizon $H$ due to non-generic configurations; }
  \label{fig-32}
\end{figure}
In Fig.    \ref{fig-32} we   have tried  to  visualize the   different
possibilities  for  embeddings  of  the  physical configuration  space
$\mathcal M$ into $\mathbb{R}^3$.   The Fig. \ref{fig-32}a)  indicates
the  restriction  of    the  gauge   fixing surface   $\Gamma_\chi   =
\mathbb{R}^3$ to the (Gribov) region defined by $r_i \geq 0$ and $\psi
\in [\:0, \pi]\:$ eliminating discrete  residual gauge symmetries. The
surfaces at $\psi=0,\pi$ are genuine  boundaries of the  configuration
space corresponding to the  non-generic configurations  with stability
group  $SO(2)$.   Since   we  still have    continuous residual  gauge
symmetries on the boundaries   $r_i=0$, this Gribov region  cannot  be
identified with the physical configuration   space $\mathcal M$   yet.
The  analogous boundary identifications as  in the case  of the ``$2\!
\times \!    2$-model'' now  yield  the physical   configuration space
$\mathcal  M$,  which   we  embed  into  $\mathbb{R}^3$  according  to
(\ref{Gl-430}).    As shown in    Fig.  \ref{fig-32}b),  the  physical
configuration space $\mathcal  M$  is homeomorphic  to the  half space
$x_2\geq 0$ with its boundary at $x_2 \!  = \!  0$ formed by collinear
configurations.  The set of coordinates $(r,\Theta, \Phi)$ is used for
the  embedding shown in Fig.  \ref{fig-32}c),   where $\mathcal M$ has
become a cone, the non-generic  collinear configurations making up its
surface.  $\mathcal M$ also includes the tip of the cone at the origin
$r=0$,  which actually  is the  only  configuration having  the entire
gauge group  $SO(3)$ as its  stability group.   Thus,  we may call the
zero configuration the ``most non-generic'' configuration.

For the new set of coordinates ($r,\Theta, \Phi$  on $\mathcal M$) the
Jacobian is
\begin{equation}
  \left| \frac{\partial(a_1^1, a_1^2, a_1^3, a_2^1, a_2^2, a_2^3)}
    {\partial (r,\Theta,\Phi,\varphi^1,\varphi^2,\varphi^3)} \right|
  = \frac{1}{8}\: r^5 \sin(4\Theta) \;.
  \label{Gl-437}
\end{equation}
Using  the Gau{\ss}   constraints  (\ref{Gl-424})  to   eliminate  all
dependence on the gauge   variant variables $\varphi^a$ we obtain  the
Hamilton operator on the physical configuration space $\mathcal M$
\begin{align}
  \hat H^{2 \! \times \! 3}_{\mathcal M} = & - \frac{g^2}{2} \!\! 
   \left( \frac{1}{r^5} \frac{\partial}{\partial r}
              r^5       \frac{\partial}{\partial r} + 
   \frac{1}{r^2 \sin(4\Theta)} \: \frac{\partial}{\partial \Theta} \: 
     \sin (4\Theta)            \: \frac{\partial}{\partial \Theta} + 
   \frac{4}{r^2\sin^2 (2\Theta)} \: \frac{\partial^2}{\partial \Phi^2}
   \right) \nonumber\\ 
  & + \frac{1}{8\,g^2}\: r^4\:\cos^2 (2\Theta)  \; .
  \label{Gl-438}
\end{align}
The metric $g_{\mathcal M}$ on $\mathcal M$ is easily calculated to be
\begin{equation}
  g_{\mathcal M} = \begin{pmatrix}
    1 &  \qquad \; 0  &  \; 0  \\ 
    0 & \qquad \; r^2 &  \; 0  \\
    0 &  \qquad \; 0  &  \; \frac{1}{4}\:r^2 \sin^2 (2\Theta)
  \end{pmatrix} \; ,
  \label{Gl-439}
\end{equation}
which  determines the scalar curvature   to be $R=6/r^2$, singular  at
$r=0$.
%
%
\subsection{Quantum mechanics}
%
%
It  is   interesting  to   note,  that   the  Yang-Mills potential  in
(\ref{Gl-438}) takes its minimum at $\Theta=\pi / 4$, which is exactly
on the boundary of  $\mathcal M$. From the discussion  of the  ``$2 \!
\times  \! 2$-model'' it  is clear, that the  present model also has a
discrete energy spectrum. Unfortunately the  quartic dependence on $r$
and the  factor $\cos^2 (2\Theta)$   make it impossible to  solve  the
Schr\"odinger  equation exactly.  However,  for a consistency check we
would like  to compare  the  energy spectra obtained by   defining the
Hamilton operator  on  the gauge fixing   surface $\Gamma_\chi$ to the
case, where  its domain is the  physical configuration space $\mathcal
M$.  We also want to  study the behavior  of the wave functions on the
boundary  of $\mathcal    M$.  Therefore we   replace  the  Yang-Mills
potential ${\mathcal   V}^{2   \!  \times \!    3}_{\scriptscriptstyle
  \text{YM}}$ with  the harmonic oscillator potential ${\mathcal V}^{2
  \! \times \! 3}_{\text{osc}} = a^a_i a^a_i / 2$.

Let us  solve the  Schr\"odinger  equation for  the  Hamilton operator
$\hat  H^{2 \!  \times \!  3}_{\text  {nc}}$ (\ref{Gl-427}) in natural
coordinates with the harmonic oscillator potential $(r_1^2+r_2^2)/2$,
\begin{equation}
  \left( \sum_{i=1}^2 \left( 
      \frac{1}{r_i^2}\frac{\partial}{\partial r_i}
                r_i^2\frac{\partial}{\partial r_i} 
    + \frac{1}{r_i^2\, \sin\psi}\,\frac{\partial}{\partial \psi}
      \,\sin\psi\, \frac{\partial}{\partial \psi}
    - \frac{r_i^2}{g^2} \right)\! 
  + \frac{2}{g^2}\, E \right) \Psi(r_1,r_2,\psi) = 0 \;.
  \label{Gl-440}
\end{equation}
The  wave  function $\Psi$ is   defined on   the gauge fixing  surface
$\Gamma_\chi$  where   $\chi_{\text{pl}}^1    =  \chi_{\text{pl}}^2  =
\chi_{\text{nc}} = 0$. Thus, the coordinates are not restricted to one
Gribov  region (i.e.  $r_i  \in  \mathbb {R}$).  Therefore, we have to
translate  the  residual symmetries (\ref{Gl-428a}-\ref{Gl-428c}) into
symmetry conditions  to be imposed  on the  wave function.  Apart from
the additional symmetry,
\begin{equation}
  \Psi(r_1,r_2,\psi) = \Psi(r_1,r_2,2\pi-\psi) \; ,
  \label{Gl-429}
\end{equation}
these are  identical  to  the ones   stated in  Section  \ref{a2} (cf.
(\ref{Gl-344}) and  (\ref{Gl-344a})).   The  energy  spectrum is  most
easily  calculated  by making  the ansatz  $\Psi(r_1,r_2,\psi) \propto
R_1(r_1) \, R_2(r_2)\,   P_l(\cos\psi)$, where $P_l(z)$  are  Legendre
polynomials.  Due to one of the symmetry conditions (\ref{Gl-344}) and
the requirement of   $\Psi$ being regular, $l$ has   to be a  positive
integer ($l=0,1,2,\ldots$).  Note that in particular $\cos(2\pi-\psi)=
\cos\psi$, such that  (\ref{Gl-429}) is  automatically satisfied.  The
radial equations are solved using Laguerre polynomials
\begin{equation}
  R_i(r_i) \propto
    r_i^l \; \mathrm{e}^{-\frac{1}{2g}r_i^2} \:
    L_{n_i}^{l+\frac{1}{2}}\!({\textstyle \frac{r_i^2}{g} })
    \qquad (i=1,2)\;,
  \label{Gl-442}
\end{equation}
characterized by the   radial quantum numbers  $n_1$  and  $n_2$ being
positive integers ($n_i=0,1,2,\ldots$).  The spectrum of $\hat H^{2 \!
  \times \! 3}_{\text {nc}}$ is given by
\begin{equation}
  E_\nu = 2 \: g \: \left(\nu +\frac{3}{2} \right) 
  \qquad\text{with}\quad \nu =n_1+n_2+l \;.
  \label{Gl-443}
\end{equation}
Hence, the ground state energy is $E_0=3\,g$ and the energy levels are
equidistant with  spacing  $\Delta  E=2\,g$. The   degeneracy of  each
energy level $E_\nu$ is
\begin{equation}
  g_\nu=\frac{1}{2} \, (\nu+1)(\nu+2) \; .
  \label{Gl-444}
\end{equation}
Like in Section \ref{a2} it can  easily be checked, that the solutions
obey     the   symmetry    conditions   (\ref{Gl-344},  \ref{Gl-344a},
\ref{Gl-429}).  Thus, after correctly normalizing these solutions with
respect to the Jacobian (\ref{Gl-423}), we accept them as the physical
states.

On the   physical configuration space  $\mathcal  M$ the Schr\"odinger
equation has the form
\begin{equation}
  \left( \!\frac{1}{r^5}\frac{\partial}{\partial r}
                    r^5 \frac{\partial}{\partial r}\!+\!\frac{16}{r^2}
   \bigg( \frac{1}{\sin\theta} \frac{\partial}{\partial \theta}
                    \sin\theta \frac{\partial}{\partial \theta} \!+\! 
   \frac{4}{r^2 \, \sin^2 \frac{\theta}{2}}
   \frac{\partial^2}{\partial \Phi^2} \! \bigg)\!
   - \frac{r^2}{g^2} \! + \! \frac{2 E}{g^2} \!
  \right) \!\Psi(r,\theta,\Phi) = 0 \, ,
  \label{Gl-445}
\end{equation}
where    we  have    replaced $\theta=4\Theta   \in   [\:0,\pi]\:$ for
simplicity.  Again a separation ansatz is in order,
\begin{equation}
  \Psi(r,\theta,\Phi) = R(r)\:Y_{lk}(\theta,\Phi)\;,
  \label{Gl-446}
\end{equation}
where the  angular   dependence    is given by    Jacobi   polynomials
$P_n^{(\alpha,\beta)}$ \cite{gradshteyn:81}
\begin{equation}
  Y_{lk}(\theta,\Phi) = \sqrt{\frac{2l+1}{\pi}} \; 
  \left(\sin \frac{\theta}{2}\right)^{|k|} \; 
  P_{\:l-|\frac{k}{2}|}^{(|k|,0)}\!\Big(\cos \theta \Big) 
  \; \mathrm{e}^{\mathrm{i} \, k \, \Phi}\; .
  \label{Gl-447}
\end{equation}
Since  $Y_{lk}$ has to be  regular on the whole physical configuration
space $\mathcal M$ (boundaries included!), the quantum number $l$ must
be a positive half-integer ($l=0,  1/2, 1,  3/2,  \ldots$) and $k$  an
integer with $k  =  -2l, -2l+2, \ldots,  2l-2,  2l$. The radial   part
of the Schr\"odinger equation yields
\begin{equation}
  R(r) \propto
  r^{4l}\; \mathrm{e}^{-\frac{1}{2g}r^2} L_n^{4l+2}(r^2/g) \;,
  \label{Gl-448}
\end{equation}
where regularity conditions, in  particular at $r \!  = \! 0$ (the tip
of the cone), require $n$ to be an integer. The energy spectrum is
\begin{equation}
  E_\nu = 2\: g \: \left( \nu+\frac{3}{2} \right)
  \quad\text{with}\quad \nu=n+2\,l \;.
  \label{Gl-449}
\end{equation}
As  expected, we  have the same  energy levels  as  for  $\hat H^{2 \!
  \times   \!    3}_{\text   {nc}}$ on    the   gauge  fixing  surface
$\Gamma_\chi$ (\ref{Gl-443}).  In addition, the degeneracy of $E_\nu$,
given by
\begin{equation}
  g_\nu = \frac{1}{2} \, (\nu+1)(\nu+2) \; ,
  \label{Gl-450}
\end{equation}
is in agreement with (\ref{Gl-444}).

Let  us finally examine the behavior of  the solutions on the boundary
of the physical configuration space $\mathcal M$.  On the gauge fixing
surface $\Gamma_\chi$, parameterized in natural coordinates, we obtain
$\Psi(r_1,r_2,0)=  R_{n_1l}(r_1)\,R_{n_2l}(r_2)$ for $\psi=0$, so that
$\Psi$ is  regular but not constant  on the boundary.  This also holds
for the non-generic  configurations  with $\psi=\pi$.  Let  us compare
this  to  the wave  function    $\Psi(r,\theta,\Phi)$ defined on   the
physical configuration space $\mathcal M$, where the boundary is given
by $\theta \! = \! \pi$.   Since the $\theta$-dependent part of $\Psi$
remains finite at  $\theta \!  = \!  \pi$,   on the boundary the  wave
function can be  any  regular function of  $r$  and $\Phi$.   In  this
respect, there is no difference between singular and regular points in
$\mathcal M$. Note,      however, that  on  the    boundary  $\partial
\Psi/\partial \theta = 0$.

Let  us finally summarize the main  results of  this section.  We have
studied the ``$2\!  \times \!3$-model'', which  may be considered as a
simplified  version of   the  $SU(2)$ Yang-Mills theory of   spatially
constant fields.  Due to the fact that  the relevant group $SO(3)$ has
nontrivial  continuous  $SO(2)$  subgroups, we   found a  new  type of
non-generic  configurations apart from  the  zero configuration  being
invariant  under  the  whole  gauge group $SO(3)$.   We   were able to
explicitly   demonstrate  that these    configurations form a  genuine
boundary of the physical configuration space $\mathcal M$, contrary to
the apparent boundary    we  encountered for  the ``$2\!    \times  \!
2$-model'', which disappeared  after performing the necessary boundary
identifications.  For a certain embedding in $\mathbb{R}^3$, $\mathcal
M$  could  be  visualized   as  a cone   with the  tip  at  the origin
corresponding to the ``most non-generic'' configuration at $\vec a_1\!
= \! \vec a_2 \! = \!  0$.  We have also  shown, that the zeros of the
Jacobian stemming from  non-generic configurations remain, although it
is possible to eliminate the gauge degrees of freedom completely, i.e.
to find  a coordinate system  covering all of  $\mathcal M$.  Again we
have verified the equivalence of  two different methods of solving the
quantum mechanical problem: Defining  the domain of  the Schr\"odinger
wave functions as the gauge  fixing surface $\Gamma_\chi$ leads to the
same     results as  the   calculation    performed  on   the physical
configuration  space $\mathcal M$,  if  in the  first case appropriate
symmetry  conditions are imposed on  the wave function.  In the latter
case we  only  had to  require   regularity of the wave  function,  in
particular on     the boundary of    $\mathcal  M$.  Unfortunately the
complicated   form of  the  Yang-Mills  potential ${\mathcal V}^{2  \!
  \times \!  3}_{\scriptscriptstyle \text{YM}}$ makes it impossible to
solve the   corresponding  Schr\"odinger  equation exactly.    It  is,
however, interesting to note, that the Yang-Mills potential ${\mathcal
  V}^{2 \!  \times   \!  3}_{\scriptscriptstyle \text{YM}}$   seems to
favor   the non-generic configurations, as   it vanishes for $\vec a_1
\times \vec a_2 = 0$.
%
%
\section{$SU(2)$ Yang-Mills theory on a cylinder}
\label{a4}
%
%
In  this last  section we  will apply our  methods to  pure Yang-Mills
theory on a cylinder   $S^1\!\times \mathbb{R}$ with   structure group
$SU(2)$.  After the   elimination  of all gauge  dependent  degrees of
freedom   this field  theoretical model    is  reduced  to a   quantum
mechanical one, thus fitting in the series of models discussed in this
paper.  The properties of  this exactly solvable model  are well-known
\cite{migdal:75,rajeev:88,langmann:93,shabanov:93a,jahn:96}, hence the
motivation for  this section is  to  obtain a better understanding  of
these results using a different approach along  the lines presented in
the  previous sections.  Our  main objective will  be to  show how the
configuration space  is reduced to  the  structure group and  finally,
after dividing out constant  gauge  transformations, to the   physical
configuration space $\mathcal M$.

So  let    us consider  $SU(2)$  Yang-Mills   theory  on  a  $1\!+\!1$
dimensional  space-time,  where space  is  compactified to a circle of
length $L$.  The Lagrangian is
\begin{equation}
  {\mathcal L} = 
  \frac{1}{2} \:
  \big(\partial_0 A_1^a - D_1^{ab}A_0^b \big)\,
  \big(\partial_0 A_1^a - D_1^{ac}A_0^c \big)
  \label{Gl-501}
\end{equation}
where  $D_1^{ab} = \delta^{ab}  \partial_1 -  \epsilon^{abc} A_1^c$ is
the covariant derivative and fields are expanded in terms of the Pauli
matrices, $A_\mu = A_{\mu}^a\  \, \sigma^a/2$.  There is no Yang-Mills
potential in $1\!+\!1$ dimensions, therefore the Hamiltonian expressed
in  terms  of color-electric  fields  $E^{1a} :=  F_{01}^a/g^2$ simply
reads
\begin{equation}
  {\mathcal H} = \frac{g^2}{2} \: E^{1a} E^{1a} 
               - A_0^a \: {\mathcal G}^a \; ,
  \label{Gl-502}
\end{equation}
where we have performed  a partial integration in  order to obtain the
Gau{\ss} constraints ${\mathcal G}^a = D_1^{ab}E^{1b}$. The Lagrangian
(\ref{Gl-501})  is    invariant    under    gauge      transformations
(\ref{Gl-101}), which we will write as \cite{christ:80}
\begin{equation}
  A_\mu^a \mapsto \tilde A_\mu^a =
  A_\mu^b \: U^{ba}[\vec \varphi \, ] + u_\mu^a[\vec \varphi \, ] \;, 
  \label{Gl-504}
\end{equation}
where   $U[\vec  \varphi(x,t)]  \in     SU(2)$   is in   the   adjoint
representation and  parameterized by three space-time dependent angles
$\varphi^1$, $\varphi^2$ and $\varphi^3$.  The translational terms  in
(\ref{Gl-504}) are  given  by  $u_\mu^a =  \mathrm{i} \,  \text{tr}(\,
U_f^{-1} \, \partial_\mu U_f \, \sigma^a)$ with $U_f = \exp(\mathrm{i}
\, \varphi^a \sigma^a/2)$ in the fundamental representation. 

Since the gauge  fields are in the   adjoint representation, they  are
invariant under constant  gauge   transformations with values   in the
center $\mathbb{Z}_2$ of the  structure  group $SU(2)$. Therefore,  we
can  replace   $SU(2)$  by  $SU(2)/\mathbb{Z}_2 =    SO(3)$,  which is
topologically nontrivial,  its fundamental group being $\mathbb{Z}_2$.
As a consequence, we  have two  types of  gauge group elements  $U \in
{\mathcal G}$, where   $\mathcal G$ is (loosely  speaking)  the set of
maps from $S^1 \!  \times \mathbb{R}$ into the  structure group. If we
require  the gauge fields  $A_\mu$ to  be periodic  in $x$, $A_\mu(0)=
A_\mu(L)$, the elements $U \in {\mathcal G}$  have to be periodic only
up to an element  of the center $\mathbb{Z}_2=\{ \pm 1\!\!\!\:\text{l}
\}$ \cite{tHooft:79}.  For small  gauge transformations (connected  to
unity)  $U$  is    periodic, $U(L)=U(0)$,  whereas   for  large  gauge
transformations, $U$ is anti-periodic, $U(L)=-U(0)$.  We can represent
any element $U \in {\mathcal G}$ as a product,
\begin{equation}
  U[\vec \varphi(x,t)] = 
  \exp( \mathrm{i} \, \varphi^a(x,t) \, t^a) \cdot
  V^n[\hat \varphi(x,t)] \; ,
  \label{Gl-504a}
\end{equation}
with  $(t^a)^{bc}  = -  \,  \mathrm{i} \,  \epsilon^{abc}$  and $n \in
\mathbb{N}$, where  the gauge parameter functions $\varphi^a(x,t)$ are
periodic in $x$ and therefore can be Fourier expanded,
\begin{equation}
  \label{Gl-504b}
  \varphi^a(x,t) = \sum_k \bar \varphi^a_k(t) \; 
    \mathrm{e}^{2\pi \mathrm{i} \, k \, \frac{x}{L}} \; .
\end{equation}
The second factor in (\ref{Gl-504a}) is given by
\begin{equation}
  \label{Gl-504c}
  V^n[\hat \varphi(x,t)] = \exp( \mathrm{i} \, n \pi \,  
    \hat \varphi(x,t) \! \cdot \! \vec \sigma \, \frac{x}{L} ).
\end{equation}
Hence for  $n$ even, $V^n$ is  periodic  and therefore belongs  to the
class  of    small  gauge   transformations,  whereas   for   $n$ odd,
$V^n(x+L)=-V^n(x)$,  so that $V^n$  and  in particular  $V:=V^1$ is  a
large   gauge transformation.     Note that  for  infinitesimal  gauge
transformations  with infinitesimal $\varphi^a(x,t)$,  $n$ has   to be
zero.

As usual we  start the reduction of  the configuration space $\mathcal
A$ by transforming  the configurations  into the Weyl-gauge,  $A_0=0$,
with
\begin{equation}
  U^W_f = {\mathcal T} \exp \big( \, \mathrm{i} \, 
  \int\limits_0^t A_0(x,\tau) \, \mathrm{d} \tau \big) \;.
  \label{Gl-505}
\end{equation}
This yields the Hamiltonian
\begin{equation}
  {\mathcal H} = \frac{g^2}{2} \: E^{1a} E^{1a} \; .
  \label{Gl-506}
\end{equation}
The     remaining     coordinates   $A_1^a(x)$   parameterize      the
pre-configuration space ${\mathcal A}_0$, which has a Euclidean metric
following     from     the Killing   form       (trace)   on the   Lie
algebra\footnote{For constant fields  the spatial manifold  reduces to
  one point   and the metric takes  the   form used  in  the preceding
  sections for our simple models.}  and the  trivial metric on   $S^1$
\cite{babelon:81}
\begin{equation}
  g_{ij}^{ab}(x,y) = \delta_{ij}\: \delta^{ab} \: \delta(x-y) \; .
  \label{Gl-507a}
\end{equation}
Furthermore, the canonical Poisson brackets are
\begin{equation}
  \big\{ A_1^a(x,t), E^{1b}(y,t) \big\}= \delta^{ab}\: \delta(x-y) \;.
  \label{Gl-507}
\end{equation}
%
%
\subsection{Coulomb gauge}
%
%
As usual, the Gau{\ss}   constraints ${\mathcal G}^a$   generate small
gauge transformations with $ n\!   = \!  0$ (\ref{Gl-504a}), where the
parameters $\varphi^a(x)$ now only depend on the space coordinate $x$.
We denote the    group of time-independent gauge  transformations   as
${\mathcal G}_0$.  Due to  the boundary conditions  on the elements of
${\mathcal G}_0$, it  is not  possible to transform  $A_1$  to zero as
would be the case, if we had chosen the  entire real line $\mathbb{R}$
as the spatial manifold \cite{langmann:94,lenz:94}.  All  we can do is
to require $A_1$ to be constant in space by imposing the Coulomb gauge
condition
\begin{equation}
  \chi^a_{\scriptscriptstyle \text{C}}(\tilde A_1) = 
  \partial_1 \tilde A_1^a = 0 \; , \quad \tilde A_1^a =: a_1^a \;.
  \label{Gl-508}
\end{equation}
This gauge condition can also be formulated in terms of the distance
functional
\begin{equation}
  \label{Gl-508a}
  F_{A_1}[U] := d^2(0,\, ^U{\!A_1}) = 
  \int\limits_0^L \mathrm{d}x \: 
  \text{tr} \left(\, ^U{\!A_1} \, ^U{\!A_1} \right) \;.
\end{equation}
Requiring  extrema  of  (\ref{Gl-508a})  to  be at  $U  \!    = \!   1
\!\!\!\:\text{l}$, the linear term in  an expansion of $F_{A_1}[U[\vec
\varphi(x)]]$ in terms of infinitesimal $\varphi^a$ yields the Coulomb
gauge   condition   \cite{franke:82}.    We   parameterize the   three
dimensional    gauge    fixing   surface   $\Gamma_{\scriptscriptstyle
  \text{C}}$   determined by   (\ref{Gl-508})  with  the   coordinates
$a_1^a$.  Like in   the models discussed  before  we  write  the gauge
fixing  transformation  as  a   transformation  from   the   Euclidean
coordinates  $A_1^a(x)$ to  gauge variant   and invariant  coordinates
$a_1^a$ and $\varphi^a(x)$
\begin{equation}
  A_1^a \big[ a_1^a, \varphi^a(x) \big] = 
  a_1^b \, \left( U^\dagger [\, \vec \varphi(x)] \right)^{ba}
  - u_1^a \left[\, \vec \varphi(x) \right] \; .
  \label{Gl-509}
\end{equation}
Likewise, looking   for  residual gauge symmetries  we   have  to find
solutions to the equation
\begin{equation}
  \partial_1 \! \left( ^{U[\vec \varphi]} a_1 \right)^a =
  a_1^b \, \partial_1 \left( U[\vec\varphi \,] \right)^{ba} +
  \partial_1 u_1^a [ \vec \varphi \,] = 0 \; .
  \label{Gl-510}
\end{equation}
If we   consider  (small)  gauge transformations  with  infinitesimal,
$\varphi^a$
\begin{equation}
  \left( ^{U[\vec \varphi]} a_1\right)^a \cong 
  a_1^a(x) - \epsilon^{abc} \, \varphi^b(x) \, a_1^c(x)
  + \partial_1 \varphi^a(x) 
  = a_1^a(x) - D_1^{ab}[a_1] \varphi^b(x) \; ,
  \label{Gl-510a}
\end{equation}
equation (\ref{Gl-510}) becomes
\begin{equation}
  - \partial_1 D_1^{ab}[a_1] \varphi^b(x) = 0 \; ,
  \label{Gl-510b}
\end{equation}
which   is  nothing but  the  equation   for  the   zero modes  of the
Faddeev-Popov          operator         $\text{FP}_{\scriptscriptstyle
  \text{C}}^{ab}(x,  y)  = \{ \chi_{\scriptscriptstyle \text{C}}^a(x),
{\mathcal G}^b(y) \}$ for the Coulomb gauge \cite{gribov:78}. In terms
of    the distance functional    (\ref{Gl-508a}),  we  can demand  the
configuration  $A_1$  to be at a  local  minimum for  $U  \!   = \!  1
\!\!\!\:\text{l}$.   This  amounts   to  the   requirement,   that the
coefficient matrix of the second term in the expansion of $F_{A_1}[U]$
has   to  be  positive   definite.   Actually,  this   matrix   is the
Faddeev-Popov   operator,  so  that  local  minima  are determined  by
requiring FP to be  positive  definite.  Hence, the zero mode equation
(\ref{Gl-510b})  defines the  boundary  of the  set of  configurations
$A_1$ which   are  at a   local minimum  of  the  distance  functional
(\ref{Gl-508a}).

Obvious solutions of    condition (\ref{Gl-510b}) are  given by  gauge
transformations  with constant $\vec   \varphi$,   which act  on   the
remaining degrees  of  freedom, $a_1^a$,  as  pure rotations.  For the
moment, however, we will not  consider constant gauge transformations,
but leave  them as global gauge symmetries.   As a consequence, in the
expansion (\ref{Gl-504b}) we  discard  the Fourier  coefficients  with
$k\! = \! 0$.

Let   us   look  for  other   solutions   of   the  zero mode equation
(\ref{Gl-510b}).  Take for  example $a_1 = r  \sigma^3 /2$ with $r \in
\mathbb{R}$, then we have the general solution
\begin{equation}
  \label{Gl-511}
  \begin{matrix}
     \quad \\  \vec \varphi(x) \\ \quad
  \end{matrix}  =
  \begin{pmatrix}
    c_1 \: \cos(r\,x) +  c_2 \: \sin(r\,x)  \\
   -c_1 \: \sin(r\,x) +  c_2 \: \cos(r\,x)  \\
    c_3 \: x
  \end{pmatrix} \;,
\end{equation}
where we have already  omitted the integration constants corresponding
to constant $\vec \varphi$. One might be tempted to conclude, that for
any   value of     $r$,   FP[$r$]     has  a   zero mode,   so    that
$\bigtriangleup_{\scriptscriptstyle   \text{FP}}     $  should  vanish
everywhere  on the  gauge  fixing surface  $\Gamma_{\scriptscriptstyle
  \text{C}}$.  However, we have to  take into account  that FP acts on
infinitesimal parameter functions  $\vec \varphi(x)$  only. Therefore,
the solutions  (\ref{Gl-511}) have to  be periodic in $x$, which leads
to the conditions $c_3 \!  = \!  0$ and $r = 2\pi\,k / L$ with integer
$k$.  Hence,   the Faddeev-Popov  operator   only has zero  modes  for
certain  values of $r$, and   we expect the corresponding  determinant
$\bigtriangleup_{\scriptscriptstyle  \text{FP}}      $  to  be nonzero
everywhere else.

\begin{figure}
  \begin{center}
    \leavevmode
    \epsfig{file=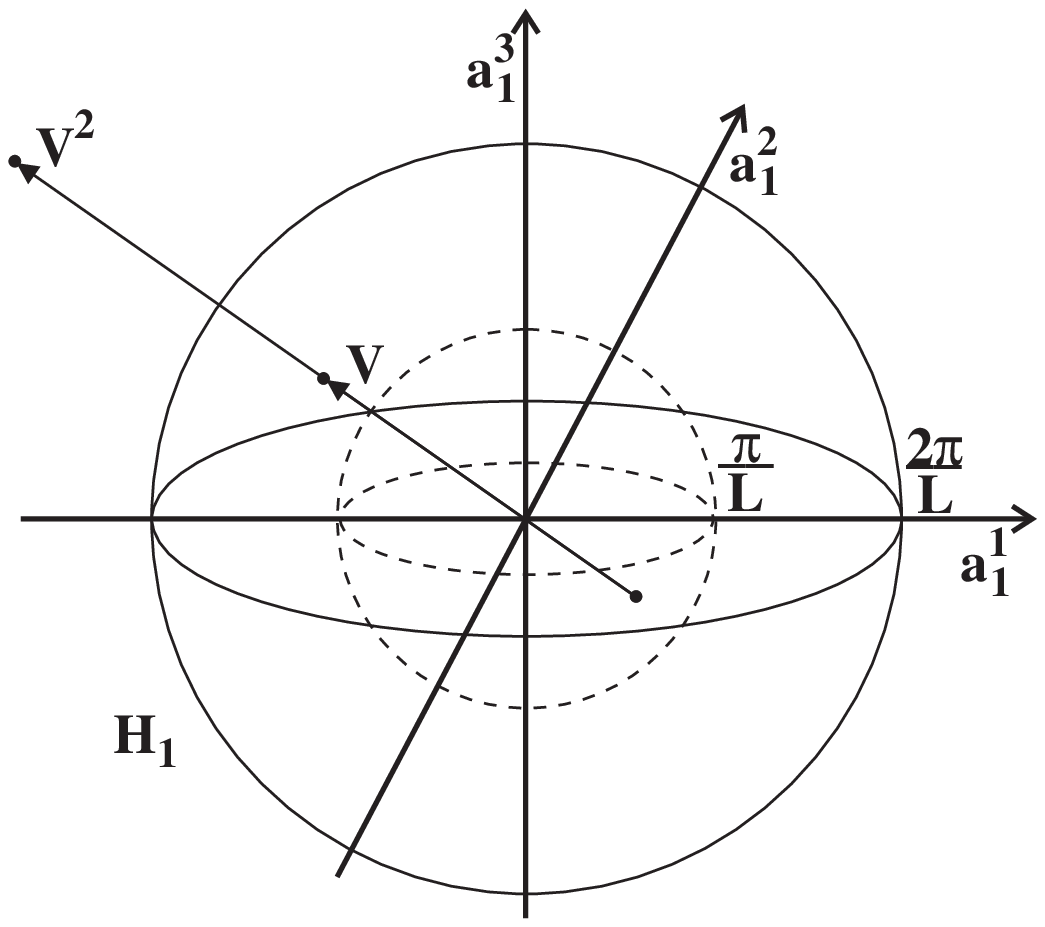,height=60mm}
  \end{center}
  \caption{The gauge fixing surface $\Gamma_{\text c}$ with the first 
    Gribov horizon ${\mathcal  H}_1$ and the discrete shifts generated
    by the large gauge   transformation $V$. Eliminating large   gauge
    transformations     completely     would   require  to    restrict
    $\Gamma_{\scriptscriptstyle \text{C}}$ to the ball with the dotted
    sphere as its boundary.}
  \label{fig-41}
\end{figure}

Instead  of  calculating other  solutions  of (\ref{Gl-510b}),  let us
determine   the  Jacobian   of   the    gauge  fixing   transformation
(\ref{Gl-509}) for infinitesimal  $\varphi^a(x)$. This  is most easily
done by plugging the Fourier expansions (\ref{Gl-504b}) and
\begin{equation}
  A_1^a(x) = \sum_k \bar A_k^a \; 
  \mathrm{e}^{2\pi \mathrm{i} \,k\frac{x}{L}} \;.
  \label{Gl-512}
\end{equation}
into the transformation (\ref{Gl-509}) and keeping only terms linear
in $\bar \varphi^a_k$,
\begin{equation}
  \bar A_k^a = a_1^a \, \delta_{k0} + 
  \epsilon^{abc} \, a_1^b \, \bar \varphi^c_k 
  + \mathrm{i} \, k \, \frac{2\pi}{L} \, \bar \varphi^a_k \; .
  \label{Gl-514}
\end{equation}
To avoid a trivial zero of the Jacobian due to the continuous residual
gauge symmetry  generated  by constant  $U$'s,  we omit  the  constant
parameters $\bar \varphi^a_0$ in the Jacobian matrix
\begin{equation}
  J = \frac { \partial (\ldots, 
    \bar A_{-1}^a, \bar A_0^a, \bar A_1^a, \ldots ) } 
    { \partial (\ldots, \, 
    \bar \varphi_{-1}^a, a_1^a\,, \bar \varphi_1^a \,, \ldots ) } \;.
  \label{Gl-515}
\end{equation}
The resulting $J$ has block structure with sub-matrices
\begin{equation}
  J_{k\neq 0}^{ab}\Big|_{\bar \varphi^a_k = 0} = 
   - \epsilon^{abc} \: a_1^c + \mathrm{i} \, k \, \frac{2\pi}{L}\, 
   \delta^{ab} \quad \text{and} \quad  J_0^{ab} = \delta^{ab} \; .
  \label{Gl-516}
\end{equation}
Therefore the Jacobian factorizes as
\begin{align}
  & \det J = \prod_k \det J_k \quad\text{with}\\
  & \det J_0=1 \quad \text{and} \quad \det J_{k\neq 0} = 
    \left( \mathrm{i} \, k \, \frac{2\pi}{L} \right)
  \left(\vec a_1 \! \cdot \vec a_1 - 
    \left(\frac{2\pi}{L} k \right)^2 \right) \; . \nonumber
  \label{Gl-517}
\end{align}
Note, that $\det   J$  is indeed  proportional  to   the Faddeev-Popov
determinant  $\bigtriangleup_{\scriptscriptstyle \text{FP}}$, which in
Fourier space has the form
\begin{equation}
  \bigtriangleup_{\scriptscriptstyle \text{FP}} =
  \prod_{k>0} \left( k \, \frac{2\pi}{L}  \right)^8
  \left( \vec a_1 \! \cdot \vec a_1 
       - \left(\frac{2\pi}{L} k \right)^2 \right)^2 = 
  \det J \cdot \prod_{k\neq 0} 
       \left( \mathrm{i} \, k  \, \frac{2\pi}{L} \right)^3
  \label{Gl-517a}
\end{equation}
corresponding to  the factorization $\det  \partial D =  \det \partial
\cdot    \det    D$.     We     find     that     $\det      J$    and
$\bigtriangleup_{\scriptscriptstyle \text{FP}}$ vanish whenever
\begin{equation}
  |\,\vec a_1|^2 = \left(\frac{2\pi}{L}\:k\right)^2 \; .
  \label{Gl-518}
\end{equation}
In particular for $a_1 =  r \sigma^3 /2$ we  recover the zeros at $r =
2\pi \, k / L$ due to the  zero modes (\ref{Gl-511}).  By analogy with
the  models studied in  the preceding  sections  we expect  the Gribov
horizons $H_k$, defined by $\det  J=0$ and labeled  by an integer $k$,
to  separate different Gribov regions which  are  related via discrete
residual gauge symmetries.  Graphically we can represent the situation
as  in  Fig.  \ref{fig-41},  where  we  have  identified the remaining
variables  $a_1^a$  with the coordinate  axes  of $\mathbb{R}^3$.  The
Gribov horizons  $H_k$   form an infinite  set of   concentric spheres
around    the   origin       of    the     gauge    fixing     surface
$\Gamma_{\scriptscriptstyle \text{C}}$ with the  first one  located at
$|\vec a_1| = 2\pi/L$. Like in Sections  \ref{a2} and \ref{a3}, we can
restrict    the   gauge   fixing surface   $\Gamma_{\scriptscriptstyle
  \text{C}}$ to one Gribov region by demanding
\begin{equation}
  \label{Gl-518a}
  |\vec a_1| < \frac{2\pi}{L} \;.
\end{equation}
Note  that, contrary to discussions  of  the Coulomb  gauge in  higher
dimensions     \cite{franke:82} via      the  distance      functional
(\ref{Gl-508a}),       the  set      of     configurations   $a_1  \in
\Gamma_{\scriptscriptstyle        \text{C}}$         defined        by
$\bigtriangleup_{\scriptscriptstyle \text{FP}}  > 0$ is not connected.
In           particular,          since        from    (\ref{Gl-517a})
$\bigtriangleup_{\scriptscriptstyle  \text{FP}} \geq 0$ everywhere  on
$\Gamma_{\scriptscriptstyle \text{C}}$, it is not sufficient to define
a  fundamental  region   via       $\bigtriangleup_{\scriptscriptstyle
  \text{FP}} > 0$.   The reason is, that  we have to take into account
the domain on which the  Faddeev-Popov  operator acts. Thus, from  the
general solutions of (\ref{Gl-510b}) we  may  only accept those  being
periodic in the spatial  coordinate $x$.  Ultimately, this  is related
to  the nontrivial  topology of  the   spatial manifold  $M=S^1$.  For
$M=\mathbb{R}$, all solutions to  (\ref{Gl-510b}) would be admissible,
so  that  $\bigtriangleup_{\scriptscriptstyle  \text{FP}} = 0$  on the
entire gauge  fixing  surface $\Gamma_{\scriptscriptstyle  \text{C}}$,
indicating that in this case the constant fields $a_1$ could be gauged
away completely.

Let us discuss  the residual gauge symmetries  in detail.  Remembering
the product  representation (\ref{Gl-504a}) of  a general  gauge group
element $U \in \mathcal G$, we can study the action of the large gauge
transformation  $V(a_1) = \exp( \mathrm{i}\,  \pi \, \hat a_1 \! \cdot
\! \vec  \sigma \,  x/L)$ and powers   $V^n$ thereof  on  an arbitrary
configuration $a_1   \in \Gamma_{\scriptscriptstyle \text{C}}$,  $\hat
a_1$ being the unit vector in the direction of $\vec a_1$.  We obtain
\begin{equation}
  \label{Gl-519}
  \left(^{V^n}{\!a}_1\right)^a = 
   a_1^b \: \left( V^n [\vec \varphi\,] \right)^{ba} + v_1^a = 
   a_1^a - \frac{2\pi}{L} \, \hat a_1^a \; .
\end{equation}
Hence, the transformed configuration, $^{V^n}{\! a_1}$, also satisfies
the gauge condition (\ref{Gl-508}),  and therefore provides a solution
to equation (\ref{Gl-510}) for  finite gauge transformations.  For $n$
even, $V^n$  is a small  gauge transformation and generates a discrete
residual gauge  symmetry,  which  translates  the configuration  $a_1$
along its direction  in color space by multiples  of $4\pi/L$.   Thus,
any configuration $a_1$ in one Gribov region has a gauge copy in every
other Gribov region,  which proves   that  all the Gribov  regions are
homeomorphic to each other.

For $n \!   = \! 1$ the  large gauge transformation $V(a_1)$ yields an
additional  symmetry     relating     two    different  configurations
\emph{within} a Gribov region to  each other (cf.  Fig. \ref{fig-41}).
In addition, under large   gauge transformations the   Gribov horizons
contain   gauge   copies   of    the   classical  vacuum   $a_1\!=\!0$
\cite{vanbaal:92a}. In order to  eliminate this residual symmetry, one
might be tempted to reduce the  gauge fixing surface further to $|\vec
a_1|  < \pi/L$.   Note,   however, that  Gau{\ss}' law   only requires
invariance  of  the wave function  under  small gauge transformations,
whereas in the  case of large gauge  transformation the  wave function
may  transform in  an   arbitrary representation of the  corresponding
symmetry group  \cite{chandar:94}.  So for the moment  let us stick to
the reduced gauge fixing  surface defined by (\ref{Gl-518a}).  We will
come back to this problem later on.

Let us have  a closer look at   the Gribov horizon $H_1$,  which again
seems to constitute  a  boundary of  the physical configuration  space
$\mathcal M$.  Since  the consecutive application  of  two large gauge
transformations $V(a_1)$ results  in a small  gauge transformation, it
is straightforward to construct the  expected residual gauge  symmetry
on the Gribov horizon.  Given  two configurations $\vec a_1$ and $\vec
a_1'$ with  magnitudes $2\pi/L$  we may transform   $a_1'$ to zero via
$V(a_1')$ and subsequently apply the gauge transformation $V(-a_1)$ to
generate a shift  from the origin to  the horizon configuration $a_1$.
Therefore, for any  two configurations  on  the Gribov  horizon, there
exists  a local small  gauge transformation, transforming one into the
other. As a consequence, all points on $H_1$ are  gauge copies of each
other. Analogously to  the models discussed   before, we can  fix this
continuous residual gauge  symmetry by  choosing an additional   gauge
condition  on  the Gribov  horizon, which  amounts to  identifying all
points on the  horizon with one point.   In fact, it is precisely this
boundary identification  that    gives the  configuration  space   the
topology  of $S^3$  or $SU(2)$, which  is  the structure group for the
case of   small gauge  transformations.  We  have  thus  recovered the
well-known result, that  the configuration  space of pure   Yang-Mills
theory  on a cylinder is  the structure group itself \cite{rajeev:88}.
Denoting elements of $SU(2)$ by $\exp(\mathrm{i} \, \vec a_1 \!  \cdot
\! \vec  \sigma/2)$,  we have in fact   introduced a chart (coordinate
system)   on  $SU(2)  \cong   S^3$,    whose coordinate   neighborhood
\cite{nakahara:92} is just the   central Gribov region $|\vec  a_1|  <
2\pi/L $. The point $ - 1\!\!\!\: \text{l} \in SU(2)$, however, is not
covered  by  this chart  indicating the  need  for a   second one.  In
contrast to the simple models discussed before,  it is not possible to
cover the configuration space with one chart only.

If  we  had required $|\vec  a_1|  <  \pi/L$ to eliminate  large gauge
transformations,  there would  only be   a  discrete  symmetry  on the
boundary at $|\vec a_1| = \pi/L$, relating  antipodal points $a_1$ and
$-a_1$ via the shift generated  by $V(a_1)$. Identifying these points,
the  configuration    space once  again   would  become  topologically
nontrivial.   In this case, the physical configuration space $\mathcal
M$ would be homeomorphic to $SO(3)$,  the relevant structure group for
large gauge transformations.

Since there is still the   global  symmetry with respect to   constant
gauge  transformations,    we  have  to  demand  appropriate  symmetry
conditions $\Psi(a)  =   \Psi(^U{\!a})$  on  the wave    functions  as
discussed before.   In the sequel we would  obtain for the Hamiltonian
on the configuration space $SU(2)$ the corresponding quadratic Casimir
operator   with       its    well-known      eigenvalues    $j\,(j+1)$
\cite{rajeev:88,marinov:79}.  Nevertheless, since we are interested in
the physical configuration  space ${\mathcal M}={\mathcal A}/{\mathcal
  G}$,  we   will  now   proceed    by  dividing  out   global   gauge
transformations, too.
%
%
\subsection{Constant gauge transformations}
%
%
So far, the configuration  space is, whatever gauge transformations we
admit,   a   smooth manifold.     Eliminating  also   constant   gauge
transformations, we anticipate the appearance  of singular points, due
to the existence  of non-generic configurations.   This is most easily
seen by applying a constant gauge transformation $U_{\text{c}}$ to the
origin $a_1\!=\!0$.  Since $U_{\text{c}}$  acts as a pure rotation, it
leaves the origin invariant, as it was the case  in our simple quantum
mechanical ``$d   \times r$-models''.  There are,  however, additional
non-constant elements of the stability group of  the origin, which may
be found with the help of the large gauge transformation $V$. Applying
$V(-a_1)$ to the origin, we get to the point $a_1$ on the first Gribov
horizon $H_1$  (cf.  Fig.  \ref{fig-41}).    Then  we can  rotate  the
configuration $a_1$ to another $a_1' \in {\mathcal  H}_1$ via a global
gauge transformation $U_{\text{c}}$ and  shift back to the origin with
$V(a_1')$.   Hence,  the  product  $V(a_1')\, U_{\text{c}}\,  V(-a_1)$
leaves the origin invariant.  Since it  is in general non-constant and
belongs  to the  class of  small  gauge  transformations, we  can thus
generate a  large  number of  elements of the  stability group  of the
origin.

Using  a    similar construction, it    is  also easy   to   see, that
configurations $a_1$ on  an arbitrary Gribov horizon ${\mathcal  H}_k$
are  invariant  under  local    gauge transformations   of  the   form
$V^k(a_1)\, U_{\text{c}}\, V^{-k}(-a_1)$, where $U_{\text{c}}$ is once
again   a   global gauge   transformation.    In   fact, these   gauge
transformations are  the   zero modes  of  the  Faddeev-Popov operator
corresponding to  solutions of equation  (\ref{Gl-510b}). For example,
choosing $a_1 = r   \sigma^3/2$ and a global   rotation $U_{\text{c}}$
with angles $\varphi^1=c^1$ and  $\varphi^2=c^2$ we recover the result
(\ref{Gl-511}). We notice that for  the Coulomb gauge, the zero  modes
of the Faddeev-Popov operator correspond  to elements of the stability
group  of configurations on the Gribov  horizons.  From our experience
from Sections \ref{a2}   and  \ref{a3} we expect that,    after having
eliminated  constant    gauge  transformations,  these   (non-generic)
configurations  will   become singular     points  of  the    physical
configuration space $\mathcal M$.

\begin{figure}
  \begin{center}
    \leavevmode
    \epsfig{file=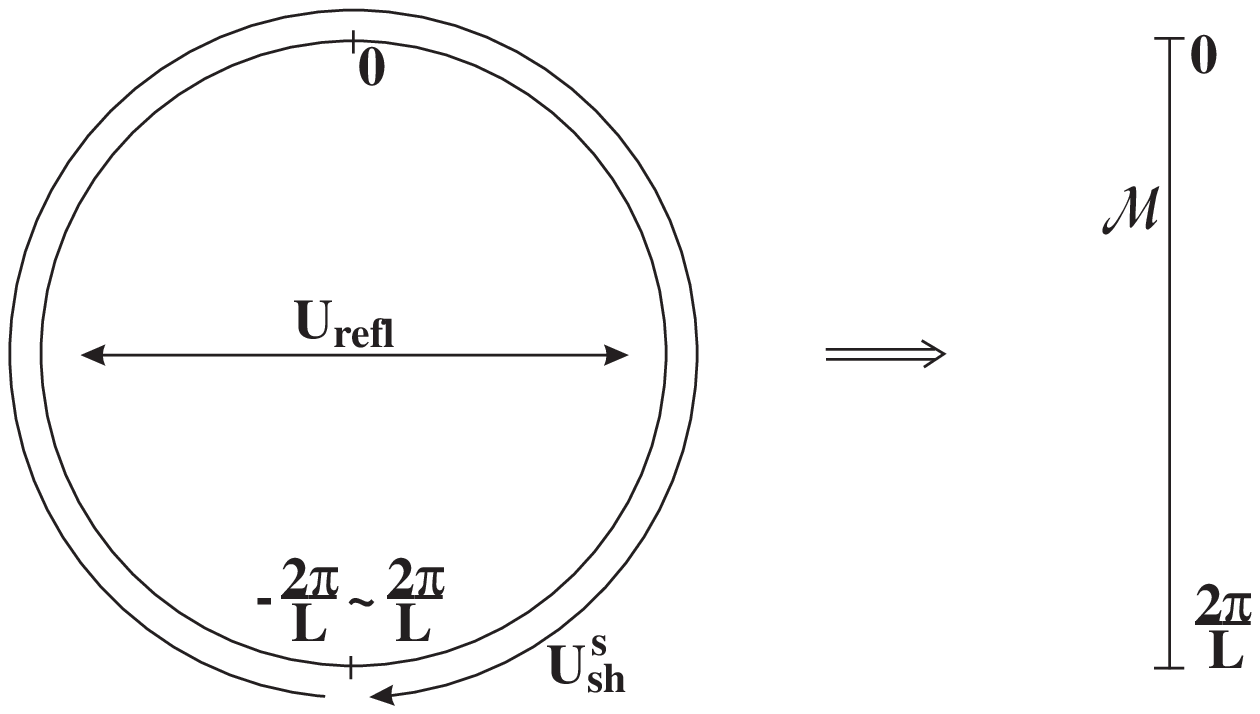,height=40mm}
  \end{center}
  \caption{Construction of the physical configuration space 
    $\mathcal M$ from the  reduced gauge fixing  surface $|\vec a_1| <
    2\pi/L$  and  the gauge   conditions (\ref{Gl-521}),  taking  into
    account the necessary identifications due to residual symmetries}
  \label{fig-42}
\end{figure}

So let us consider constant gauge transformations of the form
\begin{equation}
  a_1^a \mapsto \tilde a_1^a = a_1^b \: U^{ba} \;.
  \label{Gl-520}
\end{equation}
We will fix the corresponding gauge symmetry by demanding $\tilde a_1$
to be diagonal:
\begin{equation}
  \chi^1(\tilde a_1) = \tilde a_1^1 = 0 \quad \text{and} \quad 
  \chi^2(\tilde a_1) = \tilde a_1^2 = 0 \; . 
  \label{Gl-521}
\end{equation}
We observe that there is still a gauge symmetry left, corresponding to
rotations around    the $\hat  e_3$-axis.    However, since \emph{all}
configurations $\tilde a_1=(0,0, \tilde a_1^3 \!=:\! r)$ in  the gauge
(\ref{Gl-521})  are (at least)  invariant under these transformations,
the residual $SO(2)$ symmetry  has no further observable consequences.
Calculating the Jacobian of the infinitesimal transformation
\begin{equation}
  a_1^a = r \: \delta^{a3} + r \: \epsilon^{a3b} \, \bar \varphi_0^b
  \label{Gl-522}
\end{equation}
we obtain
\begin{equation}
  \left| \frac{\partial(a_1^1,a_1^2,a_1^3)}
  {\partial (\bar\varphi_0^1 ,\bar\varphi_0^2 , r) } \right|
  = r^2 \; .
  \label{Gl-523}
\end{equation}
Note, that (\ref{Gl-522}) can again be interpreted as a transformation
to the  gauge  invariant variable $r \!=\!|\vec  a_1|$  and the  gauge
variant $\bar\varphi_0^1$ and $\bar\varphi_0^2$.   Like in the quantum
mechanical problems discussed before we  find  a zero of the  Jacobian
for $r\!  = \!  0$ and a  discrete  residual symmetry  transforming $r
\mapsto -r$ (Weyl-reflections \cite{shabanov:93a,marinov:79}),   which
we eliminate by  demanding  $r\geq 0$.  Now   that we have  taken into
account all residual gauge symmetries, how does the resulting physical
configuration space   $\mathcal M$  look  like?  The  gauge  condition
(\ref{Gl-521}) restricts the configuration  space to a circle, due  to
the identification   of   $r=2\pi/L$   with $r=-2\pi/L$  (cf.     Fig.
\ref{fig-42}).
Identifying the points  $r$ and $-r$   reduces the circle to  a closed
interval, and in  the end we  have ${\mathcal  M}=[\,0,2\pi/L\,]$.  As
expected, the zero configuration and the configuration  with $r = 2\pi
/ L$ constitute the boundary of $\mathcal M$, as their stability group
is  larger  than   the $SO(2)$   of  the   other configurations.   The
configuration with $r=  2\pi / L $, for  instance, is  invariant under
the combination of a translation $r \mapsto r -  4\pi / L  = -r$ and a
reflection $-r\mapsto   r$.  If we   had  restricted the configuration
space  to the ball with  $|\vec a_1| <  \pi/L$, we would have obtained
${\mathcal M} = [\, 0,\pi/L \,]$, where the boundary points once again
have a larger stability group, when shifts $r \mapsto  r - 2\pi/L$ are
included.
%
%
\subsection{Quantum mechanics}
%
%
Let us turn  to  the quantum theory.  We  quantize the  theory on  the
Euclidean pre-configuration space ${\mathcal A}_0$ via
\begin{equation}
  \left[ \hat E^{1a}(x,t) , \hat A_1^b(y,t) \right] = 
  -\mathrm{i} \: \delta^{ab}\, \delta(x-y)
  \label{Gl-524}
\end{equation}
and    represent the  states  as   Schr\"odinger  wave functionals  on
${\mathcal A}_0$
\begin{equation}
  \hat E^{1a}(x) \, |\Psi\rangle \rightarrow 
  \frac{\delta}{\mathrm{i}\, \delta A_1^a(x)} \, \Psi(A) \; .
  \label{Gl-525}
\end{equation}
Thus the Hamiltonian has the form
\begin{equation}
  \hat H = -\frac{g^2}{2} \int_0^L \mathrm{d}x \: 
  \frac{\delta}{\mathrm{i}\, \delta A_1^a(x)}
  \frac{\delta}{\mathrm{i}\, \delta A_1^a(x)}
  = -\frac{g^2}{2} \bigtriangleup_{{\mathcal A}_0} \; ,
  \label{Gl-526}
\end{equation}
which, after Fourier transforming the fields becomes
\begin{equation}
  H = - \frac{g^2}{2L}\sum_k 
  \frac{\partial^2}{\partial \bar A_k^a \: \partial \bar A_k^a} \; .
  \label{Gl-528}
\end{equation}
The physical states are determined by Gau{\ss}'s law
\begin{equation}
  \hat G \: | \Psi \rangle_{\text{phys}} = 0 \; .
  \label{Gl-527}
\end{equation}
To  shorten  the discussion,  we combine transformation (\ref{Gl-522})
with (\ref{Gl-509}) to represent every  configuration $A$ by the gauge
invariant variable $r$ such  that the gauge  conditions (\ref{Gl-508})
and  (\ref{Gl-521})  are  satisfied.     In Fourier   components   the
infinitesimal gauge (fixing) transformation is
\begin{equation}
  \bar A_k^a = 
  \frac{2}{L}\, \vartheta \: \delta^{a3} \delta_{k0} + 
  \frac{2}{L}\, \vartheta  \: \epsilon^{a3b}\: \bar \varphi_k^b
  + \mathrm{i} \, k \: \frac{2\pi}{L} \, \bar \varphi_k^a\;,
  \label{Gl-529}
\end{equation}
where we have rescaled the variable $r$ to $\vartheta = r\,L/2$. Note,
that the angle $\bar \varphi^3_0$ does  not show up in (\ref{Gl-529}),
due to the residual gauge symmetry $SO(2)$, the common stability group
of  all configurations. The   Jacobian at $\bar \varphi^a_k=0$ can  be
expressed as
\begin{equation}
  \det J  := \left| \frac{\partial 
   (\ldots, \bar A_{-1}^a, \bar A_0^a, \bar A_1^a, \ldots) }
   {\partial (\, \bar \varphi^a_{k\neq 0} \,, \:
   \bar \varphi_0^1\,, \:\bar \varphi_0^2\,, \, \vartheta\,)} \right|
 = \sin^2\!\vartheta \: \left( \frac{2}{L} \prod_{k>0}
   \left(\frac{2\pi}{L} \, k \right)^{\! 2} \right)^{\! 3}  \! \! , 
  \label{Gl-530}
\end{equation}
using  the representation of $\sin \vartheta$  as  an infinite product
\cite{gradshteyn:81}.  As  before, we  find  that on  the gauge fixing
surface   $\Gamma_\chi=\mathbb{R}$, defined by   the gauge  conditions
(\ref{Gl-508})  and (\ref{Gl-521})  and  parameterized by $\vartheta$,
there  are zeros of the  Jacobian  at $\vartheta  =  z\,\pi \;\; (z\in
\mathbb{Z})$ corresponding to an infinite set of Gribov horizons.  The
Gribov regions are related by  discrete (small) gauge  transformations
$\vartheta   \mapsto  \vartheta   -  2\pi$   and  $\vartheta   \mapsto
-\vartheta$, which can  be  eliminated by restricting  $\vartheta$  to
obtain ${\mathcal  M}= [\,0, \pi\:]$.  Note,  that $\mathcal M$ always
has a boundary at  multiples of $\pi$,  since these points have larger
stability group  than a generic  configuration.  We also observe that,
like in  the models  discussed in Section   \ref{a2} and \ref{a3}, the
zeros of the Jacobian  due  to non-generic configurations remain  even
after the elimination of all residual gauge symmetries.

Taking into   account  Gau{\ss}´  law  (\ref{Gl-527}),  which requires
physical  wave functions  to    be independent of  the gauge   variant
variables $\bar \varphi_i^a$, we obtain the Hamiltonian
\begin{equation}
  H_{\mathcal M} = 
  -\frac{g^2 L}{8} \left( \frac{1}{\sin^2 \! \vartheta} \: 
   \frac{\partial}{\partial \vartheta} \: \sin^2 \! \vartheta \:
   \frac{\partial}{\partial \vartheta} \right) \; .
  \label{Gl-531}
\end{equation}
The factor   $\sin^2     \! \vartheta$  stems     from the    Jacobian
(\ref{Gl-530}) and may  be re-absorbed in the  definition  of the wave
function $\Psi(\vartheta) = \psi(\vartheta) / \sin \vartheta$. In this
case, the  Schr\"odinger equation for $SU(2)$   Yang-Mills theory on a
cylinder is simply a harmonic oscillator equation
\begin{equation}
  \label{Gl-532}
  \left( \frac{\partial^2}{\partial \vartheta^2 } + \omega^2 \right)
  \psi(\vartheta) = 0 \quad \text{with} \quad
  \omega^2 = 1 + \frac{8\:E}{g^2\,L}
\end{equation}
with its general solution
\begin{equation}
  \label{Gl-533}
  \psi(\vartheta) = c_1 \: \sin (\omega \vartheta)
                  + c_2 \: \cos (\omega \vartheta) \; .  
\end{equation}
Comparing  (for   small   gauge  transformations)  the    solutions of
(\ref{Gl-533}) on the gauge fixing surface $\Gamma_\chi$ with those on
$\mathcal  M$,   we find complete  equivalence.    For instance, since
$\Psi(\vartheta)$ has to be  regular  at every point of  $\Gamma_\chi$
and $\mathcal  M$  respectively,  $\psi(\vartheta)$ has to  vanish  at
$\vartheta = z \pi$, so that we have to discard the second solution in
(\ref{Gl-533}), and $\omega$ has to be an integer. Hence, the linearly
independent physical wave functions are
\begin{equation}
  \label{Gl-534}
  \Psi_n(\vartheta) = N \:\frac{\sin(n\vartheta)}{\sin \vartheta}
  \quad (n \in \mathbb{N})
\end{equation}
with corresponding energies
\begin{equation}
  \label{Gl-535}
  E_n = \frac{g^2 L}{8} \: (n^2 - 1)\;.
\end{equation}
The  solutions  (\ref{Gl-534})   automatically  satisfy  the  symmetry
conditions    $\Psi(\vartheta) =      \Psi(\vartheta-2\pi)$        and
$\Psi(\vartheta) = \Psi(-\vartheta)$,  needed to take into account the
residual symmetries  on $\Gamma_\chi$.  The normalization constant $N$
is  easily calculated on  ${\mathcal   M}  = [\,0,\pi\:]$,  using  the
measure which follows   from  the Jacobian  (\ref{Gl-530}),  where the
infinite volume of the gauge group must be divided out. 

Putting $j=(n-1)/2$ we  see that (\ref{Gl-535}) is indeed proportional
to the  eigenvalues $j\,(j+1)$  of  the quadratic  Casimir operator on
$SU(2)$,  where $j$ is  half-integer ($j=0,  1/2,  1, \ldots$).  Thus,
upon  including large gauge  transformations, we expect  to obtain the
same expression for  $SO(3)$, but with  $j$ being an integer.  And  in
fact, requiring  $\Psi(\vartheta)=\Psi(\vartheta-\pi)$  on  the  gauge
fixing surface  $\Gamma_\chi$ yields  the solutions (\ref{Gl-534}) for
$n$  odd,    i.e.  integer $j$.    For  the  case  of   the nontrivial
representation $\Psi(\vartheta)=-\Psi(\vartheta-\pi)$  we      get the
solutions with $n$ even. Therefore, the energy spectrum (\ref{Gl-535})
is  split into  two parts, each   representing a superselection sector
\cite{chandar:94}.  This   is  analogous to  the $\theta$-sectors   of
$U(1)$ gauge theory on the cylinder, where the group $\mathbb{Z}_2$ is
replaced   by $\mathbb{Z}$ giving rise   to  a continuous infinity  of
superselection sectors \cite{manton:85}.

Let  us  summarize the  main points  of  this section,  which has been
devoted  to  the study   of the  configuration  space of  pure $SU(2)$
Yang-Mills theory on a cylinder.  Due to the  absence of matter fields
in the  fundamental  representation,  this  model admits  large  gauge
transformations, not  connected  to unity.  Maintaining   global gauge
symmetries on the configuration  space by demanding wave  functions to
be trivial representations  of constant gauge transformations, we have
shown  how  the configuration  space can be  reduced to  the structure
group.       We    have     noticed         that      the    condition
$\bigtriangleup_{\scriptscriptstyle  \text{FP}} > 0$ is not sufficient
to pick out a connected Gribov region as one has  to take into account
the domain of $\bigtriangleup_{\scriptscriptstyle \text{FP}}$ which is
related  to the nontrivial spatial  manifold $M=S^1$.  For small gauge
transformations, the configuration space is $SU(2)$, whereas we obtain
$SO(3)$ by also dividing out large gauge transformations. Under these,
however,    the   wave   function may   transform    in  an  arbitrary
representation.    This is     most simply  implemented  by   defining
appropriate  symmetry conditions on the  wave  function defined on the
configuration space for small gauge transformations.  Note that $SU(2)
\cong S^3$ cannot be embedded  into $\mathbb{R}^3$, i.e., contrary  to
the quantum mechanical  examples  discussed in Sections \ref{a2}   and
\ref{a3}, there is no  single chart covering the entire  configuration
space.  The latter, however, is still a  manifold, and it is only when
we divide    out constant gauge   transformations  also, that singular
points  appear.   The physical configuration space   turns out to be a
closed interval, where the boundary points are singled out by having a
larger stability   group.  Notice that   the remaining one-dimensional
$SO(2)$  symmetry of  all configurations  is  directly related to  the
remaining  degree of  freedom for the  gauge  field.  As expected, the
discrete energy  spectrum  corresponds  to   the  eigenvalues of   the
quadratic Casimir operator on $SU(2)$ and  is independent of the stage
of reduction, as long as residual  gauge symmetries are implemented as
symmetry conditions on  the wave  functions.   The inclusion  of large
gauge  transformations    splits  the   spectrum  into    two  halves,
corresponding to the two possible representations of $\mathbb{Z}_2$.
%
%
\section{Conclusions}
\label{a5}
%
%
In the preceding  sections we have  studied the physical configuration
space ${\mathcal M} =   {\mathcal A}/{\mathcal G}$ of  low-dimensional
gauge  theories,  where   $\mathcal  A$     was the   space of     all
configurations, $A_\mu(x)$, and $\mathcal G$ the gauge group.  Because
there was only a finite number of gauge  invariant degrees of freedom,
we were able to  construct $\mathcal M$  explicitly and write down the
Hamiltonian in   terms   of  gauge  invariant    variables.   We  have
demonstrated    that   the  physical     quantities  are unique,  i.e.
independent of  the chosen gauge condition  or  set of gauge invariant
variables, provided that  \emph{all} (residual)  gauge symmetries  are
correctly taken into account.

\begin{figure}
  \begin{center}
    \leavevmode
    \epsfig{file=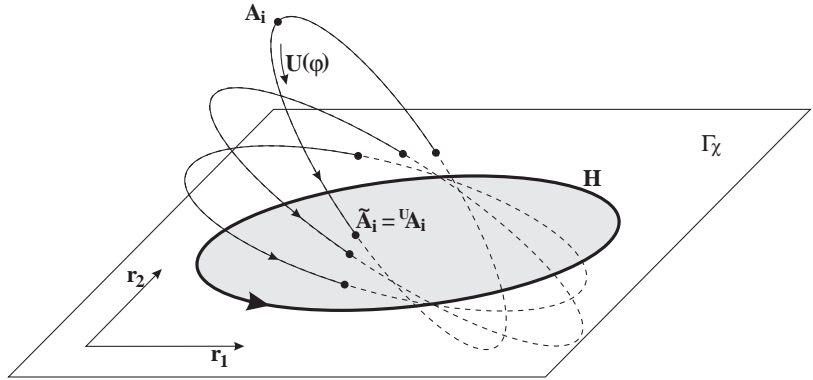,height=50mm}
    \caption{Gauge fixing surface $\Gamma_\chi$ for a fictitious 
      problem.  The shaded  area is the  reduced gauge  fixing surface
      $\tilde \Gamma_\chi$ corresponding  to  one  Gribov region   and
      bounded by a Gribov horizon $H$ consisting of one gauge orbit.}
    \label{fig-51}
  \end{center}
\end{figure}

Starting  from the  Weyl  gauge $A_0  \!   = \!   0$,  the problem  of
constructing  $\mathcal  M$  is    reduced   to the  elimination    of
time-independent gauge transformations $U  \in {\mathcal G}_0$, acting
on the pre-configuration space ${\mathcal A}_0$.  The fundamental tool
we have used  was the inverse of the  gauge fixing transformation $A_i
\mapsto \tilde  A_i     \!  =  \,^{U}\!A_i$,   mapping   an  arbitrary
configuration $A_i$ along its   orbit onto the  configuration  $\tilde
A_i$, such that $\tilde   A_i$  satisfies a suitable  gauge  condition
$\chi[\tilde  A_i\,] = 0$ (cf.   Fig.  \ref{fig-51}).  This constraint
defines  a hypersurface in  ${\mathcal A}_0$, the gauge fixing surface
$\Gamma_\chi$.     We  can  introduce    a  set  of (gauge  invariant)
coordinates $r_j$ on $\Gamma_\chi$, such that $\chi [\tilde A_i[r_j]]$
vanishes  by construction.   Taking  the inverse of   the gauge fixing
transformation, we obtain
\begin{equation}
  {\mathcal M} \times {\mathcal G}_0 \rightarrow {\mathcal A}_0 : 
  \quad (r_j, \varphi^a) \mapsto A_i[r_j, \varphi^a] := \,
  ^{U^{-1}[\varphi^a]} \! \tilde A_i[r_j] \; ,
  \label{Gl-601}
\end{equation}
where we  have parameterized the  gauge group element $U \in {\mathcal
  G}_0$ by gauge  variant time-independent angles $\varphi^a$. If this
map  was  one-to-one, we    would have constructed    a  homeomorphism
${\mathcal A}_0  \cong {\mathcal  M}  \times {\mathcal  G}_0$,  which,
according to Singer \cite{singer:78}, is not possible in general. This
is the essence of  the ``Gribov-problem''.  From (\ref{Gl-601})  it is
also easy to  see that  with every set   of gauge invariant  variables
$r_i$,  given in terms of  a map $A_i[r_j]$,  we can associate a gauge
condition $\chi$. We  just need  to describe  the surface $\Gamma  \in
{\mathcal A}_0$ parameterized via  $r_j \mapsto A_i[r_j]$ in  terms of
an equation $\chi  [A_i]  = 0$.  However,   it  may happen,  that  the
solution to $\chi  =  0$  yields  a larger hypersurface   $\Gamma_\chi
\supset \Gamma$ (cf. Section \ref{a2}).

One way to study the  map (\ref{Gl-601}), is to look  for zeros of the
corresponding Jacobian  $\det  J = | \partial    A_i / \partial  (r_j,
\varphi^a) |$, which is proportional  to the Faddeev-Popov determinant
$\bigtriangleup_{\scriptscriptstyle    \text{FP}}$   via      equation
(\ref{Gl-107}).  The zeros  of  $\det J$  form  connected sets on  the
gauge  fixing surface   $\Gamma_\chi$,   the  Gribov  horizons,  which
separate different Gribov regions  from each other.  Configurations in
one   Gribov  region are related  to   configurations  in  another via
discrete  gauge transformations such as  the shifts in (Coulomb gauge)
Yang-Mills theory on a   cylinder  (cf.  Section \ref{a4}).   One  can
eliminate this discrete  residual  gauge symmetry by restricting   the
domains of the gauge invariant  variables $r_j$, like depicted in Fig.
\ref{fig-51},   where the  shaded      area (without the    boundary!)
corresponds to one Gribov region, the ``reduced gauge fixing surface''
$\tilde \Gamma_\chi$.

For the models studied in this paper, we found that the Gribov horizon
contained residual gauge copies.  In the case of the ``$2\!  \times \!
2$''- and ``$2\!  \times \!  3$-models''  and for Yang-Mills theory on
a   cylinder (considering  small    gauge transformations  only),  the
residual symmetries relating the  gauge copies were continuous, giving
rise to complete gauge orbits on the Gribov horizon.  These continuous
residual     gauge  symmetries    have   to   be    eliminated through
\emph{additional} gauge conditions, which  imply the identification of
gauge-equivalent points   (``boundary  identifications'').    In  Fig.
\ref{fig-51}, the Gribov  horizon  consists of  one  gauge orbit only.
Choosing  one representative on  this  orbit  via  a gauge   condition
$\chi'$ amounts to the identification of all  the points of the Gribov
horizon with   this point.  It   is easy  to  see that  the  resulting
physical  configuration space $\mathcal M$  is a 2-sphere $S^2$, which
is  topologically  nontrivial, i.e.  one needs at least two coordinate
systems  (charts) to cover $S^2$.  For  instance, we can interpret the
shaded region in Fig.  \ref{fig-51}  as the coordinate neighborhood of
one chart covering  the North Pole  of  $S^2$, whereas the  South Pole
corresponding  to \emph{all}  points on  the  Gribov horizon has to be
excluded.  This is very similar to  what happens in $SU(2)$ Yang-Mills
theory on a cylinder for small  gauge transformations (ignoring global
gauge  symmetries), where     the  resulting configuration  space   is
homeomorphic     to   $S^3$.   Taking    into   account   large  gauge
transformations in addition, the symmetry relating different points on
the    Gribov   horizon    becomes    discrete.    Nevertheless,  upon
identification  of  gauge   equivalent  configurations,  the   horizon
disappears  and we  obtain  a  smooth  manifold  without boundaries or
singular points.  According to Singer \cite{singer:78}, this should be
generally true, as  long   as the gauge   group  acts  freely  on  the
configuration  space $\mathcal A$.      However, due to   the boundary
identifications, $\mathcal M$ will become topologically nontrivial.

This picture changes  drastically, when the  gauge group  does not act
freely   on  the configuration     space.    For instance,  the   zero
configuration  $A_i=0$    is   invariant  under  all  constant   gauge
transformations, so that  its stability group  is the entire structure
group.    As  we  have demonstrated   within   our  models, the   zero
configuration  becomes a singular point  of the physical configuration
space $\mathcal M$ after having divided out all gauge transformations.
In general, such non-generic configurations, i.e.  configurations with
larger stability group, form manifolds of lower dimension in $\mathcal
M$, so  that  $\mathcal M$ is  no  longer  a  manifold, but  becomes a
stratified  variety  \cite{singer:78,fuchs:96}.    In particular,  the
physical configuration space $\mathcal M$  may have a genuine boundary
of co-dimension $1$, which is  to be distinguished from the fictitious
boundary due to coordinate singularities  \cite{yabuki:96,yabuki:89a}.
Accordingly, we cannot eliminate the zeros of the Jacobian, related to
non-generic configurations.  On  the  other hand, the  zeros  stemming
from coordinate singularities are gauge (coordinate) dependent and, in
some cases (cf.  Sects.   \ref{a2} and \ref{a3}), may  even disappear.
As we have shown, however, these singular  points do not introduce any
new features into the theory.  For  example, solving our models on the
gauge fixing surface $\Gamma_\chi$  without singular points yields the
same  results as the  discussion  on the physical configuration  space
$\mathcal  M$,    if in  the first     case,  residual  symmetries are
implemented  appropriately  as   symmetry conditions    on  the   wave
functions.

It  is not clear, to  what extent these results  can be generalized to
field theoretical models.  For instance, the picture of Gribov regions
separated by Gribov horizons and  related via discrete residual  gauge
transformations   may have to be modified.    There may be more Gribov
copies \cite{vanbaal:92a}  inside the  Gribov horizon, being  not only
due to large  gauge transformations, as  was the case in  pure $SU(2)$
Yang-Mills theory  on a cylinder.  Nevertheless,  it is to be expected
that in general the physical configuration space $\mathcal M$ of gauge
field   theories  will be topologically   nontrivial,  due to boundary
identifications.  For practical calculations, this will be most easily
taken into account  by defining wave  functionals on  the gauge fixing
surface  and  implementing  the residual  gauge   freedom via symmetry
conditions on them. In  addition, the nontrivial topology of $\mathcal
M$ may modify the spectrum of the theory. For instance, in the case of
non-Abelian  gauge theories  in  $2\!   +  \!  1$ dimensions,  Feynman
argued,  that there is a  mass  gap in  the gluon   spectrum due to  a
maximal distance   on  the physical configuration   space $\mathcal M$
\cite{feynman:81}.   Another  approach  currently being  pursued is to
analyze the  influence of the nontrivial structure  of $\mathcal M$ on
the   gluon   propagator    using  the    lattice    as a    regulator
\cite{zwanziger:96}. On the  lattice, Gribov copies have been detected
numerically upon choosing    a    maximally  abelian  gauge     fixing
\cite{bali:96,hart:97} which aims at a dual superconductor scenario of
confinement.  It  is, however, not clear, how  to obtain  the physical
configuration space in this case so  that manifest gauge invariance is
still an   open  question in this   approach  \cite{shabanov:96}.  The
presented formalism might also shed some light on these issues.
\subsection*{Acknowledgement}
  The authors are  indepted  to E.~Werner  for continued interest  and
  support. They also thank S.~Shabanov for enlightening discussions.


\bibliographystyle{h-elsevier}
 
\bibliography{paper}
 
\end{document}